\documentclass[twocolumn,showpacs,preprintnumbers,amsmath,amssymb]{revtex4-1}
\usepackage{graphicx,color}
\usepackage{dcolumn}
\usepackage{bm}
\usepackage{longtable}
\usepackage{amsmath}
\usepackage{mathrsfs}
\usepackage{amsfonts}
\usepackage{textcomp}
\usepackage{esvect}
\usepackage{float}

\def\gsim {\mbox{\hbox{ \lower-.6ex\hbox{$>$}
\kern-1.12em \lower.5ex\hbox{$\sim$}\kern+.35em}}}
\def\lsim {\mbox{\hbox{ \lower-.6ex\hbox{$<$}
\kern-1.12em \lower.5ex\hbox{$\sim$}\kern+.35em}}}

\begin{document}


\title{First simultaneous measurement of sextupolar and octupolar 
        resonance driving terms in a circular accelerator from 
	turn-by-turn beam position monitors data\vspace{-0.0 cm}}

\author{A. Franchi\footnote[*]{\email{andrea.franchi@esrf.fr}}
 L. Farvacque, F. Ewald, G. Le Bec, K. B. Scheidt}
\affiliation{ESRF, Grenoble, France}

\date{\today}

\begin{abstract}\vspace{0.0 cm}
Beam lifetime in storage rings and colliders is affected by , among 
other effects, lattice nonlinearities. Their control are of great 
benefit to the dynamic aperture of an accelerator, whose 
enlargement leads in general to more efficient injection and 
longer lifetime. This article describes a procedure to evaluate and 
correct unwanted nonlinearities by using turn-by-turn beam position 
monitor data, which is an evolution of previous works on the resonance 
driving terms (RDTs). Effective sextupole magnetic errors and tilts at the 
ESRF electron storage ring are evaluated and corrected (when possible) 
by using this technique. For the first time, also octupolar RDTs could 
be measured and used to define an octupolar model for the main 
quadrupoles. 
Most of the deviations from the model observed in the sextupolar 
RDTs of the ESRF storage ring turned out to be generated by 
focusing errors rather than by sextupole errors. These results 
could be achieved thanks to new analytical formulas describing 
the harmonic content of the nonlinear betatron motion to the 
second order. For the first time, RDTs have been also used for 
beam-based calibration of individual sextupole magnets. They also 
proved to be a powerful tool in predicting faulty magnets and  
in validating magnetic models. This technique provides also a 
figure of merit for a self-assessment of the reliability of the 
data analysis.

\end{abstract}



\maketitle

\section{Introduction and motivation}
Many factors make the implementation of a magnetic 
optics in a circular accelerator  
different from the nominal one: deviations from the 
magnet calibration curves and from the ideal magnetic 
lengths, displacements from the reference position 
and axis, unknown multipole 
components, and the like. This generally results in 
machine performances below expectations: low 
beam lifetime and dynamic aperture, poor injection 
efficiency, large emittances (in lepton machines) and 
limited luminosity (in colliders). 

While an artillery of 
different methods and algorithms has been developed and 
successfully implemented  in routine operation for the 
evaluation and correction of focusing errors (linear 
optics) and betatron coupling, their extension to the 
nonlinear modelling and correction remains difficult, 
because either time-consuming or requiring diagnostic tools 
unavailable a decade ago. In most cases, such as at the ESRF 
storage ring, the correction of the nonlinear optics is 
done by trials and errors seeking heuristically longer 
lifetime. Nevertheless, the installation of beam position 
monitors (BPMs) with turn-by-turn (TbT) acquisition system in 
many circular accelerators and the parallel development 
of a theoretical formalism for the description of the 
harmonic content of the acquired data paved the way for 
more rapid and deterministic measurement and correction 
of the nonlinear optics. This paper proposes a new method 
aiming at such characteristics.

Several approaches for the evaluation of the nonlinear 
lattice model exist and  any attempt to offer a 
coverage of the pertinent literature would be incomplete. 
However, in the context 
of this paper a few works may be recalled, either for 
their proximity or because they represented milestones 
for this work. A pioneering work on the exploitation 
of TbT BPM data dates back to the early 
'90s~\cite{Caussyn}. The application of the normal 
form approach~\cite{Turchetti1,Forest1} to single-particle 
tracking data of Ref.~\cite{Bartolini1} introduced 
for the first time an explicit correspondence between 
spectral lines of TbT data and resonance driving 
terms (RDTs). A breakthrough was represented by the 
experience at the CERN Super Proton Synchrotron, 
where sextupolar RDTs along the entire ring were 
measured and used to detect faulty sextupole 
magnets~\cite{Rogelio1} and to extract strength 
and polarity of some sextupoles~\cite{Andrea-thesis,prstab_strength}. 
More recently 
independent component analysis (ICA) was applied 
to TbT BPM data for the extraction of 
lattice linear and nonlinear properties~\cite{LosAlamos}. 
Of interest are also the experimental results 
of Ref.~\cite{Bartolini2}, where the nonlinear model 
was fit to the spectral content of TbT data, 
even though not via the RDTs. 
In Ref.~\cite{RDT-PS} simultaneous 
measurements of two RDTs, one sextupolar and one 
octupolar, are reported.

The paper is structured as follows. After highlighting 
and anticipating both advantages and limitations of 
the proposed method in Sec.~\ref{MovelLimit}, the 
technique is introduced and discussed in its main 
results in Sec.~\ref{Method} (all mathematical 
derivations are put in separate appendices). The experimental 
results of the new method in evaluating the nonlinear 
model, calibrating sextupole magnets and computing 
a corrector setting are presented in 
Sec.~\ref{sec_ExpResults}. The 
evaluation of octupolar terms in the 
lattice from the same TbT data is described in 
Sec.~\ref{sec_OctDec}. 

\section{Novelties and limitations of the proposed method}
\label{MovelLimit}
The main novelties of the proposed scheme (compared 
to other methods for the measurement and correction 
of nonlinear lattice model) may be listed as follows: 
$(i)$ The harmonic analysis is performed on the 
pure position data $x$ ($y$), rather than on the 
complex signal $x-ip_x$ ($y-ip_y$) of 
Ref.~\cite{Bartolini1,Rogelio1}, hence with no  
concern about errors in the evaluation of the 
momentum $p_x$ ($p_y$) and about BPM synchronization; $(ii)$ 
The possibility of measuring at the same time 
linear combinations of all sextupolar (normal and skew) 
and most of octupolar RDTs offers a complete and simultaneous 
picture of all resonances at a given working point, 
rather than having to shift the tunes close to a single 
resonant condition to excite a specific mode, 
as in Ref.~\cite{LosAlamos}; $(iii)$  
The nonlinear problem of inferring sextupole and octupole 
strengths from the betatron beam motion is translated 
into a linear system to be inverted (with due preliminary 
precautions) when RDTs are used as observables, hence 
rendering the model fit and correction straightforward; 
$(iv)$ Last but not least, the quality of the analysis 
of sextupolar RDTs may be self-assessed, so to optimize 
the experimental conditions as well as the initial 
lattice model.

Of course, this approach suffers from some practical 
limitations too. 

First, the quality of the analysis is limited by the 
spectral resolution, here defined as the ratio 
between the amplitudes of the harmonics and the 
background noise. The resolution scales with the number of 
turns of acquired data exploitable for a Fast Fourier 
Transform (FFT). Ideally, the greater the number 
of turns with exploitable data, the higher the spectral 
resolution and hence the quality of the RDT 
measurement. Data filtering and interpolation~\cite{Laskar, TuneMeasBazzani}, 
may provide excellent resolution already with tens 
of turns, though the presence of noise~\cite{Zimmermann-book} 
and the need of detecting spectral lines whose 
amplitudes are orders of magnitude lower than the 
tune line necessitate several hundred of exploitable 
oscillation turns. 
Chromaticity and decoherence (induced by nonzero amplitude 
dependent detuning) modulate and damp the TbT signal~\cite{Chao}. 
This multi-particle effect is not contemplated 
here, the baseline model being of a beam moving rigidly 
as a single particle. Most of the operational settings 
optimized for beam lifetime and stability result in 
nonzero chromaticity and detuning with amplitude. In the case 
of the ESRF storage ring, the exploitable number of 
turns ranges from about 30 to 60 turns, depending on 
the optics put in operation, insufficient to detect 
sextupolar harmonics whose amplitude is typically 2 
or 3 orders of magnitude lower than the tune line. 
Therefore, a special optics was designed to provide 
almost zero linear chromaticity and detuning. For 
hadron machines this may be sufficient to obtain 
thousands of exploitable TbT data, as in 
Ref.~\cite{LosAlamos}. In lepton machines radiation 
damping depresses naturally the TbT signal. At the ESRF 
storage ring, the damping time being of about 2500 
turns, the signal is sufficiently depressed to 
compromise the whole measurement already after 1024 
turns (see Fig.~\ref{fig_TbT}). Usually either 256 
or 512 turns are used for the FFT, as radiation 
damping would enhance the background spectral noise. 

Second in the list of limitations is the BPM electronic 
resolution and noise. The commercial Libera 
Brilliance~\cite{Libera} BPMs installed in the ESRF 
storage ring are equipped with a standard electronic 
filter that covers several turns, resulting in TbT 
data corrupted from the neighbor turns. Even though 
a convolution may be carried out to extract clean 
TbT data, the spectral resolution remained insufficient 
for a detailed nonlinear analysis. A great improvement 
was achieved when implementing a moving-average filter 
(MAF)~\cite{MAF}. All results presented 
here are based on acquisitions carried out with this filter. 

The dedicated optics designed and implemented for 
having zero chromaticity and amplitude detuning 
sets a limit on the capability of inferring 
sextupole errors for other operational optics, 
these having different sextupole settings. A way out 
would be to evaluate the relative field error for each 
magnet and to export it in all other optics, though this 
assumes that the relative errors are constant along the 
entire range of sextupole currents (i.e. strengths). The 
situation at the ESRF storage ring turned out to be 
rather relaxed in this sense, because almost all 
sextupole errors inferred from this method lay well 
within the specified $\pm1\%$. Moreover, most of 
the deviations observed from the ideal model are 
originated by the focusing errors, through the modulated 
beta functions present in the definitions of the 
RDTs. A first correction, hence, may be carried out 
by assuming ideal sextupoles and using the best linear 
model (easy to infer with other techniques) to evaluate 
the modulated RDTs and use sextupole correctors to restore 
at the best their periodicity. 

A last limitation is linked to the linear system to be 
pseudo-inverted to extract the nonlinear model. The 
singular value decomposition (SVD) has been used for this 
purpose. The resulting model shall be then considered as 
effective and the inferred sextupole errors depend clearly 
on the numerical parameters (such as number of eigen-vectors, 
choice of weights among different possible sources of errors, 
and the like).

\section{Combined resonance driving terms from (dual-plane) BPM data}
\label{Method}

\subsection{Theoretical introduction}
\label{sec_TheoIntro}

\begin{figure}[b]
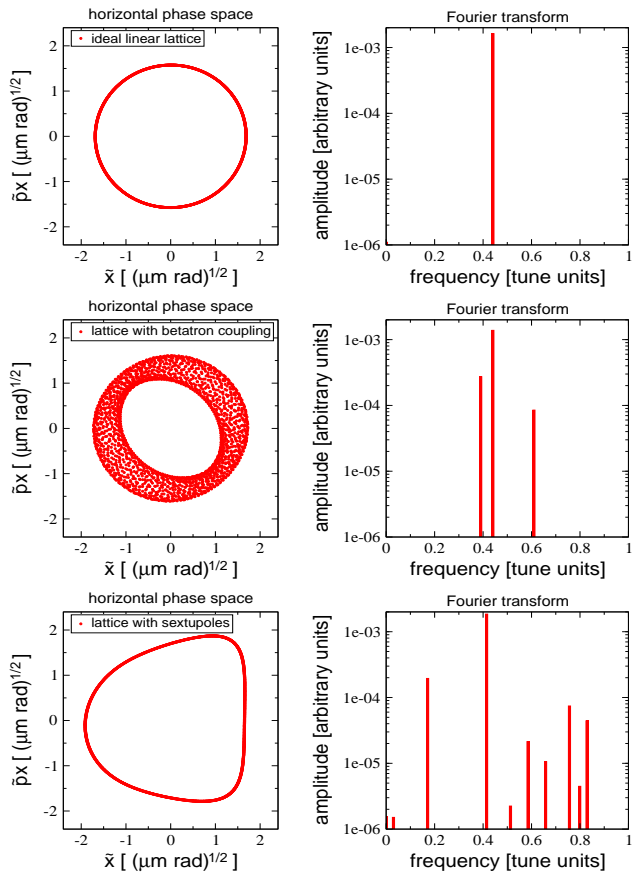
\vskip 0.4cm
  \centerline{
  \includegraphics[width=8.3cm,height=3.75cm,angle=0]
	{fig01A.eps}}\vskip 0.1cm
  \centerline{
  \includegraphics[width=8.3cm,height=3.75cm,angle=0]
	{fig01B.eps}}\vskip 0.1cm
  \centerline{
  \includegraphics[width=8.3cm,,height=3.75cm,angle=0]
	{fig01C.eps}}
  \caption{\label{fig:phsp}Three examples of phase space trajectory 
         of a displaced particle in Courant-Snyder coordinates 
         and corresponding Fourier 
         coefficients (amplitudes): ideally linear (top), with 
         betatron coupling (centre) and sextupoles (bottom). The 
         fractional parts of the tunes in this example are $Q_x=0.44$ 
         and $Q_y=0.39$.}\vskip -0.4cm
  \rule{0mm}{0mm}
\end{figure}

A charged particle circulating inside a circular accelerator 
or storage ring will execute a turn-by-turn oscillation around 
the closed orbit, if displaced transversely by a pulsed 
magnet, such as a magnet kicker. In phase space this 
oscillation describes a closed curve which reflects the 
local characteristics of the magnetic lattice. In an 
ideally linear machine this curve will be an ellipse. 
In the presence of strong sextupoles, at sufficiently large 
oscillation amplitudes the curve takes the shape of a triangle. 
With strong betatron coupling, the curves in the two 
transverse phase spaces $(x,p_x)$ and $(y,p_y)$ are 
linear combinations of the original ones. As long as the curve 
in phase space is closed and reasonably continuous (i.e. 
the motion is stable), the Fourier theorem ensures that 
it may be expressed as the sum of a series of sine or 
cosine terms (called the Fourier series), each of which 
has specific amplitude and phase coefficients known as 
Fourier coefficients. In other words, each closed curve 
may be described (after a linear transformation) as the 
superposition of circles. The harmonic content of the ideal 
ellipse in phase space will be represented by a single 
pair of amplitude and phase (the tune line): After the 
Courant-Snyder (C-S) transformation, the ellipse becomes a 
circle, whose radius corresponds to the amplitude and 
its position along the circumference is defined by the 
phase. Two additional harmonics (i.e. circles) are excited 
with betatron coupling (the tune of the other plane 
and its opposite). The triangular-shaped curve 
will instead require several harmonics to be 
properly described. The three examples are depicted 
in Fig.~\ref{fig:phsp}. Intuitively, the stronger the 
sextupoles, the more triangular is the horizontal phase 
space curve and the larger are the additional harmonics. 
It is indeed possible to correlate quantitatively (up 
to a certain precision) these deformations, and hence 
these harmonics, to the strengths of the magnets installed 
in the machine. More important, this correlation is in 
most cases linear with a precision much larger than the 
experimental resolution. The fundamental ingredient to 
establish this correlation is provided by the resonance 
driving terms (RDTs) of 
Refs.~\cite{Bartolini1,Rogelio1,Andrea-thesis,prstab_strength}. 
These may be evaluated both from analytic formulas (rather 
simple in most cases) and from the harmonic analysis 
of the phase space curves, i.e. from their Fourier 
coefficients of the TbT BPM data. In Appendix~\ref{app:1} 
the theoretical description establishing this correlation 
is recalled and extended (when needed).  The results 
of interest in the context of this paper are 
summarized in the next section.

Throughout the paper, all RDTs and machine parameters 
(linear and nonlinear) from the {\sl model} are 
evaluated by MADX-PTC~\cite{madx}. Tracking simulations 
to simulate and validate the harmonic analysis of TbT 
BPM data have been also carried out with this code. 
No significant difference has been observed in the 
RDTs when representing sextupoles either as thin 
or thick elements and when including fringe fields. 
As far as nonlinear parameters (such as amplitude 
dependent detuning and chromaticity) are concerned, 
significant differences between the three 
representations appear and only the thick-lens model 
with fringe field has been used.\vspace{-0.2cm}

\subsection{ Theoretical results}
\label{sec_CRDT}
\begin{table*}[!ht]\vskip 0.6cm
\caption{List of lines in the spectra of $\tilde{x}(N)$ and 
	$\tilde{y}(N)$ with corresponding measurable combined RDTs 
	(CRDTs) $F=|F|e^{iq_F}$ and excited resonances. 
        A horizontal (vertical) spectral 
	line $H(n_x,n_y)$ ($V(n_x,n_y)$) is located at the frequency 
	$n_xQ_x+n_yQ_y$. For each line, expressions 
	for its amplitude and phase are given. 
	The choice was made here to make use of the lines in the 
	region [0,0.5] in tune units. First-order RDTs (defined in 
	Table~\ref{tab:lattice-rdt}) are sufficient for coupling and 
        normal sextupoles, while the 
	analysis of skew sextupole terms requires a second-order 
	analysis, through the observable RDTs (ORDTs) $g_{jklm}$ 
	of Table~\ref{tab:g_jklm_SS} (The justification for such a 
	choice is given in Appendix~\ref{app:3}). Quadrupole errors 
	are to	be included in the model when computing the Courant-Snyder 
	(C-S) parameters used to evaluate $\tilde{x}(N)$, 
	$\tilde{y}(N)$ and the RDTs $f_{jklm}^{(1)}$ of 
	Table~\ref{tab:lattice-rdt}. \vspace{-0.0 cm}}
\vskip 0.0cm
\centering 
{
\begin{tabular}{c}
\end{tabular}\\
\begin{tabular}{c c l l c c}
\hline\hline\vspace{-0.25cm} \\ \vspace{0.10 cm}
                 spectral line                &
{\hskip 0.4 cm}\ amplitude{\hskip 1.2 cm}\    & 
{\hskip 0.0 cm}  phase $\phi$  {\hskip 2.4 cm}\     &
{\hskip 0.0 cm}  Combined RDT{\hskip 0.4 cm}\ &
{\hskip 0.2 cm}\ resonances {\hskip 0.2 cm}\  &
 magnetic term     \\
\hline\hline\vspace{-0.30 cm}\\\vspace{0.07 cm}
       $H(1,0)$ &$\frac{1}{2}(2I_x)^{1/2}$    &         $\psi_{x0}$                             &                                    &            & normal quadrupole \\
\hline\vspace{-0.30 cm}\\\vspace{0.07 cm} 
       $V(0,1)$ &$\frac{1}{2}(2I_y)^{1/2}$&              $\psi_{y0}$                             &                                    &            & normal quadrupole \\
\hline \\ & & & & & \vspace{-0.33cm}\\
\hline\vspace{-0.30 cm}\\\vspace{0.07 cm}
       $H(0,1)$ &    $(2I_y)^{1/2}|F_{xy}|$    &      $q_{F_{xy}}+\frac{3}{2}\pi+\psi_{y0}$      &    $F_{xy}=f_{1001}^{(1)} -f_{1010}^{(1)*}$ &(1,1),(1,-1)& skew quadrupole \\
\hline\vspace{-0.30 cm}\\\vspace{0.07 cm}
       $V(1,0)$ &    $(2I_x)^{1/2}|F_{yx}|$    &      $q_{F_{yx}}+\frac{3}{2}\pi+\psi_{x0}$      &    $F_{yx}=f_{1001}^{(1)*}-f_{1010}^{(1)*}$ &(1,1),(1,-1)& skew quadrupole \\
\hline \\ & & & & & \vspace{-0.3cm}\\
\hline\vspace{-0.30 cm}\\\vspace{0.07 cm}
     $H(-2, 0)$ &    $(2I_x)|F_{NS3}|$         &      $q_{F_{NS3}}+\frac{3}{2}\pi-2\psi_{x0}$     &    $F_{NS3}=3f_{3000}^{(1)}-f_{1200}^{(1)*}$ &(1,0),(3,0) & normal sextupole \\
\hline\vspace{-0.30 cm}\\\vspace{0.07 cm}
     $H( 0,-2)$ &    $(2I_y)|F_{NS2}|$         &      $q_{F_{NS2}}+\frac{3}{2}\pi-2\psi_{y0}$     &    $F_{NS2}= f_{1020}^{(1)}-f_{0120}^{(1)}$  &(1,-2),(1,2)& normal sextupole \\
\hline\vspace{-0.30 cm}\\\vspace{0.07 cm}
     $V(-1,-1)$ &  $(2I_x2I_y)^{1/2}|F_{NS1}|$ & $q_{F_{NS1}}+\frac{3}{2}\pi-\psi_{x0}-\psi_{y0}$&    $F_{NS1}=2f_{1020}^{(1)}-f_{0111}^{(1)*}$ &(1,-2),(1,0)& normal sextupole \\
\hline\vspace{-0.30 cm}\\\vspace{0.07 cm}
     $V( 1,-1)$ &  $(2I_x2I_y)^{1/2}|F_{NS0}|$ & $q_{F_{NS0}}+\frac{3}{2}\pi+\psi_{x0}-\psi_{y0}$&    $F_{NS0}=2f_{0120}^{(1)}-f_{0111}^{(1)}$  &(1,-2),(1,0)& normal sextupole \\
\hline \\ & & & & & \vspace{-0.33cm}\\
\hline\vspace{-0.30 cm}\\\vspace{0.07 cm}
     $V( 0,-2)$ &    $(2I_y)|F_{SS3}|$         &      $q_{F_{SS3}}+\frac{3}{2}\pi-2\psi_{y0}$    &    $F_{SS3}=3g_{0030}-g_{0012}^*$  &(0,1),(0,3) & skew sextupole \\
\hline\vspace{-0.30 cm}\\\vspace{0.07 cm}
     $V(-2, 0)$ &    $(2I_x)|F_{SS2}|$         &      $q_{F_{SS2}}+\frac{3}{2}\pi-2\psi_{x0}$    &    $F_{SS2}=g_{2010,V}-g_{0210}^*$ &(2,-1),(2,1)& skew sextupole \\
\hline\vspace{-0.30 cm}\\\vspace{0.07 cm}
     $H(-1,-1)$ &  $(2I_x2I_y)^{1/2}|F_{SS1}|$ & $q_{F_{SS1}}+\frac{3}{2}\pi-\psi_{x0}-\psi_{y0}$&    $F_{SS1}=2g_{2010,H}-g_{1101}^*$&(2,-1),(0,1)& skew sextupole \\
\hline\vspace{-0.30 cm}\\\vspace{0.07 cm}
     $H( 1,-1)$ &  $(2I_x2I_y)^{1/2}|F_{SS0}|$ & $q_{F_{SS0}}+\frac{3}{2}\pi+\psi_{x0}-\psi_{y0}$&    $F_{SS0}=g_{1110}-2g_{2001}^*$  &(2,-1),(0,1)& skew sextupole \\
\hline
\end{tabular}}
\rule{0mm}{0.1cm}
\vskip 2.0cm
\label{tab:line-selection}
\caption{Formulas to evaluate combined RDTs (CRDTs) from the 
	secondary lines in the spectra of $\tilde{x}(N)$ 
	and $\tilde{y}(N)$ assuming properly calibrated 
	BPMs, turn-by-turn oscillations without 
	decoherence and quadrupole errors included  
	in the C-S parameters.\vspace{-0.0 cm}}
\centering 
{
\begin{tabular}{c}
\end{tabular} \\ 
\begin{tabular}{l l l}
\hline\hline\vspace{-0.25cm} \\ \vspace{0.10 cm}
   Combined RDT{\hskip 1.4 cm}\    &{\hskip 1.6 cm}\ amplitude {\hskip 3.2 cm}\  & {\hskip 0.7 cm}  phase $q_F$ {\hskip 0.0 cm}             \\
\hline\hline\vspace{-0.30 cm}\\\vspace{0.07 cm}
$F_{xy}=|F_{xy}|e^{iq_{F_{xy}}}$   &   $|F_{xy}|=|H(0,1)|/[2|V(0,1)|]$           &  $q_{F_{xy}}=\phi_{H(0,1)}-\phi_{V(0,1)}-\frac{3}{2}\pi$ \\
\hline\hline\vspace{-0.30 cm}\\\vspace{0.07 cm}
$F_{yx}=|F_{yx}|e^{iq_{F_{yx}}}$   &   $|F_{yx}|=|V(1,0)|/[2|H(1,0)|]$           &  $q_{F_{yx}}=\phi_{V(1,0)}-\phi_{H(1,0)}-\frac{3}{2}\pi$ \\
\hline \\ & &  \vspace{-0.33cm}\\
\hline\vspace{-0.30 cm}\\\vspace{0.07 cm}
$F_{NS3}=|F_{NS3}|e^{iq_{F_{NS3}}}$&   $|F_{NS3}|=|H(-2,0)|/[4|H(1,0)|^2]$       &  $q_{F_{NS3}}=\phi_{H(-2,0)}+2\phi_{H(1,0)}-\frac{3}{2}\pi$ \\
\hline\vspace{-0.30 cm}\\\vspace{0.07 cm}
$F_{NS2}=|F_{NS2}|e^{iq_{F_{NS2}}}$&   $|F_{NS2}|=|H(0,-2)|/[4|V(0,1)|^2]$       &  $q_{F_{NS2}}=\phi_{H(0,-2)}+2\phi_{V(0,1)}-\frac{3}{2}\pi$ \\
\hline\vspace{-0.30 cm}\\\vspace{0.07 cm}
$F_{NS1}=|F_{NS1}|e^{iq_{F_{NS1}}}$&   $|F_{NS1}|=|V(-1,-1)|/[4|H(1,0)||V(0,1)|]$&  $q_{F_{NS1}}=\phi_{V(-1,-1)}+\phi_{H(1,0)}+\phi_{V(0,1)}-\frac{3}{2}\pi$ \\
\hline\vspace{-0.30 cm}\\\vspace{0.07 cm}
$F_{NS0}=|F_{NS0}|e^{iq_{F_{NS0}}}$&   $|F_{NS0}|=|V( 1,-1)|/[4|H(1,0)||V(0,1)|]$&  $q_{F_{NS0}}=\phi_{V( 1,-1)}-\phi_{H(1,0)}+\phi_{V(0,1)}-\frac{3}{2}\pi$ \\
\hline \\ & &  \vspace{-0.33cm}\\
\hline\vspace{-0.30 cm}\\\vspace{0.07 cm}
$F_{SS3}=|F_{SS3}|e^{iq_{F_{SS3}}}$&   $|F_{SS3}|=|V(0,-2)|/[4|V(0,1)|^2]$       &  $q_{F_{SS3}}=\phi_{V(0,-2)}+2\phi_{V(0,1)}-\frac{3}{2}\pi$ \\
\hline\vspace{-0.30 cm}\\\vspace{0.07 cm}
$F_{SS2}=|F_{SS2}|e^{iq_{F_{SS2}}}$&   $|F_{SS2}|=|V(-2,0)|/[4|H(1,0)|^2]$       &  $q_{F_{SS2}}=\phi_{V(-2,0)}+2\phi_{H(1,0)}-\frac{3}{2}\pi$ \\
\hline\vspace{-0.30 cm}\\\vspace{0.07 cm}
$F_{SS1}=|F_{SS1}|e^{iq_{F_{SS1}}}$&   $|F_{SS1}|=|H(-1,-1)|/[4|H(1,0)||V(0,1)|]$&  $q_{F_{SS1}}=\phi_{H(-1,-1)}+\phi_{H(1,0)}+\phi_{V(0,1)}-\frac{3}{2}\pi$ \\
\hline\vspace{-0.30 cm}\\\vspace{0.07 cm}
$F_{SS0}=|F_{SS0}|e^{iq_{F_{SS0}}}$&   $|F_{SS0}|=|H( 1,-1)|/[4|H(1,0)||V(0,1)|]$&  $q_{F_{SS0}}=\phi_{H( 1,-1)}-\phi_{H(1,0)}+\phi_{V(0,1)}-\frac{3}{2}\pi$ \\
\hline
\end{tabular}}
\vskip 0.0cm
\label{tab:line-rdt}
\end{table*}

In this section the correlation between the harmonic content 
of an {\sl ideal} TbT oscillation in the two transverse 
planes, $x$ and $y$, and the RDTs is discussed. For 
{\sl ideal} it is meant here a free oscillation of an 
instantaneously displaced particle beam without any damping 
(from radiation, chromaticity, amplitude-dependent 
detuning, and the like) and with perfectly 
calibrated BPMs. Forced oscillations induced 
by resonant devices, such as AC dipoles, require a 
different description~\cite{RogelioAC,SimonAC}.

\begin{table}[!t]
\caption{Formulas to calculate first-order RDTs from the 
	lattice model. The magnet integrated strengths 
	(MADX definition) are $J_1$ (m$^{-1}$),$K_2$ 
	and $J_2$ (m$^{-2}$) for skew quadrupoles, 
	normal and skew sextupoles respectively. The 
	C-S parameters $\beta$ and 
	$\phi$ are evaluated from the linear 
	lattice model with quadrupole errors (i.e. 
	beta-beating) included. $\Delta\phi_{w}$ is the 
	phase advance between the magnet $w$ and the 
	location where the RDTs are computed (BPM). 
	$Q_{x,y,}$ denote the linear betatron tunes, or 
        eigen-tunes if coupling may not be 
        neglected~\cite{prstab_coup}.
	\vspace{-0.0 cm}}
\vskip 0.0cm
\centering 
{
\begin{tabular}{c}
\end{tabular} \\ 
\begin{tabular}{l r}
\hline\hline\vspace{-0.25cm} \\ \vspace{0.0 cm}
{\hskip 2.4 cm} RDT    & {\hskip 0.0 cm}  resonance and {\hskip 0.0 cm} \\
                       & {\hskip 0.0 cm}  magnetic term {\hskip 0.0 cm} \\
\hline\hline\vspace{-0.20 cm}\\\vspace{0.17 cm}
$f^{(1)}_{1001}=\displaystyle\frac{\sum\limits_w J_{w,1}\sqrt{\beta_x^w\beta_y^w} 
                    e^{i(\Delta\phi_{w,x} - \Delta\phi_{w,y})}}
	       {4\left[1-e^{2\pi i(Q_x-Q_y)}\right]}$ & 
\begin{tabular}{r} (1,-1) skew \\ quadrupole \end{tabular}\\
\hline\vspace{-0.20 cm}\\\vspace{0.17 cm}
$f^{(1)}_{1010}=\displaystyle\frac{\sum\limits_w J_{w,1}\sqrt{\beta_x^w\beta_y^w} 
                    e^{i(\Delta\phi_{w,x} + \Delta\phi_{w,y})}}
	       {4\left[1-e^{2\pi i(Q_x+Q_y)}\right]}$ & 
\begin{tabular}{r} (1, 1) skew \\ quadrupole \end{tabular}\\
\hline \\ &  \vspace{-0.3cm}\\ 
\hline \vspace{-0.20 cm}\\\vspace{0.17 cm}
$f^{(1)}_{3000}=-\displaystyle\frac{\sum\limits_w K_{w,2}(\beta_x^w)^{3/2} 
                    e^{i(3\Delta\phi_{w,x})}}
	       {48\left[1-e^{2\pi i(3Q_x)}\right]}$ & 
\begin{tabular}{r} (3,0) normal \\ sextupole \end{tabular}\\
\hline\vspace{-0.20 cm}\\\vspace{0.17 cm}
$f^{(1)}_{1200}=-\displaystyle\frac{\sum\limits_w K_{w,2}(\beta_x^w)^{3/2} 
                    e^{i(-\Delta\phi_{w,x})}}
	       {16\left[1-e^{2\pi i(-Q_x)}\right]}$ & 
\begin{tabular}{r} (1,0) normal \\ sextupole \end{tabular}\\
\hline\vspace{-0.20 cm}\\\vspace{0.17 cm}
$f^{(1)}_{1020}=\displaystyle\frac{\sum\limits_w K_{w,2}\sqrt{\beta_x^w}\beta_y^w
                    e^{i(\Delta\phi_{w,x}+ 2\Delta\phi_{w,y})}}
	       {16\left[1-e^{2\pi i(Q_x+2Q_y)}\right]}$ & 
\begin{tabular}{r} (1,2) normal \\ sextupole \end{tabular}\\
\hline\vspace{-0.20 cm}\\\vspace{0.17 cm}
$f^{(1)}_{0120}=\displaystyle\frac{\sum\limits_w K_{w,2}\sqrt{\beta_x^w}\beta_y^w
                    e^{i(-\Delta\phi_{w,x}+ 2\Delta\phi_{w,y})}}
	       {16\left[1-e^{2\pi i(-Q_x+2Q_y)}\right]}$ & 
\begin{tabular}{r} (1,-2) normal \\ sextupole \end{tabular}\\
\hline\vspace{-0.20 cm}\\\vspace{0.17 cm}
$f^{(1)}_{0111}=\displaystyle\frac{\sum\limits_w K_{w,2}\sqrt{\beta_x^w}\beta_y^w
                    e^{i(-\Delta\phi_{w,x})}}
	       {8\left[1-e^{2\pi i(-Q_x)}\right]}$ & 
\begin{tabular}{r} (1,0) normal \\ sextupole \end{tabular}\\
\hline \\ &  \vspace{-0.3cm}\\ 
\hline\vspace{-0.20 cm}\\\vspace{0.17 cm}
$f^{(1)}_{0030}=-\displaystyle\frac{\sum\limits_w J_{w,2}(\beta_y^w)^{3/2} 
                    e^{i(3\Delta\phi_{w,y})}}
	       {48\left[1-e^{2\pi i(3Q_y)}\right]}$ & 
\begin{tabular}{r} (0,3) skew \\ sextupole \end{tabular}\\
\hline\vspace{-0.20 cm}\\\vspace{0.17 cm}
$f^{(1)}_{0012}=-\displaystyle\frac{\sum\limits_w J_{w,2}(\beta_y^w)^{3/2} 
                    e^{i(-\Delta\phi_{w,y})}}
	       {16\left[1-e^{2\pi i(-Q_y)}\right]}$ & 
\begin{tabular}{r} (0,1) skew \\ sextupole \end{tabular}\\
\hline\vspace{-0.20 cm}\\\vspace{0.17 cm}
$f^{(1)}_{2010}=\displaystyle\frac{\sum\limits_w J_{w,2}\beta_x^w\sqrt{\beta_y^w}
                    e^{i(2\Delta\phi_{w,x}+ \Delta\phi_{w,y})}}
	       {16\left[1-e^{2\pi i(2Q_x+Q_y)}\right]}$ & 
\begin{tabular}{r} (2,1) skew \\ sextupole \end{tabular}\\
\hline\vspace{-0.20 cm}\\\vspace{0.17 cm}
$f^{(1)}_{2001}=\displaystyle\frac{\sum\limits_w J_{w,2}\beta_x^w\sqrt{\beta_y^w}
                    e^{i(2\Delta\phi_{w,x}- \Delta\phi_{w,y})}}
	       {16\left[1-e^{2\pi i(2Q_x-Q_y)}\right]}$ & 
\begin{tabular}{r} (2,-1) skew \\ sextupole \end{tabular}\\
\hline\vspace{-0.20 cm}\\\vspace{0.17 cm}
$f^{(1)}_{1101}=\displaystyle\frac{\sum\limits_w J_{w,2}\beta_x^w\sqrt{\beta_y^w}
                    e^{i(-\Delta\phi_{w,y})}}
	       {8\left[1-e^{2\pi i(-Q_y)}\right]}$ & 
\begin{tabular}{r} (0,1) skew \\ sextupole \end{tabular}\\
\hline
\end{tabular}}
\rule{0mm}{0.0cm}
\label{tab:lattice-rdt}
\end{table}

In Table~\ref{tab:line-selection} the spectral lines 
of the signals $\tilde{x}(N)=x(N)/\sqrt{\beta_x}$ 
and $\tilde{y}(N)=y(N)/\sqrt{\beta_y}$, where $\beta$ 
denotes the Courant-Snyder (C-S) parameter, are listed  
together with the corresponding RDTs. Higher-order 
octupolar lines  are analyzed in Sec.~\ref{sec_OctDec}. 
As discussed in Appendix~\ref{app:2}, for the 
evaluation of the RDTs the complete phase space curve 
$(x,p_x)$ and $(y,p_y)$
generated by the TbT oscillation is necessary, which in 
turn requires the combination of signals from 2 synchronized 
BPMs. By analyzing its projection on the $x$ and 
$y$ axis, i.e. by using single-BPM TbT data, 
RDTs are no longer measurable. However, their 
linear combinations, the combined RDTs (CRDTs) 
$F_{xy}$, $F_{yx}$, $F_{NS}$ and $F_{SS}$ are still observables. 
CRDTs are defined in the fourth column of 
Table~\ref{tab:line-selection} and are derived in 
Appendix~\ref{app:4}.
In Table~\ref{tab:line-rdt} formulas to infer their 
amplitudes and phases from the spectral lines are 
reported, whereas analytic formulas for the computation 
of first-order RDTs, and hence of the CRDTs, from the 
lattice model are listed in Table~\ref{tab:lattice-rdt}. 
\\ \ \\

Before entering in the perilous terrain of higher 
orders, it is worthwhile to define what actually 
{\sl order} means. in Appendix~\ref{app:1} it is 
shown how the nonlinear betatron motion may be 
described in terms of truncated Lie series: the 
degree of precision (and difficulty) of the description 
is related to the order at which these series are 
truncated. RDTs depends on the strengths of the 
corresponding magnets $\vec{K}=(\delta K_1,J_1,K_2,J_2,K_3, ...)$
where $\delta K_1$ is the quadrupole error field not 
included in the computation of the C-S parameters, 
$J_1$ is the skew quadrupole field, $K_2$ and $J_2$ 
is the normal and skew sextupole fields, $K_3$ refers 
to the octupole and so on. First-order RDTs 
$f^{(1)}$ result from 
Lie series truncated to the first term, are 
generated by the specific corresponding magnet, as in 
Table~\ref{tab:lattice-rdt}, and scale linearly with 
their strengths. Schematically they may be represented 
by the following chart
\begin{equation}
\begin{array}{l l c l}
\hbox{focusing errors} &f^{(1)}&\longleftarrow &\delta K_1 \\
\hbox{betatron coupling}&f^{(1)}&\longleftarrow &J_1 \\
\hbox{normal sextupole}&f^{(1)}&\longleftarrow &K_2 \\
\hbox{skew sextupole}  &f^{(1)}&\longleftarrow &J_2 \\
\hbox{normal octupole} &f^{(1)}&\longleftarrow &K_3  \ .\\
\end{array}
\end{equation}
When the Lie series are truncated to the second order, 
cross-products between the magnet strengths appear and 
the picture becomes more complicated for the second-order 
RDTs $f^{(2)}$
\begin{equation}\hskip -0.1cm
\begin{array}{l l c l}
\hbox{focusing errors} &f^{(2)}&\longleftarrow&\ J_1 \otimes  J_1\\
\hbox{betatron coupling}&f^{(2)}&\longleftarrow&\ J_1\otimes \delta K_1\\
\hbox{normal sextupole}&f^{(2)}&\longleftarrow&K_2\otimes \delta K_1
                                               \ ,\  J_1\otimes J_2   \\
\hbox{skew sextupole}  &f^{(2)}&\longleftarrow&\ J_2\otimes \delta K_1
                                               \ ,\  J_1\otimes K_2   \\
\hbox{normal octupole} &f^{(2)}&\longleftarrow&K_3\otimes \delta K_1
                                               \ ,\  K_2\otimes K_2 \ ... \\
\end{array}\hskip -0.5cm
\end{equation}

The above scheme simplifies considerably under three 
reasonable assumptions. First, if the C-S parameters 
($\beta$ and $\phi$) used in  Table~\ref{tab:lattice-rdt} 
are evaluated from the lattice model including 
focusing errors, $\delta K_1\equiv0$. Second, no strong skew 
sextupole is installed or powered in the machine and 
$J_1$ is generated by slightly tilted normal sextupoles, 
$K_2>>J_2$ and $J_1\otimes J_2\simeq0$, betatron coupling being 
also weak. Third, coupling is assumed to be weak so that 
$J_1\otimes J_1\simeq0$. The above RDTs then reduce to 
\begin{equation}\hskip -0.1cm\label{eq:scheme3}
\begin{array}{l l c l c l l}
\hbox{focusing errors} &f^{(1)}&=         &0  &,\quad f^{(2)}&=&0\\
\hbox{betatron cop-ling}&f^{(1)}&\leftarrow&J_1&,\quad f^{(2)}&=&0\\
\hbox{normal sextupole}&f^{(1)}&\leftarrow&K_2&,\quad f^{(2)}&=&0\\
\hbox{skew sextupole}  &f^{(1)}&\leftarrow&J_2&,\quad f^{(2)}&\leftarrow&J_1\otimes K_2\\
\hbox{normal octupole} &f^{(1)}&\leftarrow&K_3&,\quad f^{(2)}&\leftarrow&K_2\otimes K_2\\
\end{array}\hskip -0.5cm
\end{equation}

These considerations (detailed mathematical derivations 
may be found in Appendix~\ref{app:1}) indicate that the 
first-order analytic formulas for coupling and normal sextupole RDTs 
of Table~\ref{tab:lattice-rdt} are valid also to the second 
order, provided that the used C-S parameters ($\beta$ 
and $\phi$) are evaluated from the lattice model including 
focusing errors. By doing so, first-order beta-beating RDTs 
are automatically zero. 

Second-order terms are instead to be computed and 
included in the evaluation of skew sextupole RDTs, 
which are excited to the first order by $J_2$ 
(introduced by tilted sextupoles and/or displaced octupoles), 
and to the second order by the cross-product between 
coupling and normal sextupoles, $K_2\otimes J_1$. In machines 
with strong focusing, such as light sources, strong normal 
sextupole ($K_2$), even if multiplied by low (i.e. well 
corrected) coupling ($J_1$), render the second-order 
contribution to the skew sextupole RDTs comparable to 
that of the first order, i.e. $K_2\otimes J_1\simeq J_2$. 
Similar considerations apply for octupolar RDTs.

Another complication appearing when second-order terms 
are to be taken into account is that RDTs $f=f^{(1)}+f^{(2)}$ 
are no longer observables from the harmonic analysis 
of turn-by-turn position data. The Observable RDTs (ORDTs) 
$g_{jklm}$ may be written as 
\begin{equation}
g_{jklm}=f^{(1)}_{jklm}+f^{(2)}_{jklm} +
         \mathbf{L}\left(f^{(1)}_{pqrt}\otimes f^{(1)}_{tuvz}\right)\ ,
\end{equation}
where $\mathbf{L}$ is a linear function and 
$jklm\ne pqrt\ne tuvz$. For this reason 
in Table~\ref{tab:line-selection} the ORDTs $g_{jklm}$ replace 
the first-order $f^{(1)}_{jklm}$ in the four lines 
corresponding to skew sextupole harmonics. Formulas for 
the computation of the ORDTs from the lattice model 
are derived in Appendix~\ref{app:1} and may be easily 
implemented numerically. An example may help clarifying 
the nature of $\mathbf{L}$ and revealing a counterintuitive 
feature: The skew sextupole ORDT $g_{0030}$ reads
\begin{eqnarray}\label{eq:g0030}
g_{0030}&=&f_{0030}^{(1)}+f_{0030}^{(2)}
		   -\frac{i}{3}\left[f_{1010}^{(1)}f_{0120}^{(1)}-
				     f_{1001}^{(1)*}f_{1020}^{(1)}\right]
\ ,\quad\\
&&\hspace{0.3cm} \uparrow \hspace{1.0cm} \uparrow\hspace{2.8cm}\uparrow
\nonumber \\
&&\hspace{0.2cm} J_2\hspace{0.4cm} J_1\otimes K_2 
                  \hspace{1.8cm} J_1\otimes K_2 \nonumber \\
\nonumber \\
&&f_{0030}^{(2)}=i\frac{\tilde{h}^{(2)}_{0030}+\hat{h}_{0030}}
			  {1-e^{2\pi i(3Q_y)}}\ ,
\end{eqnarray}
where $\tilde{h}^{(2)}(s)_{0030}$ and $\hat{h}_{0030}$ 
are computed in Eqs.\eqref{eq:tildeh2_0030} and 
~\eqref{eq:hath2_0030} respectively, whereas all other first order 
RDTs $f^{(1)}$ are defined in Table~\ref{tab:lattice-rdt}. 
The counterintuitive feature exhibited by the additional 
term $\mathbf{L}\left(f^{(1)}\otimes f^{(1)}\right)$ is that 
a skew sextupole resonance, the (0,3) in the case of $g_{0030}$, 
may be excited even in the absence of skew sextupole 
sources $f_{0030}^{(1)}=0$ and far away from the resonant 
condition $1-e^{2\pi i(3Q_y)}\simeq 1$ and $f_{0030}^{(2)}\simeq 0$, 
for example in the presence of large coupling and 
close to the difference resonance 
$|f_{1010}^{(1)}|\ll |f_{1001}^{(1)}|\sim 1$, 
since 
$|g_{0030}|\sim \frac{1}{3}|f_{1020}^{(1)}|$ 
and the normal sextupole RDT $f_{1020}^{(1)}$ may 
be arbitrarily large.

In summary, the Combined RDTs (CRDTs) $F$ are measurable 
from turn-by-turn position data (Table~\ref{tab:line-rdt}). 
They may be also computed from the model (fourth column 
of Table~\ref{tab:line-selection}) via the first-order RDTs 
(Table~\ref{tab:lattice-rdt}) or, if second-order
contributions may not be neglected, via the Observable 
RDTs (ORDTs) of Appendix~\ref{app:1}.
The logical scheme to be 
followed in the nonlinear lattice modelling discussed 
in this paper is the following:
\begin{table}[!h]
\vskip 0.3cm
\centering 
\begin{tabular}{l}
\hspace{0.0 cm}TbT single-BPM (dual plane) data   \\ \vspace{0.1 cm}
\hspace{1.0 cm}$\Bigg\downarrow$            Fourier Transform       \\ \vspace{0.1 cm}
\hspace{0.0 cm}harmonics $H( n,m)$, $V( n,m)$  \\ \vspace{0.1 cm}
\hspace{1.0 cm}$\Bigg\downarrow$            Table~\ref{tab:line-rdt}\\ \vspace{0.1 cm}
\hspace{0.0 cm}measured CRDTs $F$            \\ \vspace{0.1 cm}
\hspace{1.0 cm}$\Bigg\downarrow$             \\ \vspace{0.1 cm}
\hspace{0.0 cm}difference  $\xrightarrow[]{\hbox{linear\  system\ }}$ 
\begin{tabular}{c}
new lattice model \\ or calibration
\end{tabular}
\\ \vspace{0.1 cm}
\hspace{1.0 cm}$\Bigg\uparrow$               \\ \vspace{0.1 cm}
\hspace{0.0 cm}model CRDTs $F$               \\ \vspace{0.1 cm}
\hspace{1.0 cm}$\Bigg\uparrow$ Tables~\ref{tab:lattice-rdt},\ref{tab:line-selection}, 
                Appendix~\ref{app:1}\\ \vspace{0.1 cm}
\hspace{0.0 cm}initial lattice model          \\ \vspace{0.1 cm}
\end{tabular} 
\rule{0mm}{0.1cm}
\end{table}

\subsection{Experimental precautions}
\label{sec_precautions}
An important condition necessary to ensure the 
applicability of the single-BPM TbT data analysis is 
that monitors are dual-plane. In fact, in 
order to extract the CRDTs the knowledge of 
amplitude ($2I$) and phase ($\psi_0$) of both 
tune lines is mandatory , which in turn requires 
the possibility of measuring TbT data 
in both planes at the same BPM. Another requirement 
is that the beam does not experience decoherence, 
otherwise additional factors are to be included, as 
discussed in Ref.~\cite{Rogelio1}. A further condition 
is that BPMs have no tilt or calibration error.
Uncalibrated BPMs would provide TbT data in the form 
of $\tilde{x}(N)^{BPM}=\eta_x\tilde{x}(N)$, where 
$\eta_x\ne1$. An example may help explaining 
their impact on the formulas of 
Table~\ref{tab:line-rdt}. A hypothetical signal 
$\tilde{x}(N)$s contains two harmonics, the tune 
$H(1,0)$ and a sextupolar $H(-2,0)$, in the form of 
\begin{eqnarray}
\tilde{x}(N)&=&\frac{1}{2}(2I_x)^{1/2}e^{i(2\pi Q_xN+\psi_{x0})} + 
	\hskip 1.15cm \longleftarrow\hskip  0.2cm \ H(1,0)\nonumber \\
	&&(2I_x)F_{NS3}e^{i\left[-4\pi Q_xN-2\psi_{x0}\right]}\ .
	\hskip 0.8cm\longleftarrow\hskip -0.1cm \ H(-2,0) \nonumber
\end{eqnarray}
$|F_{NS3}|$ may be easily inferred from the amplitude 
of $H(-2,0)$ according to $|F_{NS3}|=|H(-2,0)|/[4|H(1,0)|^2]$  
(third row of Table~\ref{tab:line-rdt}). An uncalibrated 
BPM would instead generate
\begin{eqnarray}
\tilde{x}(N)^{BPM}&\rightarrow&\eta_x\left\{ \frac{1}{2}
			(2I_x)^{1/2}e^{i(2\pi Q_xN+\psi_{x0})} + 
	\right. \nonumber \\ 
	&&\left.(2I_x)F_{NS3}e^{i\left[2\pi (-2Q_x)N-2\psi_{x0}\right]}
	\right\}\ .\nonumber
\end{eqnarray}
The above normalization would carry the calibration 
factor, corrupting the CRDT measurement, since 
$|H(-2,0)|/[4|H(1,0)|^2]=|F_{NS3}|/\eta_x$. The impact 
of BPM tilts on the harmonic analysis is similar, 
though less obvious, and is assumed here to be negligible.

The dependence of the RDTs (hence of the CRDTs 
and, in turn, of the spectral lines) on the phase 
advances $\Delta\phi$ between the observation point 
(the BPM) and the magnets results in the dependence 
on the longitudinal position along the ring. Ideally, 
they shall modulate according to the machine periodicity. 
This feature is of particular interest, for the detection 
of malfunctioning magnets, as reported in 
Ref.~\cite{Rogelio1}, or for a direct measurement of 
their strengths and polarities~\cite{Andrea-thesis,prstab_strength}.

On the other hand, the amplitude of the tune lines 
$|H(1,0)|$ and $|V(0,1)|$ shall be constant along 
the ring, as indicated by the first two lines of 
Table~\ref{tab:line-selection} . There it is indeed assumed 
that focusing errors are already included in the model 
(i.e. in the beta functions used to normalize the 
TbT BPM data) and hence the beta-beating RDTs $f_{2000}$ 
and $f_{0020}$ of Eqs.~\eqref{eq:def_f2000}-\eqref{eq:def_f0020} 
are zero. 
If the tune line amplitudes measured at all BPMs 
are not constant, three are the possible causes to 
be evaluated and removed prior to any further analysis 
involving other spectral lines. 

First, the linear lattice model used to 
evaluate the beta functions at the BPMs is incorrect. This 
shall result in non-zero quadrupolar RDTs $f_{2000}$ 
and $f_{2000}$, in a modulation of the beta functions 
(beta-beating) and in wrong phase advances between BPMs. 
In Appendix~\ref{app:3} it is shown how the beta-beating 
is correlated to the RDTs and how to infer it from the 
tune line amplitudes measured at each BPM.  Even if elegant 
and handy, such formulas are of limited applicability for 
several reasons: $(i)$ it is impossible to disentangle 
beta modulation from BPM calibration factors or tilts, $(ii)$ 
quadrupoles RDTs are not measurable from the harmonic analysis 
of the single-BPM TbT data, $(iii)$ second and higher order RDTs 
do modify the tune line amplitudes introducing a dependence 
on the position along the ring. Fortunately, none of these
elements affects the phase of the tune line, and the 
difference between its value at two consecutive BPMs 
corresponds to the betatron phase advance. Hence by 
comparing and fitting the measured BPM phase advance, 
a reliable focusing model may be built and used to 
evaluate correctly the beta functions at the BPMs. 
As discussed in Sec.~\ref{sec_OctDec}, some octupolar 
CRDTs may indeed perturb the phase of the tune lines, 
though this effect may be controlled by a careful 
choice of the initial excitation (i.e. kicker 
strength). It is worthwhile reminding that the fit of 
the BPM phase advance shall be accompanied by a fit 
of the measured dispersion function for a reliable model.

Second, if a modulation persists even after fitting the BPM 
phase advance (and dispersion), it shall 
be verified whether second-order sextupole terms may introduce 
a dependence on the longitudinal position. As discussed in 
Appendix~\ref{app:3}, this perturbation depends on the 
initial oscillation amplitude (or the action), i.e. on the 
kicker strength, and may be evaluated from preliminary 
single-particle tracking simulations. For the measurement 
procedure discussed here, it is hence necessary to limit 
the kicker strength so to keep the natural modulation of 
the tune line amplitude induced by sextupoles well 
below the desired experimental resolution. This contrasts  
with the desire of having a large signal-to-noise ratio 
between spectral noise and amplitude of the sextupolar 
lines, which would require a strong initial excitation. 
Eventually a trade off shall be found with the help of 
tracking simulations and measured data at different kicker 
strengths. 

The third source of tune line amplitude modulation 
is represented by BPM calibration factors $\eta\ne1$ and 
tilts. Assuming that the latter are negligible compared 
to other sources of errors, the only way to infer $\eta$ 
and remove them before carrying out the spectral analysis 
is to impose constant tune line amplitude at all BPMs 
after having verified the previous two points. 
The Fourier transform redone after scaling the 
measured data by $\eta$ may then be used for the 
analysis of the nonlinear model.

A last interesting figure to evaluate the quality of the 
TbT BPM data and of its harmonic analysis is represented 
by a cancellation condition that shall be satisfied by  
some measured CRDTs. According to the definitions of the 
normal sextupole terms of Table~\ref{tab:line-selection} 
it is straightforward to prove that
\begin{equation}\label{eq:F0}
F_0=2\Re\{F_{NS2}\}-\Re\{F_{NS1}\}+\Re\{F_{NS0}\} \equiv 0
\end{equation}
anywhere in the ring: the closer $F_0$ is to zero, the 
more reliable is the harmonic analysis described here. 
Note that the measured $F_0$ along the ring depends only 
on: $(i)$ the linear lattice model ($\beta$ functions 
used to normalize the BPM TbT data); $(ii)$ the BPM 
calibration factors, spatial and spectral resolution, 
electronic noise and the like; $(iii)$ initial beam 
excitation, i.e. kicker strength. Luckily enough, it 
does not depend on the nonlinear lattice model, being 
valid for any sextupole setting.

\subsection{From normal sextupole CRDTs to sextupole errors}
\label{sec_RDT2K2}
Let us assume to have acquired $N$ TbT data 
from $N_{BPM}$ BPMs. From the harmonic analysis of 
Table~\ref{tab:line-rdt} it is possible to infer 
the four normal sextupole CRDTs $F_{NS3}$, $F_{NS2}$, 
$F_{NS1}$ and $F_{NS0}$, at all BPMs. Being these 
all complex quantities, a $N_{BPM}\cdot4\cdot2$ vector
$\vv{F}_{NS,meas}$ may be defined containing the 
measured real and imaginary parts of the four CDRTs
\begin{eqnarray}
\vv{F}_{NS,meas}&=&(\Re\{F_{NS3}\}_{1}, ...\ ,\Re\{F_{NS3}\}_{N_{BPM}}, 
\nonumber \\
              &&  \ \Im\{F_{NS3}\}_{1}, ...\ ,\Im\{F_{NS3}\}_{N_{BPM}}, 
\nonumber \\
              &&  \ ...,
\nonumber \\
              &&  \ \Im\{F_{NS0}\}_{1}, ...\ ,\Im\{F_{NS0}\}_{N_{BPM}})
\ . \label{eq_FNS_FFT}
\end{eqnarray}
An equivalent vector may be defined from the 
model (C-S parameters including focusing 
errors,$\beta$ and $\phi$, and sextupole 
integrated gradients, $K_2$): The RDTs $f_{3000}$, 
$f_{1200}$, $f_{1020}$, $f_{0120}$ and $f_{0111}$ 
may be evaluated from Table~\ref{tab:lattice-rdt}, 
to be used for the evaluation of the model CRDTs 
through the fourth column of Table~\ref{tab:line-selection}. 
Being all passages from the integrated strengths 
to the CRDTs linear, the model vector may be 
defined as
\begin{eqnarray}
\vv{F}_{NS,mod}&=&\mathbf{M_{NS}}\vv{K}_2\ , \label{eq_FNS_lattice}
\end{eqnarray}
where $\vv{K}_2$ contains all $N_{sext}$ known sources 
of sextupolar field along the ring, and $8\cdot N_{BPM}
\times N_{sext}$ matrix $\mathbf{M_{NS}}$ depends on the linear 
lattice only, through the beta functions, the phase 
advances between BPMs and sextupoles and the tunes. 
The difference between measured and model CRDTs then 
reads
\begin{eqnarray}
\vv{F}_{NS,meas}-\vv{F}_{NS,mod}&=&\mathbf{M_{NS}}\vv{\Delta K}_2
\ , \label{eq_FNS_diff}
\end{eqnarray}
where $\vv{\Delta K}_2$ contains the 
$N_{sext}$ sextupole field errors to be inferred 
after pseudo-inverting (via SVD, for instance) the 
above linear system. Note how the choice of defining 
the CRDTs vector with the real and imaginary parts 
ensures the linearity of the system, whereas the 
choice of decomposing these complex quantities in 
amplitude and phases, as in Ref.~\cite{Bartolini2}, 
would have introduced an 
unnecessary and avoidable nonlinearity in the problem, 
thus requiring a nonlinear minimization 
routine for the evaluation of $\vv{\Delta K}_2$. 

The linear system of Eq.~\eqref{eq_FNS_diff} may be also 
used for the calibration of an individual sextupole. 
The optics model in this case is necessary for the 
evaluation of the $(8\cdot N_{BPM})\times 1$ 
{\sl matrix} $\mathbf{M}$ only, whereas the left hand 
side contains two CRDT vectors measured with two 
different sextupole strengths, i.e. 
\begin{eqnarray}
\vv{F}_{NS,meas}(K_2)-\vv{F}_{NS,meas}(K_2+\delta K_2)=
                 \mathbf{M_{NS}}\delta K_2
\ . \hskip 0.7cm\label{eq_FNS_calib}
\end{eqnarray} 
After pseudo-inverting the above system, the inferred 
$\delta K_2$ may be compared with the expected value 
from the change of current in the sextupole and its 
calibration curve. The latter may be reconstructed and 
compared with the magnetic model (or measurement) by 
repeating the procedure at different $\delta K_2$. 
Again, the problem of calibrating a nonlinear magnetic 
element is reduced to a linear system when looking 
at the CRDTs.

\subsection{From normal sextupole CRDTs to sextupole correction}
\label{sec_RDT2K2}
The natural extension of Eq.~\eqref{eq_FNS_diff} 
is a system for the evaluation of a sextupole 
corrector setting. If all sextupole magnets have 
independent power supplies and other sources of 
sextupolar fields (such as fringe fields in bending 
magnets) may be neglected, the pseudo-inversion of 
Eq.~\eqref{eq_FNS_diff} provides already the 
set of corrections $\vv{\delta K}_2$ to insert in 
the sextupoles. If this is not the case and only 
$N_{cor}$ corrector 
sextupoles are available (which is the case at the 
ESRF storage rings, with 224 sextupoles grouped in 7 
families and only 12 sextupole correctors available), 
the system then reads
\begin{eqnarray}
\vv{F}_{NS,meas}-\vv{F}_{NS,ref}&=&\mathbf{M_{NS,cor}}\vv{K}_2^{cor}
\ , \label{eq_FNS_cor}
\end{eqnarray}
where $\vv{K}_2^{cor}$ contains the strengths of the $N_{cor}$  
sextupole correctors to be evaluated after pseudo-inverting 
the above system, and the $8\cdot N_{BPM}\times N_{cor}$ 
matrix $\mathbf{M_{NS,cor}}$ depends on the beta functions and the 
phase advances between BPMs and the correctors. Note 
that $\vv{F}_{NS,ref}$ (reference or desired vector) has 
been used in place of $\vv{F}_{NS,mod}$ (model vector). 
The difference may be of importance for those machines 
(like the ESRF storage ring) where beta-beating is 
corrected up to some percents and special insertion 
optics break up significantly the machine 
natural periodicity. $\vv{F}_{NS,ref}$ may refer hence to  
the ideal lattice without any insertion optics, whereas 
$\vv{F}_{NS,mod}$ refers to the actual model including 
insertion optics (if any) and all known lattice errors. 
It is assumed here that the best CRDT correction shall be 
based on the ideal vector $\vv{F}_{NS,ref}$, rather than 
on $\vv{F}_{NS,mod}$. 
Once more, the problem of correcting nonlinear magnetic 
elements is reduced to a linear system, 
Eq.~\eqref{eq_FNS_cor}, to be pseudo-inverted.

\subsection{From skew sextupole CRDTs  to sextupole tilts}
\label{sec_RDT2SKEWSEXT}
The procedure described in Sec.~\ref{sec_RDT2K2}
may be repeated, with some precautions, for 
the evaluation of sextupole tilts. From 
the harmonic analysis of the TbT BPM data and 
Table~\ref{tab:line-rdt} it is possible to infer 
the four skew sextupole CRDTs $F_{SS3}$, $F_{SS2}$, 
$F_{SS1}$ and $F_{SS0}$, at all BPMs, and cast them 
in a $N_{BPM}\cdot4\cdot2$ vector $\vv{F}_{SS,meas}$ 
\begin{eqnarray}
\vv{F}_{SS,meas}&=&(\Re\{F_{SS3}\}_{1}, ...\ ,\Re\{F_{SS3}\}_{N_{BPM}}, 
\nonumber \\
              &&  \ \Im\{F_{SS3}\}_{1}, ...\ ,\Im\{F_{SS3}\}_{N_{BPM}}, 
\nonumber \\
              &&  \ ...,
\nonumber \\
              &&  \ \Im\{F_{SS0}\}_{1}, ...\ ,\Im\{F_{SS0}\}_{N_{BPM}})
\ . \label{eq_FSS_FFT}
\end{eqnarray}
The evaluation of the corresponding vector from 
the model require a preliminary step. As discussed in 
Sec.~\ref{sec_CRDT}, contrary to normal sextupole 
terms, it is not possible to ignore the second-order 
contribution to the skew sextupole CRDTs due to 
cross-terms between coupling and normal sextupole 
RDTs, see Eq.~\eqref{eq:scheme3}. In other words, 
ideal skew sextupoles CRDTs shall all be zero, but 
even with perfectly upright sextupoles they may 
be nonzero because any residual coupling in the 
machine {\sl transfers} normal sextupole spectral 
harmonics in the other plane, hence generating 
skew sextupole harmonics and corresponding CRDTs. 
Not taking into account this natural contribution 
would corrupt the evaluation of sextupole tilts. 
For this reason new analytic formulas 
valid to second order have been derived. They are 
presented in Appendix~\ref{app:1} and shall be 
used for the evaluation of the skew sextupole 
ORDTs $g_{jklm}$, and hence of the CRDTs 
$F_{SS}$. Provided that betatron coupling is 
well modelled, the second-order contribution to 
the CRDTs $F_{SS}$ may be then computed and any 
difference between the model vector $\vv{F}_{SS,mod}$
and the measured one $\vv{F}_{SS,meas}$ will 
depend on the sextupole tilts only, i.e. 
\begin{eqnarray}
\vv{F}_{SS,meas}-\vv{F}_{SS,mod}&=&\mathbf{M_{SS}}\vv{J}_2
\ , \label{eq_FSS_diff}
\end{eqnarray}
where the angles $\theta$ can be extracted from 
$J_2$ according to 
\begin{eqnarray}\label{eq_FSS_tilt}
\theta=-\frac{1}{3}\arcsin{\left(\frac{J_2}{K_2}\right)}\ ,
\end{eqnarray}
since the skew sextupole strength introduced 
by a tilted sextupole is $J_2=-K_2\sin{(3\theta)}$. 

It is worthwhile mentioning that measuring skew 
sextupolar CRDTs will be unavoidably more difficult 
than extracting normal sextupole CRDTs, the 
latter scaling with the (strong) sextupole gradients, 
while the former scale with the (small) sextupole 
tilts. Depending on the lattice configuration and 
coupling correction, $\vv{F}_{SS}$  
may be one or two orders of magnitude lower than 
$\vv{F}_{NS}$.

In machine with skew sextupole correctors, the 
system of Eq.~\eqref{eq_FNS_cor} may be modified as 
\begin{eqnarray}
\vv{F}_{SS,meas}&=&-\mathbf{M_{SS,cor}}\vv{J}_2^{cor}
\ , \label{eq_FSS_cor}
\end{eqnarray}
where the reference vector $\vv{F}_{SS,ref}$ in this case 
is zero. In machine without skew sextupole correction, 
skew quadrupoles may be used to correct, along with 
coupling, skew sextupole CRDTs, via the second-order 
contributions $f_{jklm}^{(2)}$. This later option, however, 
would require to solve a more complex (but still linear 
in the skew quadrupole strengths) system not discussed here.

\section{Experimental results}
\label{sec_ExpResults}

\subsection{Preparing measurements}
\label{sec_PrepaMeas}
The TbT acquisition of the commercial Libera-Brilliance 
BPMs installed in the ESRF storage ring requires  
some care in its preparation. Even though the 
measurement is rather quick, a series of machine 
studies were carried out between 2010 and 2011 for 
the correct setup.

First of all, in the case of the ESRF storage ring 
the ideal beam filling pattern is the 1/3, i.e. with a 
train of about 330 bunches separated by about 660 empty 
RF buckets. The interest of this pattern is threefold. 
First, it allows a straightforward synchronization among 
all 224 BPMs and the bunch train, by using the booster 
(whose circumference is exactly 1/3 the one of the 
storage ring) and fast monitors in the transfer line. 
Second, both the horizontal and the vertical kickers may 
by synchronized so to have the bunch train on their 
flattop: The measured invariants and CRDTs would 
be corrupted if part of the beam experiences the 
kicker pulse rise and/or fall.
Third, this filling mode maximizes the BPM resolution, 
the beam signal being integrated over 1/3 of the 
revolution frequency, whereas a few-bunch pattern 
would provide intense peak signal with a noisier 
integrated value.

Another parameter to be adjusted to minimize the BPM 
noise and to prevent a saturation of the signal (and 
hence a corrupted harmonic analysis) is the beam 
current against the amplification factor 
of the acquisition system. In order to have 
the best dynamic range from a few $\mu$A to 200~mA, 
the ESRF BPMs apply a variable signal amplification. 
The special optics implemented for this measurement 
with almost zero linear chromaticity and amplitude 
dependent detuning may sustain up to 10~mA before 
losing the beam. However, in order to maximize the 
BPM resolution during the signal digitization, 
measurements are carried out with 7~mA for a kicker 
excitation 
so to displace horizontally the beam from 1.3 to 
5.5~mm. With a $\beta$ function of 36.6~m this 
corresponds to an invariant between about 
$10^{-4}$ and $4\times10^{-4}$~m$^{1/2}$.
Higher beam current would result in a different 
amplification range with lower digitization resolution, 
whereas stronger excitations risk to saturate the BPM 
signal. 

As mentioned in Sec.~\ref{sec_precautions} the 
kicker strengths shall be a trade-off between 
spectral resolution (the stronger, the better) and 
second-order terms affecting the invariance of the 
tune line amplitude (the weaker, the better). 
Measurements are usually repeated at different 
kicker strengths. For the final analysis of sextupolar CRDTs
only the data set of 50 repeated measurements 
with both invariants $\simeq2.0\times10^{-4}$~m$^{1/2}$ 
is used (corresponding to displacement of 
about 2.5~mm horizontally and 1.0~mm vertically, with 
$\beta$ functions of 36.6 and 6.0 m). 
For the octupolar model, instead, data with 
stronger excitations are used.
Note that hereafter, unless specified otherwise, 
the invariant is referred to as $1/2\sqrt{2I}$ so to 
be equal to the tune line amplitude (see first two 
rows in Table~\ref{tab:line-selection}).

As discussed in Sec.~\ref{MovelLimit}, even though 
more than 2000 turns may be stored by the BPMs and 
the weak decoherence ensures about 1000 turns of 
exploitable data, it will be shown how the best 
results are obtained when only 256 turns are used 
for the analysis of sextupolar CRDTs (512 turns 
for the octupolar).

\subsection{Linear analysis of TbT BPM data}
\label{sec_LinMeas}

\begin{figure}[b]
\rule{0mm}{5mm}
\centerline{
  \includegraphics[width=8.5cm,angle=0]
  {fig02.eps}}
\vskip -0.2cm
  \caption{\label{fig_BPMPhAdMeas} (Color) Example of measured BPM phase 
        advance deviation from the models: ideal (blue), with errors 
        inferred from ORM measurement (red) and with errors after fitting 
        the measured BPM phase advance and dispersion (green). Data 
        correspond to the fifth row of Table~\ref{tab:linear-anal}}
\vskip 0.1cm
\rule{0mm}{0mm}
\centerline{
  \includegraphics[width=8.5cm,angle=0]
  {fig03A.eps}}
\vskip -0.23cm
\centerline{
  \includegraphics[width=8.5cm,angle=0]
  {fig03B.eps}}
\vskip -0.2cm
  \caption{\label{fig_TunelineMeas} (Color) Example of measured variation of 
	the horizontal (top) and vertical (bottom) tune-line 
	amplitudes along the ESRF storage ring. Data correspond to the 
        last row of Table~\ref{tab:linear-anal} (average over 50 acquisitions).}
\vskip -0.5cm
\end{figure}

The starting linear optics model used to evaluate the 
expected CRDTs and the $\beta$ functions at the BPMs 
(necessary for the construction of TbT signals to be analyzed 
$\tilde{x}(N)=x(N)/\sqrt{\beta_x}$ and 
$\tilde{y}(N)=y(N)/\sqrt{\beta_y}$) 
is based on the preliminary measurement and fit of 
the orbit response matrix (ORM) and 
dispersion~\cite{prstab_esr_coupling}. 
To this end, only quadrupole and dipole relative 
errors are introduced, with a cutoff in the fit  
so not to overcome the specifications of $10^{-3}$ for 
the quadrupoles. No other source of errors is 
introduced at this stage. As figure of merit to 
assess the quality of the linear lattice model, the 
BPM phase advances are used: The smaller the deviation 
from the theoretical values, the better the model. The BPM 
phase is nothing else than the phase of the tune line 
(i.e. the first harmonic of the FFT of the TbT data) 
plus an arbitrary initial phase which depends on the 
trigger of the acquisition system. Assuming that all 
BPMs are well synchronized, the difference between the 
tune phases at two monitors shall correspond to the BPM
phase advance, the arbitrary phase being canceled out. 
At each pair of BPMs along the ring, the difference 
between measured and model phase advances may be 
evaluated and plotted. Its root-mean-square 
(rms) value over all BPMs may be also computed and used as 
single figure of merit. In Fig.~\ref{fig_BPMPhAdMeas} 
an example is shown. If the ideal model without errors 
is used, large deviations are observed, with an rms of 
about 11~mrad. When the linear error model computed 
from the standard ORM measurement is used, the residual 
rms drops to about 7.5~mrad, with some remaining 
spikes in the vertical plane. On top of the ORM model, 
a response matrix on the BPM phase advance and dispersion 
was built and applied to the measured values to 
further reduce the deviation with the measured data. 
Indeed alternative models may be defined so to have a 
residual rms deviation below 1~mrad. It is interesting 
to note how when going from the ideal model, to the one 
from the ORM, until the one(s) matching the measured BPM 
phase advances, the rms variation along the 
ring of the tune line amplitude, i.e. of the invariant, is 
also reduced, as displayed in Table~\ref{tab:linear-anal}. 

\begin{table*}[!t]
\caption{Kicker strengths and corresponding invariants for the 8 sets
         of TbT BPM data (50 repeated acquisitions). The rms modulation 
         of the tune line amplitude (i.e. of the invariant) and the rms
         deviation from the model BPM phase advance are measured with 
         three different linear lattice models: ideal with no error, with 
         errors inferred from precedent ORM measurement and fit, and from 
         the harmonic analysis of TbT data. Horizontal 
         and vertical data for the BPM phase advance are merged together 
         when computing the rms deviation.
	\vspace{-0.0 cm}}
\vskip 0.0cm
{
\begin{tabular}{c c|c|c|c}
\hline\hline 
\hspace{1.7cm} & \hspace{1.9cm} & 
\hspace{0.9cm}ideal lattice model\hspace{0.85cm}\ & 
\hspace{0.34cm}model with errors (ORM)\hspace{0.33cm}\ &  
\hspace{0.35cm}model with errors (TbT)\hspace{0.45cm}\ 
\\ \hline
\end{tabular}
\begin{tabular}{c|c|c c|c c|c c}
\begin{tabular}{c} (H,V)     \\kicker        \\ strength       \\(A,\ kV)       \end{tabular}&
\begin{tabular}{c} (H,V) mean\\invariant     \\(m$^{1/2}$)\\$\times10^{-4}$\end{tabular}&
\begin{tabular}{c} (H,V) rms \\err. invariant\\(m$^{1/2}$)\\$\times10^{-6}$\end{tabular}&
\begin{tabular}{c} BPM       \\phase advance \\rms error  \\$\times10^{-4}$(rad)\end{tabular}&
\begin{tabular}{c} (H,V) rms \\err. invariant\\(m$^{1/2}$)\\$\times10^{-6}$\end{tabular}&
\begin{tabular}{c} BPM       \\phase advance \\rms error  \\$\times10^{-4}$(rad)\end{tabular}&
\begin{tabular}{c} (H,V) rms \\err. invariant\\(m$^{1/2}$)\\$\times10^{-6}$\end{tabular}&
\begin{tabular}{c} BPM       \\phase advance \\rms error  \\$\times10^{-4}$(rad)\end{tabular}\\
\hline\hline\vspace{-0.25cm} \\ \vspace{0.0 cm}
\ ( \ 50,\ 1.0 )\ &\ ( 0.52, 0.89 )\ \ &( 0.9, 1.6 )&111&( 0.7, 1.0 )& 75&( 0.4, 0.4 )& 16 \\ \hline
\ ( 100,\ 1.0 )\  &\ ( 1.00, 0.90 )\ \ &( 1.7, 1.6 )&110&( 1.3, 1.0 )& 74&( 0.7, 0.4 )& 13 \\ \hline
\ ( 100,\ 1.5 )\  &\ ( 1.00, 1.37 )\ \ &( 1.8, 2.4 )&110&( 1.3, 1.5 )& 74&( 0.7, 0.7 )& 14 \\ \hline
\ ( 200,\ 1.5 )\  &\ ( 1.99, 1.42 )\ \ &( 3.5, 2.5 )&111&( 2.4, 1.6 )& 76&( 1.4, 0.6 )& 11 \\ \hline
\ ( 200,\ 2.1 )\  &\ ( 2.01, 1.98 )\ \ &( 3.5, 3.5 )&111&( 2.4, 2.3 )& 75&( 1.3, 0.9 )&\ 9 \\ \hline
\ ( 250,\ 2.7 )\  &\ ( 2.55, 2.58 )\ \ &( 4.5, 4.5 )&111&( 3.0, 3.3 )& 77&( 1.7, 1.5 )& 12 \\ \hline
\ ( 400,\ 2.5 )\  &\ ( 4.11, 2.52 )\ \ &( 7.5, 4.8 )&119&( 4.8, 4.0 )& 91&( 3.3, 2.2 )& 35 \\ \hline
\ ( 400,\ 3.0 )\  &\ ( 4.14, 2.98 )\ \ &( 7.5, 6.0 )&118&( 4.8, 5.0 )& 91&( 3.5, 3.0 )& 36 \\ \hline
\end{tabular}}
\rule{0mm}{0.0cm}
\label{tab:linear-anal}
\end{table*}

It is also worthwhile noticing how the residual rms error on the 
BPM phase advance has a minimum for the fifth data set of 
Table~\ref{tab:linear-anal}, with 
a kicker excitation so to generate an invariant of about 
$2\times10^{-4}$~m$^{1/2}$. Larger excitations seem 
to induce greater errors that could 
not be correlated to linear lattice elements. Octupolar 
terms affecting the tune lines may be the source of this 
larger discrepancy.

\begin{figure}[b]
\vskip 0.4cm
\centerline{
  \includegraphics[width=8.5cm,angle=0]
  {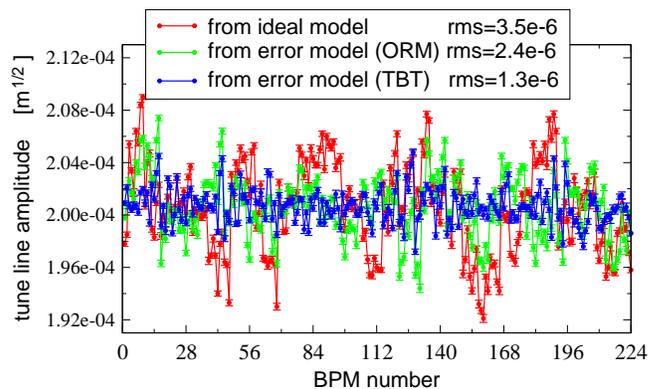}}
\vskip -0.2cm
  \caption{\label{fig_TuneLineMeas6} (Color) Example of measured horizontal 
        tune line amplitude (i.e. invariant) modulation along the ESRF ring 
        when using different lattice models: ideal (blue), with errors 
        inferred from ORM measurement (red) and from TbT data plus 
        dispersion (green). Rms values correspond to the fifth row of 
        Table~\ref{tab:linear-anal}}
\vskip -0.2cm
\end{figure}

As far as the tune line amplitude variation along 
the ring is concerned, all data set with excitation 
lower than $2.6\times10^{-4}$~m$^{1/2}$ 
show a modulation below $1\%$ (with the TbT model). 
According to tracking simulation results of 
Table~\ref{tab:track-scan},  second-order sextupolar 
terms contribute for a fraction of this modulation 
only. When comparing modulation observed in 
single-particle tracking (Fig.~\ref{fig_Tuneline}) and 
that measured with the strongest excitation 
(Fig.~\ref{fig_TunelineMeas}), even though a 
similar pattern may be guessed in the horizontal plane, 
the fast modulation observed in the vertical plane 
is unexpected. This indeed can be explained by small 
differences in the signal attenuation induced by BPM 
cables of different lengths (an effect visible in 
the {\sl sum} BPM signal). 
For sake of completion effective BPM gains have been 
computed so to cancel the tune line amplitude modulation 
corresponding to the fifth data set of 
Table~\ref{tab:linear-anal}, though they turned out to 
play a limited role in the nonlinear analysis. 

\begin{figure}
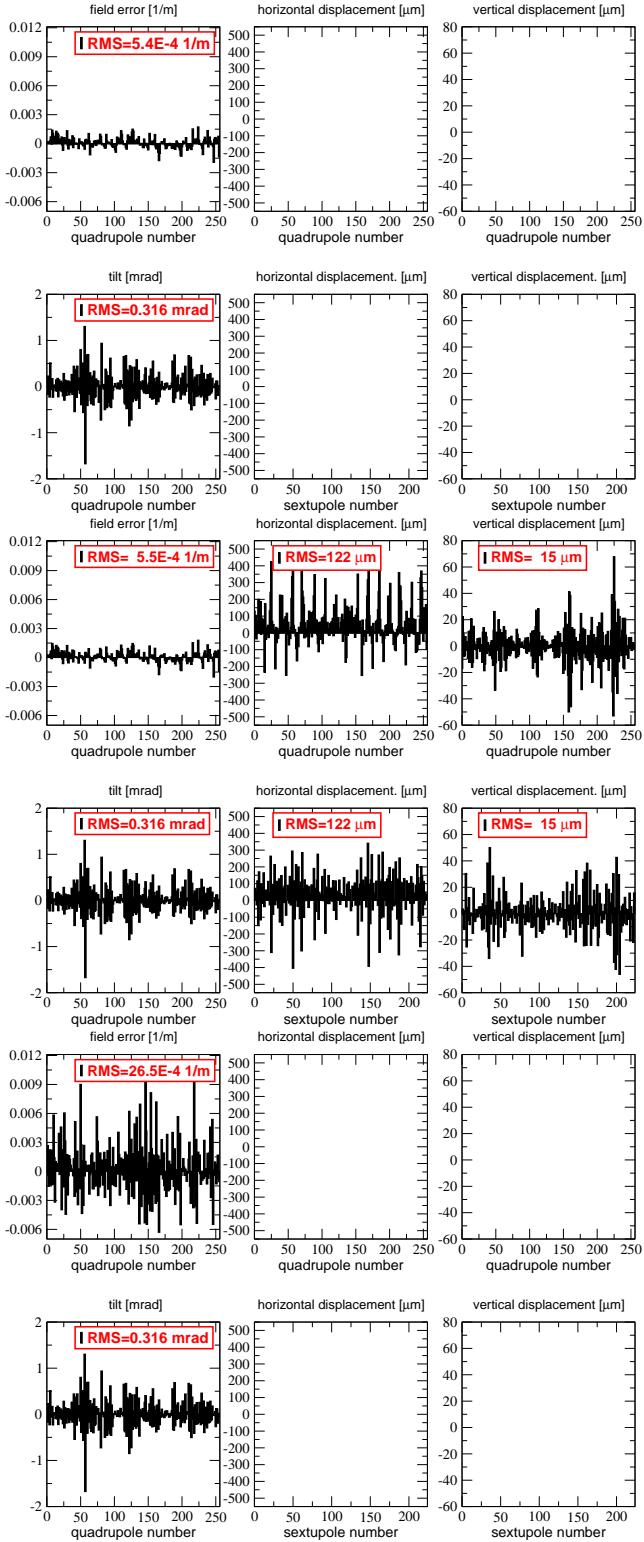

\rule{0mm}{0mm}
\centerline{
  \includegraphics[width=8.5cm,angle=0]
  {fig06A.eps}}
\rule{0mm}{3mm}
\centerline{
  \includegraphics[width=8.5cm,angle=0]
  {fig06B.eps}}
\rule{0mm}{3mm}
\centerline{
  \includegraphics[width=8.5cm,angle=0]
  {fig06C.eps}}
\vspace{-0mm}
  \caption{\label{fig_LinLattModel}(Color) Top: lattice error model inferred 
  from ORM measurement with quadrupole errors only (dipole errors are not 
  displayed here), BPM phase advance rms error of 7.5~mrad. 
  Centre: lattice error model with magnet displacements (rms error 0.9~mrad). 
  Bottom: lattice model with equivalent rms error of 0.9~mrad, 
  but (large) quadrupole errors only.}
\rule{0mm}{0mm}
\end{figure}

A brief digression on the effective linear lattice model 
inferred from the the BPM phase advance (and dispersion) 
response matrix is worth to be introduced. The matrix has been 
built having different error sources: besides quadrupole 
field errors and rotation (for coupling), quadrupole and 
sextupole displacements (longitudinal, horizontal and 
vertical) have been introduced along with longitudinal 
dislocations of BPMs. It turned out that longitudinal 
displacements have a negligible impact on the BPM phase 
advances, whereas this is not the case for the other 
two axes. The best SVD fit was obtained with the error 
model displayed in the centre plot of Fig.~\ref{fig_LinLattModel}. 
The corresponding rms values are compatible with specifications 
and well within the results from survey 
measurements~\cite{NoteSurvey}. Such a model, however, 
is difficult to handle, because magnet displacement would 
require the distorted orbit to be taken into account, 
complicating the overall analysis. Nevertheless, the same 
SVD fit could be obtained with quadrupole field errors only, 
at the price of increasing unrealistically their values, 
well above their specifications (bottom plot of 
Fig.~\ref{fig_LinLattModel}).  Nonetheless, this 
has the advantage of not introducing any complication 
with the orbit and large quadrupole errors are to be 
considered as representative of the sextupole displacement 
too. The nonlinear analysis discussed in the next section 
is based on this later baseline linear model. 

\begin{figure}[b]
\rule{0mm}{3mm}
\centerline{
  \includegraphics[width=8.5cm,angle=0]
  {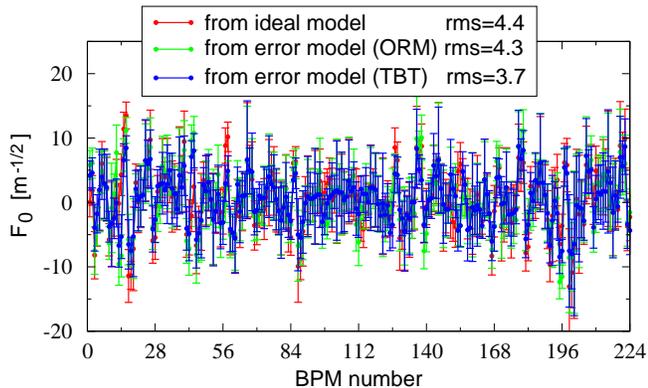}}
  \caption{\label{fig_F0Meas} (Color)
	Example of measured $F_0$ along the ESRF storage ring when 
        different linear models ($\beta$ functions at BPMs) are used: 
        ideal lattice(blue), with errors inferred from ORM measurement 
        (red) and TbT data plus dispersion (green). Data correspond to the 
        fifth row of Table~\ref{tab:linear-anal} (average over 50 
        acquisitions).}
\vskip -3mm
\end{figure}

Even though not used for the determination of linear 
lattice model, it is interesting to note how the 
CRDTs cancellation condition defined by $F_0$ of  
Eq.~\eqref{eq:F0} is improved when the BPM phase advance 
is corrected. In general, the closer $F_0$ is to zero 
among all BPMs, the more accurate is the harmonic analysis. 
$F_0$ is evaluated with the CRDTs measured from the 
sextupolar spectral lines according to 
Table~\ref{tab:line-rdt}. Its rms values (for the 
fifth data set of 
Table~\ref{tab:linear-anal}) goes from 4.4~m$^{-1/2}$ 
with the ideal linear model, to 4.3~m$^{-1/2}$  
with the ORM error model, and eventually to 3.2~m$^{-1/2}$ 
with the model based on the 
BPM phase advance, even though large fluctuations 
among the 50 acquisitions are present, as shown 
in Fig.~\ref{fig_F0Meas}.

In conclusion, the linear analysis of the tune line 
(amplitude and phase) provided an effective linear 
model that improves considerably the quality of the 
harmonic analysis. The deviation between model and 
measured BPM phase advance is reduced by a factor 
greater than ten. The modulation of the tune line 
(i.e. of the invariant) is reduced from $2\%$ to 
less than $1\%$. The cancellation term $F_0$ is 
reduced by about $20\%$. Effective BPM gains could 
be inferred from the tune line amplitude modulation.

\subsection{Nonlinear analysis of TbT BPM data}
\label{sec_NonLinMeas}
\begin{figure*}
\rule{0mm}{0mm}
\centerline{
  \includegraphics[width=17.0cm,angle=0]
  {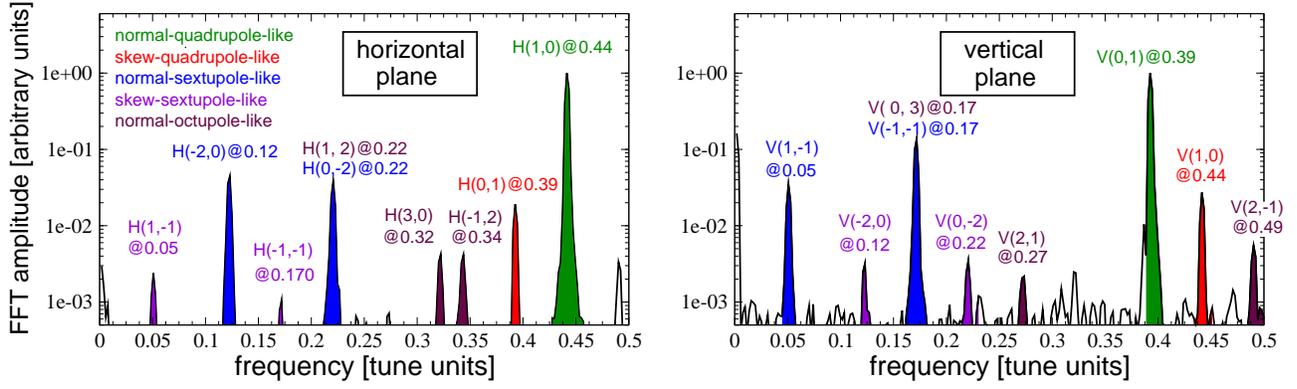}}
\vspace{-0mm}
  \caption{\label{fig_FFT_example}(Color) Example of horizontal (left) and 
           vertical (right) spectral lines inferred 
           from the FFT of TbT BPM data measured after exciting transversely 
           the beam with the strongest kicker strengths of 
           Table~\ref{tab:Nonlinear-F0_1}. A special sextupole setting was 
           used to reduce detuning with amplitude and chromaticity and to 
           enhance the normal sextupole spectral lines. Higher-order  
           octupolar-like lines  (in maroon) are also excited and 
           measurable, though they may overlap the sextupolar-like, 
           corrupting the CRDTs measurement and resulting in large 
           $F_0$. \vspace{0.0cm}}
\end{figure*}
\begin{figure*}
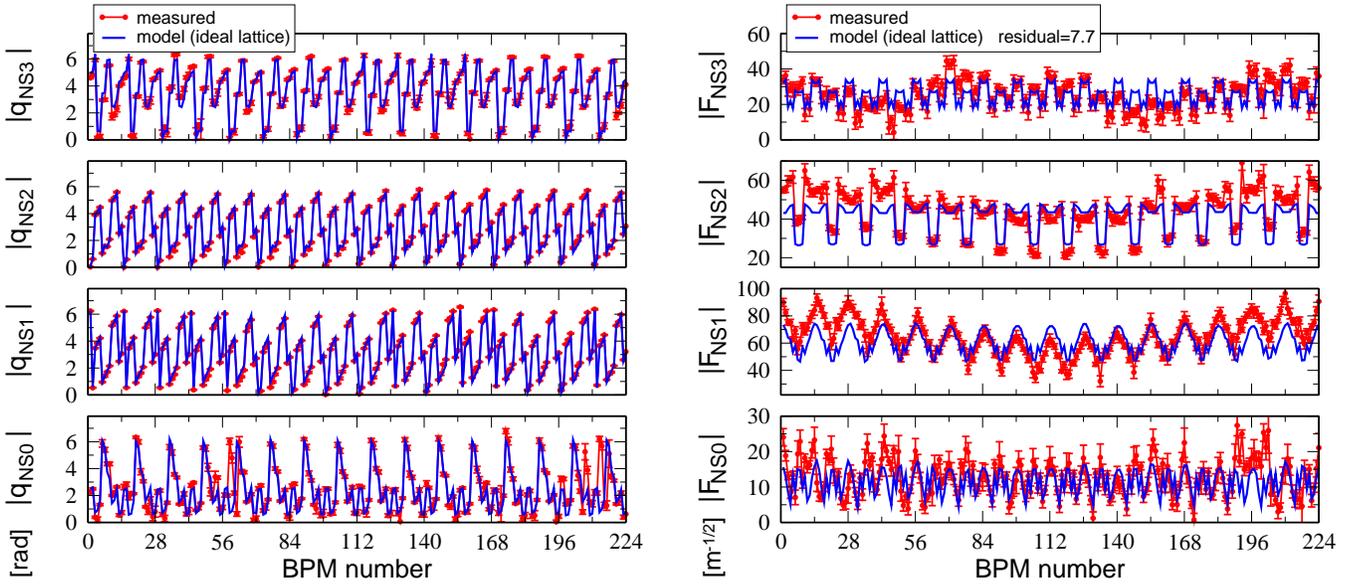

\rule{0mm}{0mm}
\centerline{
  \includegraphics[width=8.5cm,angle=0]
  {fig08A.eps} \hspace{0.6cm}
  \includegraphics[width=8.5cm,angle=0]
  {fig08B.eps}}
\vspace{-0mm}
  \caption{\label{fig_CRDTPhaseMeasID}(Color) Red: CRDT phases (left) and 
           amplitudes (right) measured at the ESRF storage ring (with 
           a special dedicated optics). Blue: the same quantities computed 
           from the ideal perfect lattice. The residual for this model is 
           $\mathcal{R}=7.7$~m$^{-1/2}$.\vspace{-0.4cm}}
\end{figure*}

\begin{table}[b]
\caption{Measured rms variation along the ESRF storage ring of  
         $F_0$ against different lattice models (i.e. $\beta$ 
         functions at BPMs) for all acquired data sets. The mean 
         value is between -0.4 and 0.8 depending on the set (50 
         acquisition each, FFT over 256 turns).\vspace{-0.0 cm}}
\vskip 0.0cm
{
\begin{tabular}{c c|c}
\hline\hline 
\hspace{1.5cm} & \hspace{2.15cm} & 
\hspace{0.65cm}rms $F_0$ (m$^{-1/2}$) \hspace{0.65cm}\ 
\\ \hline
\end{tabular}
\begin{tabular}{c|c|c c c}
\begin{tabular}{c} (H,V)     \\kicker   \\ strength       \\(A,\ kV)       \end{tabular}&
\begin{tabular}{c} (H,V) mean\\invariant\\m$^{1/2}$\\$\times10^{-4}$\end{tabular}&
\begin{tabular}{c} ideal     \\lattice  \\model           \\ \end{tabular}&
\begin{tabular}{c} error     \\model    \\(ORM)           \\ \end{tabular}&
\begin{tabular}{c} error     \\model    \\(SVD)           \\ \end{tabular}\\
\hline\hline\vspace{-0.25cm} \\ \vspace{0.0 cm}
\ ( \ 50,\ 1.0 )\ &\ ( 0.52, 0.89 )\ \ & 8.4 & 8.8& 8.5 \\
\ ( 100,\ 1.0 )\  &\ ( 1.00, 0.90 )\ \ & 5.6 & 5.9& 5.4 \\
\ ( 100,\ 1.5 )\  &\ ( 1.00, 1.37 )\ \ & 4.8 & 5.0& 4.4 \\
\ ( 200,\ 1.5 )\  &\ ( 1.99, 1.42 )\ \ & 4.4 & 4.4& 3.9 \\
\ ( 200,\ 2.1 )\  &\ ( 2.01, 1.98 )\ \ & 4.4 & 4.3& 3.7 \\
\ ( 250,\ 2.7 )\  &\ ( 2.55, 2.58 )\ \ & 5.1 & 4.9& 4.3 \\
\ ( 400,\ 2.5 )\  &\ ( 4.11, 2.52 )\ \ & 7.8 & 7.5& 7.5 \\
\ ( 400,\ 3.0 )\  &\ ( 4.14, 2.98 )\ \ & 7.9 & 7.6& 7.6 \\\hline
\end{tabular}}\vspace{-0.5cm}
\label{tab:Nonlinear-F0_1}
\end{table}

The effective linear lattice error model discussed in 
the previous section may now be used to evaluate the modulated 
C-S parameters necessary to normalize the TbT data ($\beta$ 
functions) and to evaluate the model CRDTs from 
Tables~\ref{tab:lattice-rdt} and \ref{tab:line-selection}.  
The measured CRDTs are instead inferred from the sextupolar 
lines $H(-2,0)$, $H(0,-2)$ and $V(\pm1,-1)$ according to 
Table~\ref{tab:line-rdt}. These lines are represented in blue 
in the example of measured spectra of Fig.~\ref{fig_FFT_example}. 
As for the linear analysis, a data 
set among those acquired at different kicker strengths is to 
be chosen, considering that a too weak excitation may affect 
the measurement because of the low signal-to-noise ratio, 
whereas a too strong excitation may introduce high-order 
terms not included in the present analysis.
As discussed in Sec.~\ref{sec_precautions}, the nonlinear 
analysis presented here is valid as long as the CRDT 
residual $F_0$ of Eq.~\eqref{eq:F0} is much lower than the 
average CRDT amplitudes. The rms values of $F_0$ (computed after 
averaging among the 224 BPMs and the 50 acquisitions) 
for each data set and for different linear lattice models 
are reported in Table~\ref{tab:Nonlinear-F0_1}. 
Because of the ESRF storage ring tune working point ($Q_x=36.440$ 
and $Q_y=13.390$) two sextupolar lines, $H(0,-2)$ and $V(-1,-1)$, 
necessary for the evaluation of as many CRDTs 
receive a contribution from higher-order octupolar terms, 
as shown in Fig.~\ref{fig_FFT_example}. 
A complete list of the octupolar spectral lines may be found 
in Ref.~\cite{Bartolini1}. If not 
taken into account or avoided, this superposition will corrupt 
the CRDT measurement and the overall nonlinear model. Even 
though it is not possible to disentangle the individual contributions,  
the CRDT residual $F_0$ may be evaluated: the closer to zero, the 
less detrimental are the octupolar-like terms. Another way out 
may be found by slightly detuning the machine so to ensure 
a sufficient separation between these lines: By setting for 
example $Q_x=36.438$ and $Q_y=13.385$ a separation of 0.022 in 
tune units would be assured. In the acquired data (at nominal 
tunes) the mean value of $F_0$ is always 
very close to zero (between -0.4 and 0.8~m$^{-1/2}$, 
depending on the data set). The lowest rms $F_0$ 
(3.7~m$^{-1/2}$) is obtained 
for an excitation corresponding to a mean invariant (in both 
planes) of $2.0\times10^{-4}$~m$^{1/2}$. The CRDTs 
measured from this data set are then used for the nonlinear 
lattice modelling.

\begin{figure*}
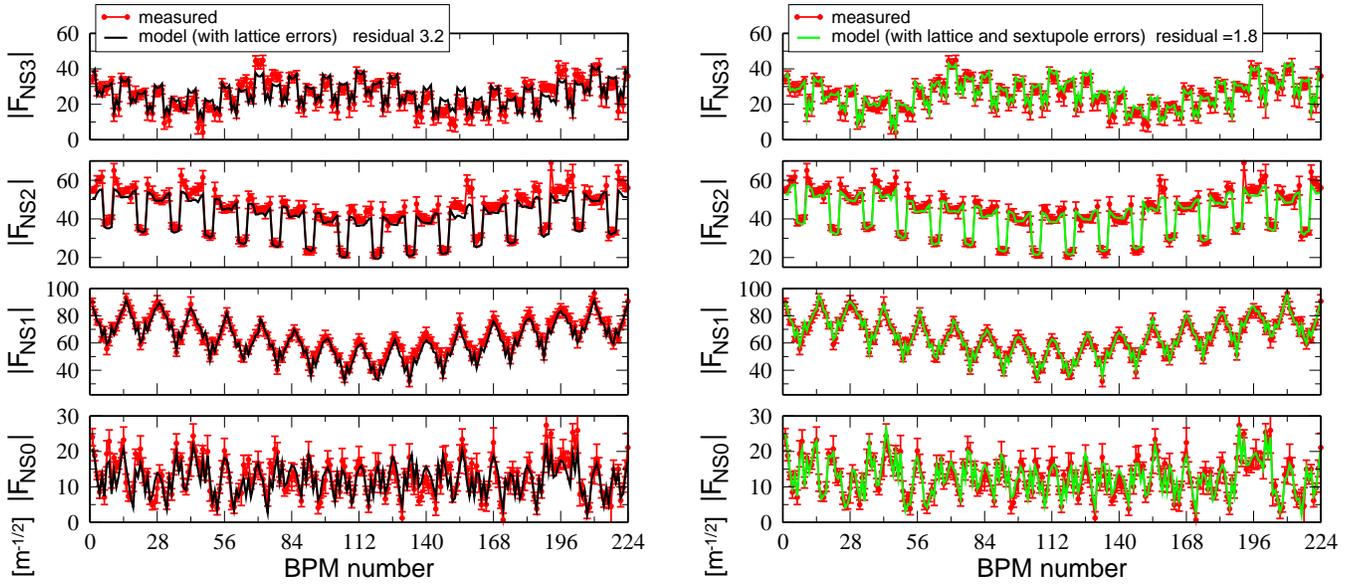

\rule{0mm}{0mm}
\centerline{
  \includegraphics[width=8.5cm,angle=0]
  {fig09A.eps} \hspace{0.6cm}
  \includegraphics[width=8.5cm,angle=0]
  {fig09B.eps}}
  \caption{\label{fig_CRDTAmpliMeasER}(Color) Amplitudes of normal sextupole 
           CRDTs (red curves) 
           measured at the ESRF storage ring (with a special dedicated 
           optics). Left: comparison with the model curves obtained from 
           the lattice model comprising  ideal sextupoles and focusing 
           errors and coupling (residual $\mathcal{R}=3.2$~m$^{-1/2}$). 
           Right:  comparison with the the same linear model after 
           introducing sextupole errors and sextupolar components 
           in the main bending magnets ($\mathcal{R}=1.8$~m$^{-1/2}$). 
           \vspace{-0.4cm}}
\rule{0mm}{0mm}
\end{figure*}

As for the linear analysis, a figure of merit needs to 
be defined to quantify the goodness of the sextupole 
model. The difference between the measured CRDT vector 
and the one for model may be used to this end. The rms 
value of this vector, i.e. the residual $\mathcal{R}$, 
provides a figure of merit for the model,
\begin{eqnarray}
\hbox{residual\ }\mathcal{R}&=&\sqrt{<|\vv{F}_{NS,meas}-\vv{F}_{NS,mod}|^2>}\\
&=&\sqrt{\frac{1}{N}\sum_{i=1}^N{(F_{NS,meas}(i)-F_{NS,mod}(i))^2}}
\nonumber \ , \label{eq_FNS_residual}
\end{eqnarray}
where $N=8\cdot N_{BPM}$ (the 4 CRDTs are complex 
quantities, separated in real and imaginary parts, 
see Eq.~\eqref{eq_FNS_FFT}). 
It is worthwhile noticing that $\mathcal{R}$ depends 
greatly on the lattice: Different machines and optics 
may result in different residuals for a similarly good 
sextupole model. In order to compare different storage 
rings or settings, the residual may be normalized by the 
mean CRDT amplitude.

A convenient way to display measured and model CRDTs 
is to separate their phase and amplitude. In the left plot of 
Fig.~\ref{fig_CRDTPhaseMeasID} the measured CRDT phases 
are shown together with the corresponding expectation 
from the ideal model of the ESRF storage ring. 
The agreement is already remarkable, 
as are the small statistical error bars. When comparing 
the CRDT amplitude instead (right plot of Fig.~\ref{fig_CRDTPhaseMeasID}), 
the 16-fold periodicity of the ideal curves is 
modulated in the measured CRDTs. The overall residual is 
$\mathcal{R}=7.7$~m$^{-1/2}$. 

Interestingly, in an earlier analysis based on a 
sextupole calibration curve measured in 2001 the initial 
residual was about $10\%$ higher ($\mathcal{R}=8.4$~m$^{-1/2}$) 
and the magnetic errors inferred from 
the measured CRTDs yielded suspiciously large 
errors in some sextupole families. This curve was indeed 
obtained after cycling a spare sextupole to 250~A. On the 
other hand, those installed in the ring (mechanically and 
magnetically identical to the spare one, with a few 
exceptions) are grouped in seven families (3 chromatic 
and 4 harmonic) subjected to different cyclings: 
close to 250~A for the chromatic sextupoles, between 
120 and 150~A for the harmonics. 
These facts triggered a new campaign of magnetic 
measurements in 2012, reproducing in the laboratory 
the different cyclings. These new calibration curves are 
used since then for this analysis, resulting in the 
smaller initial residual of $\mathcal{R}=7.7$~m$^{-1/2}$ 
and in much lower errors in the harmonic sextupoles. This 
was the first evidence of the predicting power of this 
new technique. 

The observed modulation of the CRDT amplitudes on top of 
the 16-fold periodicity is greatly reproduced after 
introducing in the model the linear lattice errors, as 
shown in the left plot of Fig.~\ref{fig_CRDTAmpliMeasER}. The 
modulated $\beta$ functions and phase advances seem to be 
an important source, the residual $\mathcal{R}$ dropping by  
more than $60\%$ to 3.2~m$^{-1/2}$. It should be mentioned 
that the ESRF storage ring suffers from large $\beta$ beating  
(about $2-3\%$ rms) compared to more recent third generation 
light sources (well below $1\%$), because only 32 corrector 
quadrupoles are available for its correction (out of 256 
magnets grouped in 6 families). The modulation of the CRDT 
amplitudes in other machines, then, shall depend greatly on 
their level of $\beta$ beating. As far as the ESRF storage 
ring is concerned, this second-order effect of quadrupole 
errors on sextupole resonances generates in some regions 
CRDTs about $50\%$ (up to $100\%$ for the vertical CRDT 
$F_{NS0}$) larger than in an ideal machine (see right plot 
of Fig.~\ref{fig_CRDTPhaseMeasID}). This may have a considerable 
impact on the dynamic aperture of the ring (the larger 
the CRDTs, the greater the phase space distortion), 
independently on any possible sextupole error.

\begin{figure}
\rule{0mm}{0mm}
\centerline{
  \includegraphics[width=8.5cm,angle=0]
  {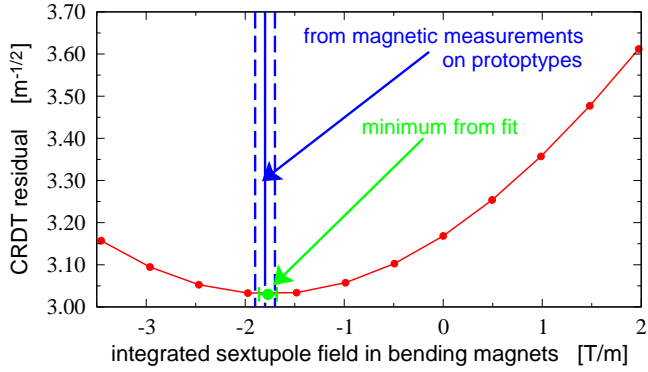}}
  \caption{\label{fig_CRDT_Residual}(Color) Dependence of the residual 
    $\mathcal{R}$ on the sextupolar component in bending magnets, where 
    the blue line indicates the field measured on two prototypes. The 
    residual is computed before introducing errors in the 224 sextupoles.}
\rule{0mm}{0mm}
\end{figure}

Before performing the SVD pseudo-inversion of the linear 
system of Eq.~\eqref{eq_FNS_diff} to infer sextupole 
errors, the measured CRDTs are compared to the model 
after introducing sextupolar fields in the 64 main 
bending magnets. Measurements performed on prototypes in 
the early 90s~\cite{Laurent-MeasurDipoles} indicated an 
integrated sextupole field of $-1.8\pm0.1$ T/m. The 
plot of Fig.~\ref{fig_CRDT_Residual} shows how the residual 
$\mathcal{R}$ reaches a minimum with 
a field of $-1.77\pm0.09$ T/m, 
The agreement is excellent, again confirming the reliability 
of this measurement.
Interestingly, when solving the system of 
Eq.~\eqref{eq_FNS_diff} to evaluate the sextupole errors, 
the average error per family is reduced by about one order of 
magnitude after introducing the -1.77 T/m sextupole field 
in all 64 bending magnets, as reported in Table~\ref{tab-ErrFam1}. 

\begin{table}
\caption{ Mean and rms relative sextupole field errors (per family) 
          inferred from the CRDTs measurement. The first column 
          corresponds to the bare machine, whereas the second 
          column is obtained after inserting a sextupole 
          field in the main 64 bending magnets. This improves 
          the overall residual $\mathcal{R}$ (see 
          Fig.~\ref{fig_CRDT_Residual}) and lowers the average
          errors in the 224 sextupoles by about one order of 
          magnitude. \vspace{-0.0 cm}}
\vskip 0.0cm
{
\begin{tabular}{c}
\hline \hline
\vspace{-0.2cm}\\
\hspace{0.0cm}
relative error [$\%$] (average $\pm$ rms) per family
\hspace{0.0cm}\ 
\vspace{-0.25cm}\\ \\ \hline
\end{tabular}
\begin{tabular}{r|c|c}
\hspace{0.3cm}family\ \ &\ \ with baseline\hspace{0.32cm}\ &\ \ with -1.77 T/m\hspace{0.32cm}\ \\
                        &\   model                  \   &\  in bendings    \   \\
\hline\hline\vspace{-0.25cm} \\ \vspace{0.0 cm}
 S4\ \ & -0.12\ $\pm$\ 0.79& -0.02\    $\pm$\ 0.80\\
 S6\ \ &\ 0.20\ $\pm$\ 0.43& -0.02\    $\pm$\ 0.43\\
S13\ \ & -0.14\ $\pm$\ 0.34&\ 0.02\    $\pm$\ 0.34\\
S19\ \ &\ 0.04\ $\pm$\ 0.11&$\sim$0.0\ $\pm$\ 0.11\\
S20\ \ & -0.07\ $\pm$\ 0.11&\ 0.01\    $\pm$\ 0.11\\
S22\ \ &\ 0.24\ $\pm$\ 0.41& -0.03\    $\pm$\ 0.41\\
S24\ \ & -0.09\ $\pm$\ 0.41&\ 0.01\    $\pm$\ 0.42\\
\hline
\end{tabular}}\vspace{0.5cm}
\label{tab-ErrFam1}
\end{table}

\begin{figure}
\rule{0mm}{0mm}
\centerline{
  \includegraphics[width=8.5cm,angle=0]
  {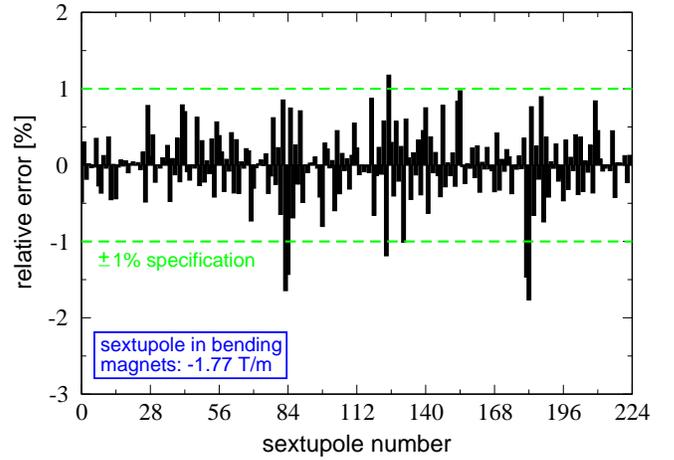}}
\rule{0mm}{0mm}
\vspace{-5mm}
  \caption{\label{fig_SextErrFit1}(Color) Errors in the 224 sextupoles 
    as computed from the (SVD) pseudo-inversion of the system in 
    Eq.~\eqref{eq_FNS_diff}. Quality specifications for those magnets 
    are at $\pm1\%$. The SVD is performed with 26 (out of 220) 
    eigen-values and sextupoles are assumed to be perfectly positioned 
    longitudinally. The residual with this set is $\mathcal{R}=1.8$~m$^{-1/2}$.
    For each family, mean and rms errors are reported in 
    the last column of Table~\ref{tab-ErrFam1}.}
\rule{0mm}{0mm}
\end{figure}

However, in both cases the sextupole family S4 seems to 
have a larger rms spread ($0.8\%$) compared to other 
families (between $0.1\%$ and $0.4\%$). The magnets 
exhibiting the largest deviations turned out to be the 
only four (recently installed) independent sextupoles, with 
a relative error of about $-1.5\%$. These are shorter than 
the standard ones (20 cm instead of 40 cm) and have been 
displaced by 10 cm with respect to the nominal position 
so to lengthen two straight sections. New positions 
and calibration curves have been included in the model (the 
integrated strength shall be the same of the standard S4 
sextupoles). A more detailed analysis on 
these magnets is reported in Sec.~\ref{sec_SextCalib}. 
When introducing the sextupole field in the bending magnets 
and fitting the 224 magnet errors by inverting the system 
of Eq.~\eqref{eq_FNS_diff}, the residual drops to 
$\mathcal{R}=1.8$~m$^{-1/2}$: 
The corresponding CRDT amplitudes are displayed in the right 
plot of Fig.~\ref{fig_CRDTAmpliMeasER}, while the errors 
(mean and rms of each family) are reported in the last column 
of Table~\ref{tab-ErrFam1}: The great majority of the 
sextupole errors is well below the $\pm1\%$ specification, 
as displayed in Fig.~\ref{fig_SextErrFit1}.

Of course, results vary according to the parameters for the 
SVD pseudo-inversion of Eq.~\eqref{eq_FNS_diff}. All the 
results shown in this section are derived by cutting the 
number of eigen-vectors to 26 (out of 224): more vectors 
induce a limited reduction of less than 0.1~m$^{-1/2}$ 
in the final residual $\mathcal{R}$ at the expenses of large 
sextupole errors, well beyond $1\%$. 

The sextupole field errors of Fig.~\ref{fig_SextErrFit1} 
are computed assuming that all 224 sextupoles are 
placed at their nominal longitudinal position along 
the ring. Any displacement along $s$, e.g. the 
circumference, would change the CRDTs through the $\beta$ 
functions and phases $\phi$ (see Table~\ref{tab:lattice-rdt}) 
because of the dependence of the latter on $s$. Part of the 
field errors attributed to the magnets by the CRDTs may 
actually be a combination of magnetic imperfection and longitudinal 
displacement. Periodic metrological surveys and adjustments 
are carried out at the ESRF storage ring twice per year. 
However, they ensure a state-of-the-art alignment on the 
transverse plane only. A longitudinal alignment was carried 
out in the early 90s during the installation only, with 
an accuracy estimated in the mm range. 
To understand the contribution of longitudinal 
displacements $\delta s$ on the CRDTs and, hence, on the 
equivalent field errors in the sextupole, the response 
matrix of Eq.~\eqref{eq_FNS_diff} was extended, namely 
\begin{eqnarray}
\vv{F}_{NS,meas}-\vv{F}_{NS,mod}&=&\mathbf{M_{NS}}
\left(\begin{array}{c} w_K\cdot \vv{\delta K}_2 \\  
                       w_s\cdot \vv{\delta s}
      \end{array}\right)
\ , \label{eq_FNS_diff2}
\end{eqnarray}
where $\vv{\delta s}$ contains the $N_{sext}=224$ sextupole 
longitudinal displacements, whereas $w_K$ and $w_s=1-w_K$ 
are introduced to weigh out the two contributions. 
$\mathbf{M_{NS}}$ is now a $8\cdot N_{BPM} \times 
2\cdot N_{sext}$ response matrix. The above system has 
been inverted with three different sets of weights. 
It has been observed that by allowing rms longitudinal 
displacement up to about 3~mm the magnetic errors drop by 
about $15\%$.

\begin{figure*}
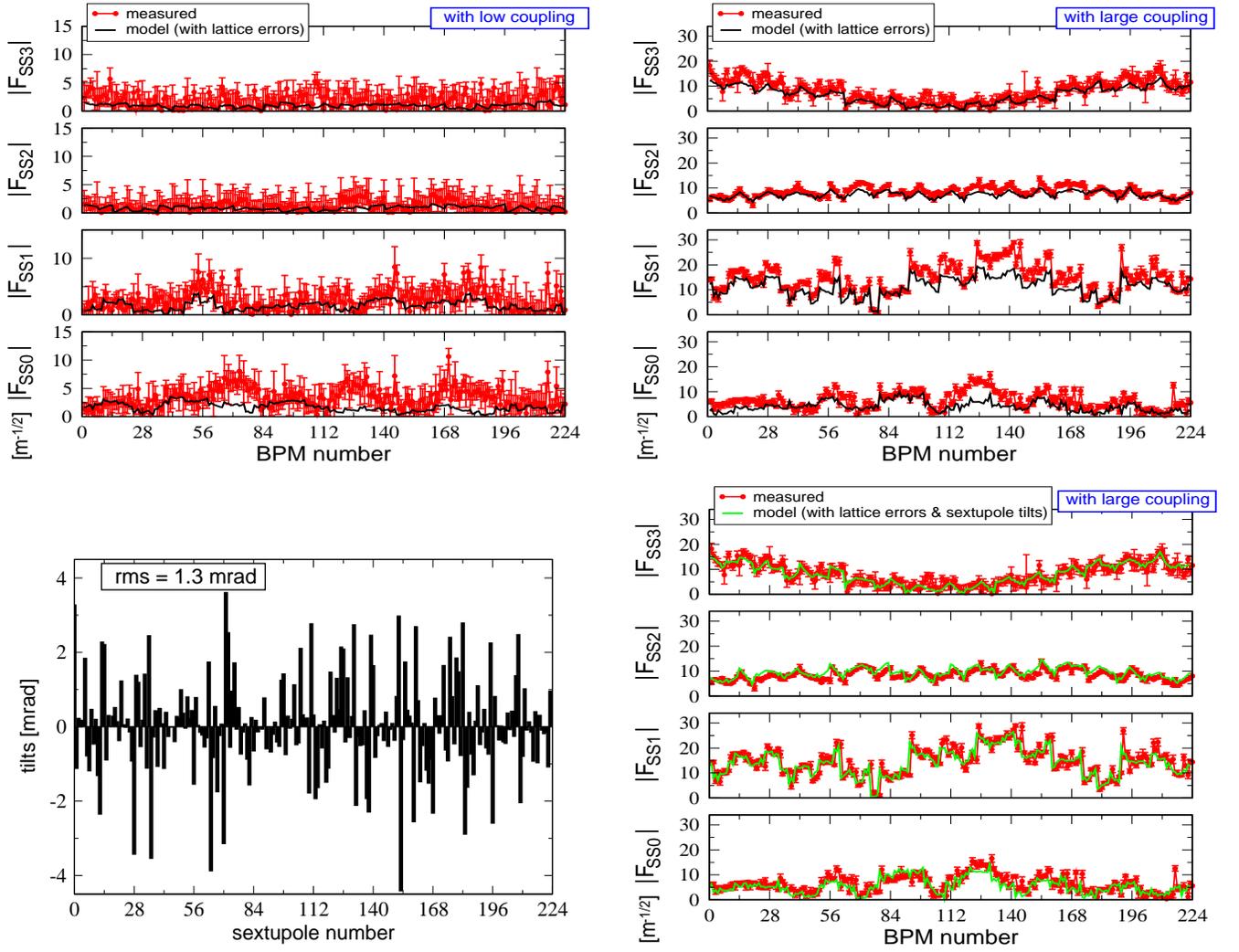

\rule{0mm}{0mm}
\centerline{
  \includegraphics[width=8.5cm,height=6.7cm,angle=0]
  {fig12A.eps}\hspace{0.6cm}
  \includegraphics[width=8.5cm,height=6.7cm,angle=0]
  {fig12B.eps}}
  \vspace{0.3cm}
\centerline{\hspace{0.2cm}
  \includegraphics[width=8.0cm,angle=0]
  {fig12C.eps}\hspace{0.95cm}
  \includegraphics[width=8.5cm,height=6.7cm,angle=0]
  {fig12D.eps}}
\vspace{-0mm}
  \caption{\label{fig_CRDTAmpliSkewSext} (Color) Top: amplitude of skew 
           sextupole CRDTs measured at the ESRF storage ring (red) and 
           from the model with no sextupole rotation (black), with nominal 
           low betatron coupling (left) and with larger coupling excited 
           by a well-calibrated skew quadrupole (right). 
           Bottom left: tilts the 224 sextupoles as computed from the (SVD) 
           pseudo-inversion of the system of Eq.~\eqref{eq_FSS_diff} 
           and from Eq.~\eqref{eq_FSS_tilt}. 
           The SVD is performed with 35 (out of 224) eigen-values. 
           Bottom right: model (green) and measured (red) CRDTs after 
           introducing these tilts.\vspace{-0.4cm}}
\rule{0mm}{0mm}
\end{figure*}

In Fig.~\ref{fig_FFT_example} four spectral lines 
excited by skew sextupolar terms, $H(\pm1,-1)$, 
$V(-2,0)$ and $V(0,-2)$ are visible, hence measurable. 
Since no physical skew sextupole is installed in the 
machine, they are generated only by tilts of the main 
normal sextupoles (or their residual skew components) 
and by the cross product between residual coupling 
and normal sextupoles. The former is a first-order 
contribution, i.e. it can be well described by the 
RDTs $f_{jklm}$ of Table~\ref{tab:lattice-rdt}, 
whereas the latter is a second-order term that can 
be correctly described only by the second-order RDTs 
$g_{jklm}$ of Table~\ref{tab:line-selection}, to be 
evaluated from the model as described in 
Table~\ref{tab:g_jklm_SS} of Appendix~\ref{app:1}.  
From these four spectral lines, skew sextupole CRDTs 
$F_{SS}$ can be measured after applying the corresponding 
formulas of Table~\ref{tab:line-rdt}. However, because 
of the low coupling and small sextupole rotations, the lines 
are about one order of magnitude lower than those excited 
by normal sextupoles (see Fig.~\ref{fig_FFT_example}). 
Hence, the signal-to-noise ratio becomes an issue and 
the measurement of the skew sextupole CRDTs is less 
reliable, as can be seen from the top left plot of 
Fig.~\ref{fig_CRDTAmpliSkewSext} where the error bars 
(evaluated from the statistics over 50 acquisitions) 
are of the same order of magnitude of the baseline 
skew sextupole CRDTs.  In order to enhance the 
signal-to-noise ratio of the four lines, another series 
of measurements have been taken after introducing 
large coupling, by powering a well-calibrated skew 
quadrupole corrector. By doing so, the second-order 
contribution to $\vv{F}_{SS,mod}$ of Eq.~\eqref{eq_FSS_diff}
from $g_{jklm}$ become dominant, though 
it can be evaluated from the model with great accuracy. 
The four skew sextupole spectral lines are now 
enhanced well above the noise level and the 
CRDT measurement become more reliable, with smaller 
error bars (in relative terms), as shown in the top right 
plot of Fig.~\ref{fig_CRDTAmpliSkewSext}.  As done for the 
sextupole field error, the system of Eq.~\eqref{eq_FSS_diff} 
may be pseudo-inverted via SVD and an effective model of 
sextupole tilts may be inferred from Eq.~\eqref{eq_FSS_tilt}. 
In the bottom left plot of Fig.~\ref{fig_CRDTAmpliSkewSext} 
a model obtained with 35 out of 224 eigen-values is displayed, 
with the corresponding CRDTs shown in the bottom 
right plot of the same figure.

\begin{figure}
\centerline{
  \includegraphics[width=4.2cm,height=5.25cm,angle=0]
  {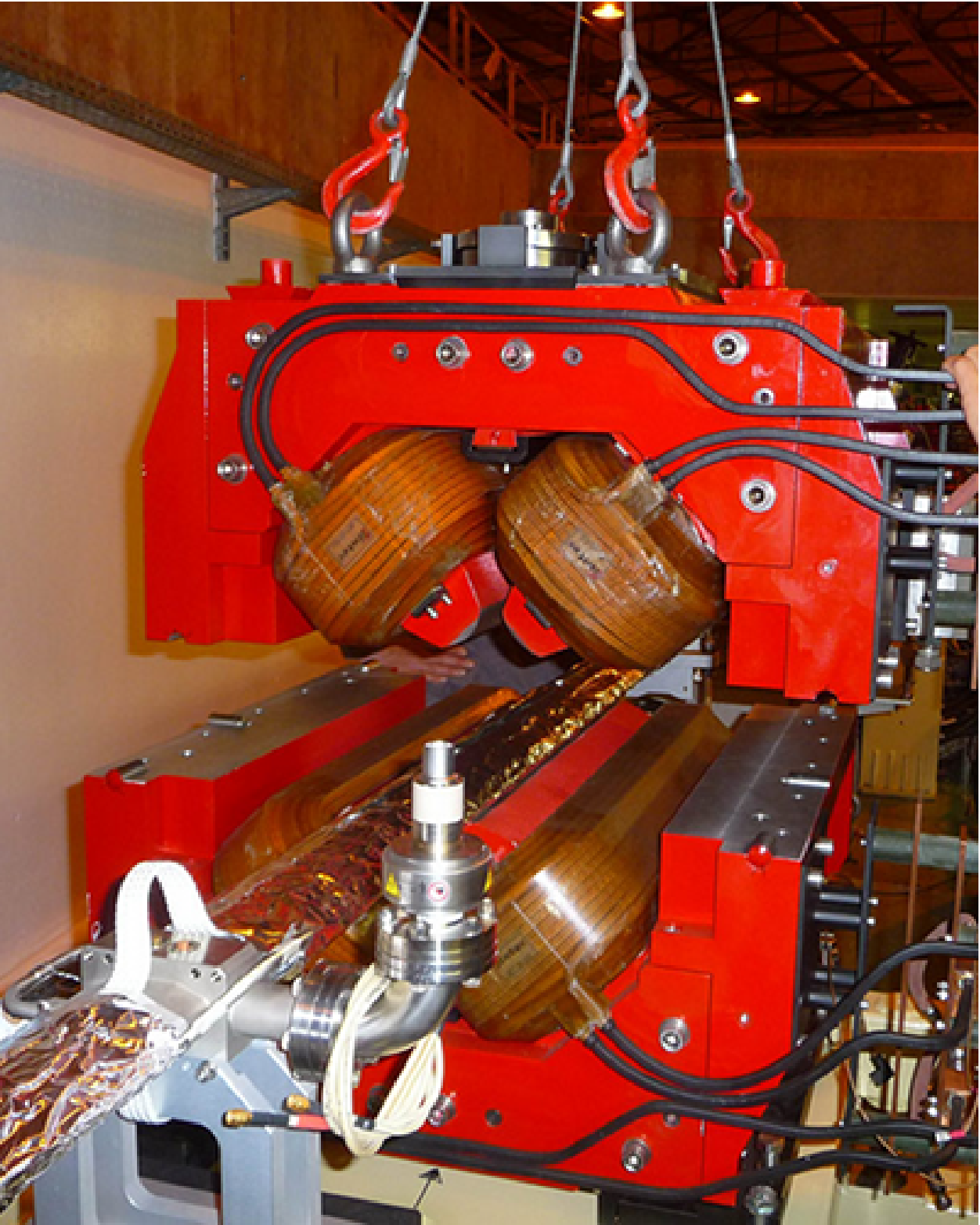}
  \includegraphics[width=4.2cm,height=5.25cm,angle=0]
  {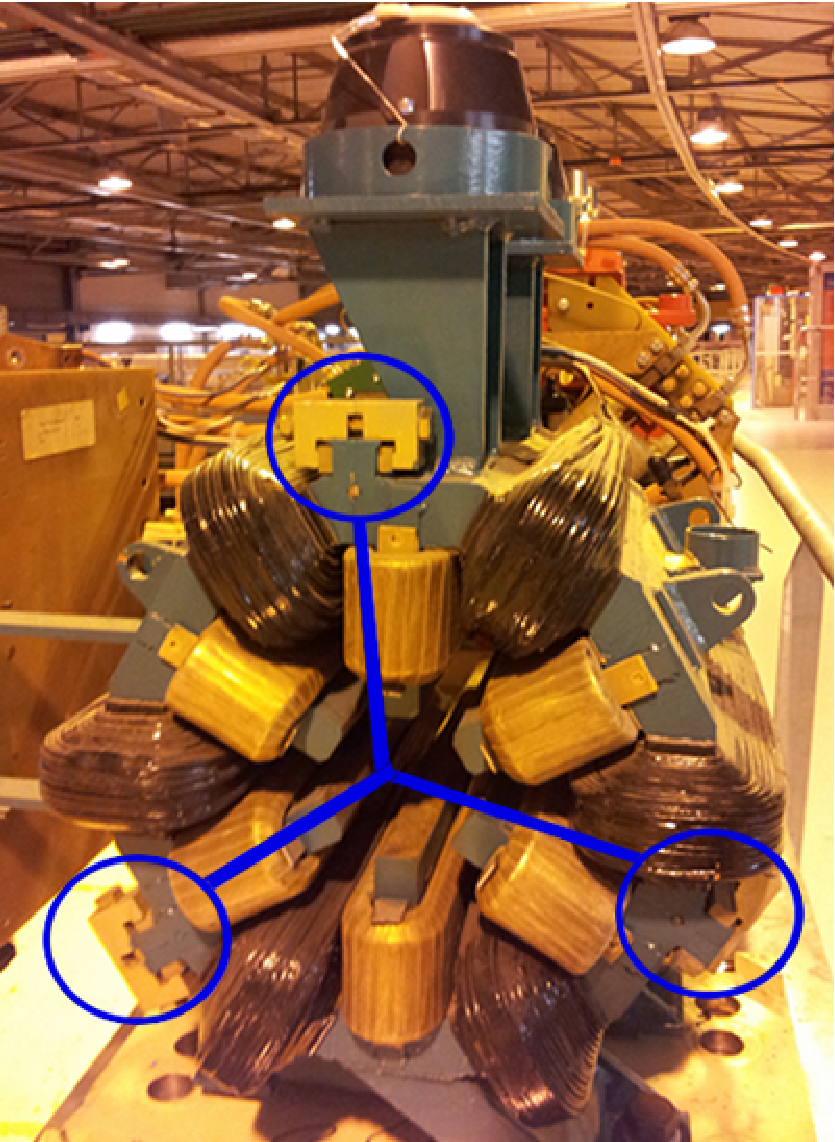}}
\vspace{-0.0cm}
  \caption{\label{fig_QuadSext}(Color) Mechanical assemblies of the ESRF 
           magnets: Quadrupoles are separated in two halves (left), whereas 
           three yokes are screwed together to create sextupoles (right) 
           with survey monuments being placed on one arc only.}
\vskip 0.5cm
\end{figure}

The resulting rms sextupole rotation is of about 1.3~mrad, 
with some magnets exceeding 
$\pm3$~mrad. Again, results vary according to the 
number of eigen-values. However, metrological surveys 
ensure rms girder rotations well below 0.1~mrad and it is 
hard to believe that magnets installed onto them may 
account for more than 1~mrad. An  alternative 
explanation for such large skew sextupole components has 
been proposed and is currently being investigated. During 
magnetic measurements of new quadrupole magnets a large skew 
quadrupole component was measured each time the two halves 
of the yoke were not properly aligned (see left picture of 
Fig.~\ref{fig_QuadSext}): For that specific 
magnet, as a rule of thumb, a horizontal displacement 
between the two yokes of 100 $\mu$m 
would generated a skew quadrupole field, 
corresponding to about 1~mrad equivalent rotation. 
This sensitivity is suspected to be of importance for 
sextupoles too, these being actually separated in three 
yokes (right picture of Fig.~\ref{fig_QuadSext}). 
Numerical simulations and magnetic measurements 
are planned to validate this conjecture. If confirmed, 
hence, the large skew sextupole components observed in the 
ESRF storage ring may be induced by the limited precision 
in the assembly of the three sextupole yokes, rather than 
by the physical magnet rotation as a whole.

\subsection{Correction of sextupolar CRDTs}
\label{sec_NonLinCorr}

\begin{figure*}
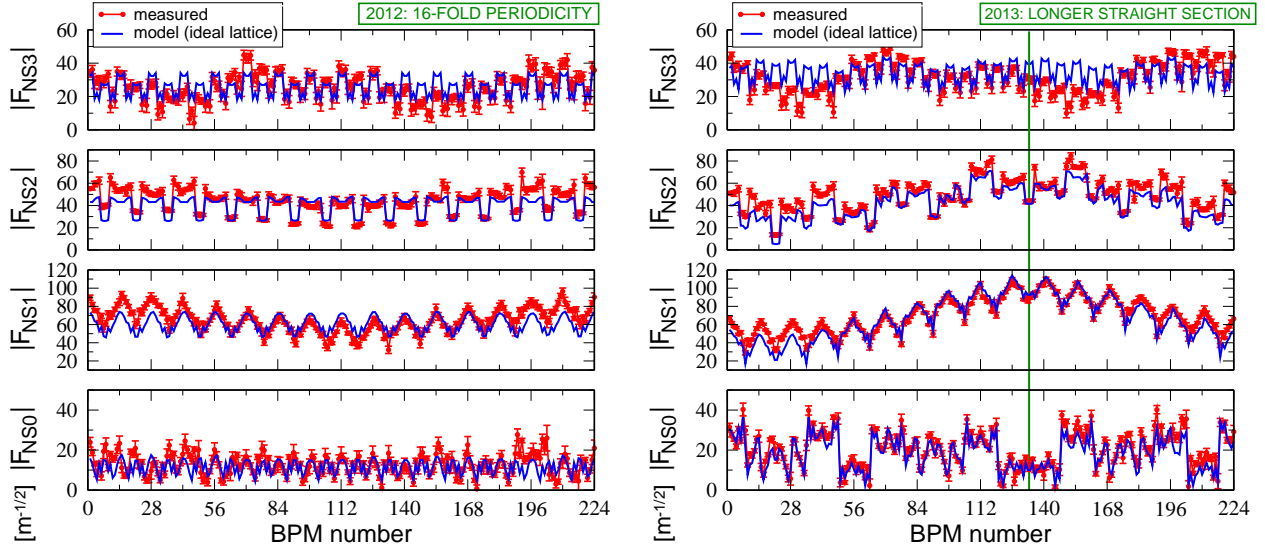

\centerline{
  \includegraphics[width=8.5cm,angle=0]
  {fig14A.eps}
  \includegraphics[width=8.5cm,angle=0]
  {fig14B.eps}}
  \caption{\label{fig_NS_CRDTCompare}(Color) Measured (red) and ideal 
    (blue) normal sextupole CRDTs of the fully 16-fold periodic ESRF 
    storage ring (left, 2012) and of the new configuration with a longer 
    straight section which breaks up the symmetry (right, 2013). The 
    position of this longer section (and of the corresponding insertion 
    optics) is represented by the vertical green line. Even in absence 
    of focusing errors the CRDTs (i.e. the blue curve of the right 
    plot) are substantially modified by this insertion.}
\end{figure*}

\begin{figure}[t]
\rule{0mm}{0mm}
\centerline{
  \includegraphics[width=8.5cm,angle=0]
  {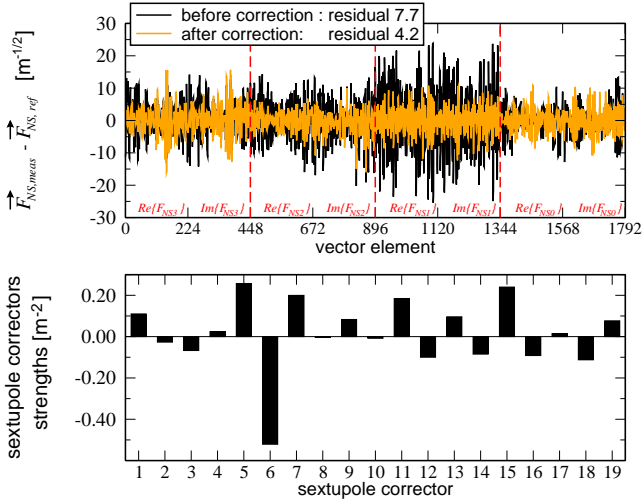}}
  \caption{\label{fig_NS_CRDTCorr1} (Color) Example of normal 
    sextupole CRDT correction. Top: difference between the measured 
    and the reference CRDT vectors before (black) and after (orange) 
    correction. Bottom: integrated strengths of the corresponding 19 
    available sextupole correctors.}
\rule{0mm}{0mm}
\end{figure}

After building up a realistic sextupolar error model, 
the natural further step would be to use the measured 
CRDTs ($\vv{F}_{NS,meas}$) and the desired one 
($\vv{F}_{NS,ref}$), insert them in the linear 
system of Eq.~\eqref{eq_FNS_cor}, and pseudo-invert it 
to evaluate the strengths of the available sextupole 
correctors. The desired CRDTs are those assuming an 
ideal lattice, with neither errors (linear and nonlinear) 
nor betatron coupling. If special insertions 
spoiling the machine periodicity are 
present, they may induce large CRDT modulations. This is 
the case of the ESRF as of 2013, when a straight section 
was lengthened from 5 to 7 m and a special insertion optics 
was implemented. The impact on the normal sextupole 
CRDTs is strong, as shown in Fig.~\ref{fig_NS_CRDTCompare}. 
When correcting the CRDTs, the reference vector $\vv{F}_{NS,ref}$ 
corresponding to the fully periodic machine (i.e. without 
the insertion) was chosen.  
The result of such an exercise is shown in 
Fig.~\ref{fig_NS_CRDTCorr1}, where the difference vector 
$\vv{F}_{NS,meas}$- $\vv{F}_{NS,ref}$ is plotted before 
and after the correction, together with the corresponding 
strengths of the 19 available sextupole correctors.

Even if elegant and efficient (the rms residual is 
almost halved with just 19 correctors out of 224 
sources of errors), this approach is rather inconvenient. 
The reason is three-fold. First, the sextupole error 
model may depend on the implemented optics. The one used 
for these measurements has been specifically tuned so to 
have high spectral resolution but would be impractical 
for operational purposes. On the other hand, it would be 
impossible to extract an error model from the operational 
settings with the same accuracy, because of their larger 
chromaticity and detuning. Depending on the filling mode, 
chromatic sextupoles may vary considerably, and so may 
their errors, being the magnets close to saturation. Second, 
while focusing errors and coupling may be evaluated and 
corrected routinely from ORM measurements, the setup for 
the TbT BPM acquisition is less obvious and difficult to 
perform with the same regularity, even admitting that a 
sufficient spectral resolution may be found. Third, the 
special sextupole setting would not permit to verify whether 
the CRDT correction is beneficial to the beam lifetime, 
the chromaticity being too low for storing more than 10~mA, 
whereas 200~mA are usually delivered to users. 

\begin{table}[b]
\vskip 0.3cm
\caption{Summary of the residual $\mathcal{R}$ of 
         Eq.~\eqref{eq_FNS_residual} between the measured 
         normal sextupole CRDTs and different lattice models.
         \vspace{-0.0 cm}}
{
\begin{tabular}{l|c}
\hline\hline \vspace{-0.25cm} & \\ \vspace{0.0 cm}
Model characteristics  &  residual $\mathcal{R}$      \\
                       &  [m$^{-1/2}$] \\
\hline\hline \vspace{-0.25cm} & \\ \vspace{0.0 cm}
ideal lattice + 2001 sextupole calibration &  8.4 \\ \hline
\vspace{-0.25cm} & \\ \vspace{0.0 cm}
ideal lattice + 2012 sextupole calibration &  7.7 \\ \hline
\vspace{-0.25cm} & \\ \vspace{0.0 cm}
lattice with focusing errors               & \\
\ \ \ + 2012 sextupole calibration         &  3.2 \\ \hline
lattice with focusing errors               & \\
\ \ \ + 2012 sextupole calibration         & 1.8 \\
\ \ \ + sextupole errors                   & \\ \hline
\hline
\end{tabular}}\vspace{0.0cm}
\label{tab-Residual}
\end{table}

A rapid look at the evolution of the residual $\mathcal{R}$ 
as a function of the lattice models may provide a handy, 
though not perfect, solution. Table~\ref{tab-Residual} 
summarizes this dependence. Because of the relatively 
large persistent $\beta$-beating in the ESRF storage ring, 
the main contribution to the the measured CRDTs ($\sim 50\%$) 
originates from focusing errors, whereas only $25\%$ seems 
to stem from sextupole errors and a further $25\%$ is 
not accounted for (possibly because of higher-order terms 
affecting the sextupolar spectral lines, as shown in 
Fig.~\ref{fig_FFT_example}). This means that a detailed 
linear lattice model would describe the true 
CRDTs with an accuracy sufficient for their rough correction, 
even ignoring sextupole errors. 
This corresponds to use and correct the black 
(model) curves in the left plot of 
Fig.~\ref{fig_CRDTAmpliMeasER} instead of the (measured) 
red curves: Even if not perfect, the model CRDTs may already be
used for an effective and rapid correction. 
This may be a peculiarity of machines like the one at the 
ESRF. For more recent accelerators with independent magnets, 
and hence with much lower residual focusing errors, this 
may not be the case.

To verify this conjecture, a test during machine-dedicated 
time was carried out. The software application for the 
measurement of the ORM and correction of focusing and 
coupling errors was modified to compute the normal 
sextupole CRDTs $F_{NS}$ from the lattice model ( i.e. 
from Tables~\ref{tab:lattice-rdt} and~\ref{tab:line-selection}), 
including all errors. These were inserted in the 
$\vv{F}_{NS,meas}$ vector, whereas the ideal C-S 
parameters were used to compute the ideal (and desired) 
CRDTs,r $\vv{F}_{NS,ref}$. The 
system of Eq.~\eqref{eq_FNS_cor} was then 
pseudo-inverted to infer the strengths of the 
sextupole correctors. This procedure was applied to a 
standard multibunch optics optimized for a train of 
868 bunches (7/8 of the storage ring circumference) 
filled with 200~mA (i.e. 0.23~mA per bunch). To enhance 
the Touschek effect which is believed to be presently the 
main limitation for the lifetime, another filling 
pattern with 192 bunches, each of about 1~mA, was stored. 
With all sextupole correctors turned off and a vertical 
emittance $\epsilon_y=7.1$ pm ($\epsilon_x=4$ nm), a 
lifetime of 16.2 hours was measured. After applying the 
standard manual optimization of the four sextupolar 
resonance stopbands, the lifetime reached 24.2 hours. When 
the strength computed automatically from Eq.~\eqref{eq_FNS_cor} 
were used, the measured lifetime was 22.4 hours. This 
proved the effectiveness of the method, which can be used 
for a first correction from scratch before trimming the 
correctors by hand with the standard ESRF procedure. 
Nevertheless, it turned out to be not yet optimal. 
Another test was carried out to a different optics with higher 
chromaticity (to stabilize higher charges per bunch) clearly 
showed a poorer performance of the CRDT correction 
compared to the standard procedure. The reason for this 
dependence on the implemented optics has not been yet 
understood.


\subsection{Beam-based calibration of independent sextupoles via CRDTs}
\label{sec_SextCalib}

\begin{figure}
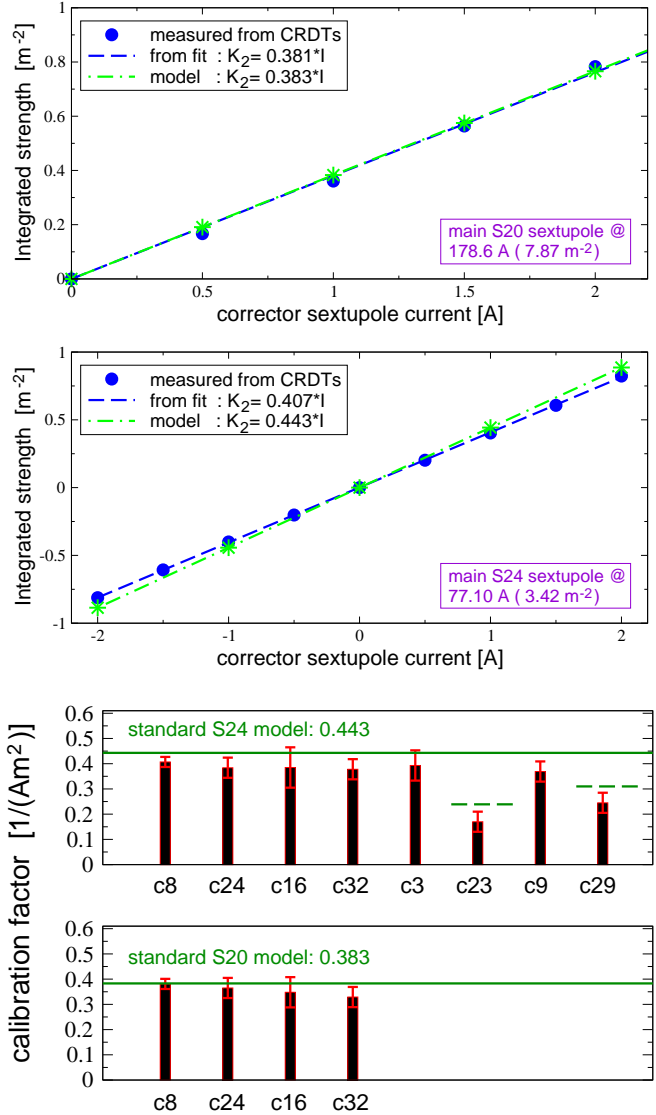

\rule{0mm}{0mm}
\centerline{
  \includegraphics[width=8.5cm,angle=0]
  {fig16A.eps}}
\vskip 0.2cm
\centerline{
  \includegraphics[width=8.5cm,angle=0]
  {fig16B.eps}}
\vskip 0.4cm
\centerline{
  \includegraphics[scale=0.45,angle=0,trim= 0mm 0mm 00mm 0mm,clip]
  {fig16C.eps}}
  \caption{\label{fig_CalibCorr1} (Color) Beam-based calibration of 
           sextupole correctors (trim coils) from measurements of CRDTs 
           and Eq.~\eqref{eq_FNS_calib}. For two correctors in cell 8 
           of the ESRF storage ring the curves have been measured in 
           detail with several points: S20C8 is hosted in a chromatic
           sextupole (top plot) close to saturation, hence with a 
           coefficient lower than in the correctors installed in the 
           harmonic sextupoles, such as S24C8 (centre plot), which are 
           still in the linear regime. The measured coefficients (i.e. 
           the slopes  of the above plots) for all twelve trim coils 
           are reported in the bottom chart.}
\rule{0mm}{0mm}
\end{figure}

As discussed in the last paragraph of Sec.~\ref{sec_RDT2K2}, 
the measurement of CRDTs $F_{NS}$ may be repeated 
for different gradients of individual sextupoles or 
correctors to obtain a calibration curve, (integrated) 
strength {\sl Vs} current. The ESRF storage ring comprises 
of 224 sextupoles mostly grouped in seven families, each 
sharing a common power supply. Six new short magnets 
recently installed to lengthen three straight sections are 
fed independently, and two standard sextupoles 
have been stripped out and paired on a single power supply. The 
correction of sextupolar resonances is carried out by twelve 
correctors (trim coils) hosted in as many magnets of the 
families \verb S24  (eight) and \verb S20  (four). Even though 
magnetically identical, because of the different working 
points, the (stronger) chromatic \verb S20  are close 
to saturation, whereas the (weaker) harmonic \verb S24  are 
still in the linear regime. This results in trim 
coils with different calibration curves, i.e. 
the slopes $C_f$ of the linear curve $K_2=C_f\cdot I$, where 
$I$ denotes the imparted current.

By measuring the CRDTs it is then possible to validate 
the calibration of both the trim coils and the six 
independent short sextupoles \verb S4 . 
The first check is necessary for the effectiveness of the 
automatic resonance correction discussed in the previous 
section, whereas the latter is to confirm the hypothesis
raised in Sec.~\ref{sec_NonLinMeas} of having short 
sextupoles weaker than expected.

\begin{figure}
\rule{0mm}{0mm}
\centerline{
  \includegraphics[width=8.5cm,angle=0]
  {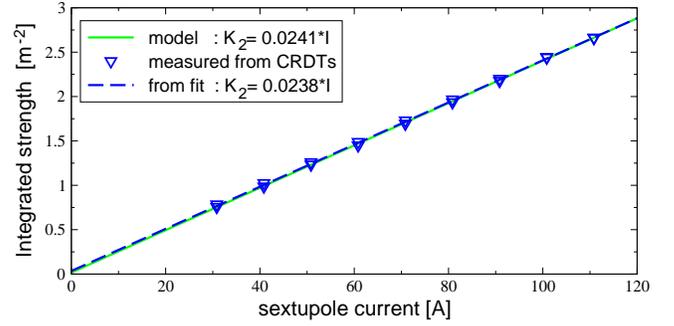}}
\vskip -0.3cm
  \caption{\label{fig_CalibS4C15} (Color) Beam-based calibration of 
           short independent sextupole S4C15 from measurements of CRDTs 
           and Eq.~\eqref{eq_FNS_calib}.}
\rule{0mm}{0mm}
\end{figure}

For two trim coils in cell 8 (\verb S20C8  and \verb S24C8 ) the 
measurement has been carried out with several points and the 
calibration coefficient is obtained from a linear fit 
of the curve $K_2=C_f\cdot I$. 
As shown in the upper plots of Fig.~\ref{fig_CalibCorr1} 
the agreement with the model is nearly perfect for the 
corrector \verb S20C8 , and compatible within $10\%$ for \verb S24C8 . 
The same measurement was then repeated for the other ten 
correctors. Results are reported in the bottom chart of 
Fig.~\ref{fig_CalibCorr1}. Because of the limited 
available machine time, the linear fit for the 
remaining ten correctors was carried out with two 
points only, at $\pm1$~A. Between each measurement, 
magnets were not cycled and the persistence of some 
remaining magnetic fields could not be excluded. If these 
arguments may explain the poorer agreement between 
model and measured coefficients for these ten correctors, 
they cannot justify the large discrepancy observed for 
two trim coils, in \verb S24C23  and \verb S24C29  respectively. 
As far as the corrector in cell 23 is concerned, the 
explanation is rather simple: As of 2013 the standard 
40 cm long main magnet has been replaced by a new short
type. As a result, the calibration for its corrector 
is almost halved ($C_f$ from $0.443$ to $0.239$~A$^{-1}$m$^2$), 
consistent with the measured value. In the case of the 
trim coils of \verb S24C29 , an inconsistency was indeed found 
between the power supplies (recent 2-A types) and one of 
the two driver cards (still set for an old 5-A type). 
This incongruity resulted in a corrector about $30\%$  
weaker than normal ($C_f$ from $0.443$ to $0.310$~A$^{-1}$m$^2$), 
again consistent with measurements. Following 
these findings, the software driving the trim coil 
\verb S24C23  has been updated (it was not when testing 
the automatic resonance correction), and the driver 
card of \verb S24C29  was updated. 

Results from the same measurement carried out for one 
of the six new (short and independent) sextupoles in cell 
15, \verb S4C15 , are reported in Fig.~\ref{fig_CalibS4C15}. 
This beam-based calibration ($C_f=0.0238\pm0.002$~A$^{-1}$m$^2$) 
is compatible with magnetic 
measurement ($C_f=0.0241$).

\section{Going beyond: octupolar spectral lines, CRDTs 
         and modelling}
\label{sec_OctDec}

Figure~\ref{fig_FFT_example} reveals the existence of 
four spectral lines well above noise level (and hence 
measurable) excited by octupolar CRDTs, namely $H(3,0)$, $H(-1,2)$, 
$V(2,1)$ and $V(2,-1)$. As shown in the same figure and 
described in Table~\ref{tab:line-selection-octu}, 
two additional lines are excited by these terms, 
$H(1,2)$ and $V(0,3)$. However, the latter are 
indistinguishable from the sextupolar lines $H(0,-2)$ 
and $V(-1,-1)$, respectively, because of the tune 
working point. Hence the corresponding CRDTs $F_{NO4}$ 
and $F_{NO5}$ of Table~\ref{tab:line-selection-octu} 
are not observable in our case. 
The best setting for the ESRF storage ring turned 
out to be the last of Table~\ref{tab:linear-anal}, 
with the horizontal kicker fired at 400~A and the 
vertical at 3.0~kV. The 
formulas of Table~\ref{tab:line-rdt-octu} to infer the 
octupolar CRTDs from the spectral lines have 
been corrected in the analysis by using the tune line 
amplitudes averaged among all 224 BPMs, i.e. 
$|H(1,0)|\rightarrow <|H(1,0)|>$ and 
$|V(0,1)|\rightarrow <|V(0,1)|>$ in order to reduce the 
impact of octupolar terms on the tune line (see 
discussion in the last part of Appendix~\ref{app:4}).

\begin{figure}
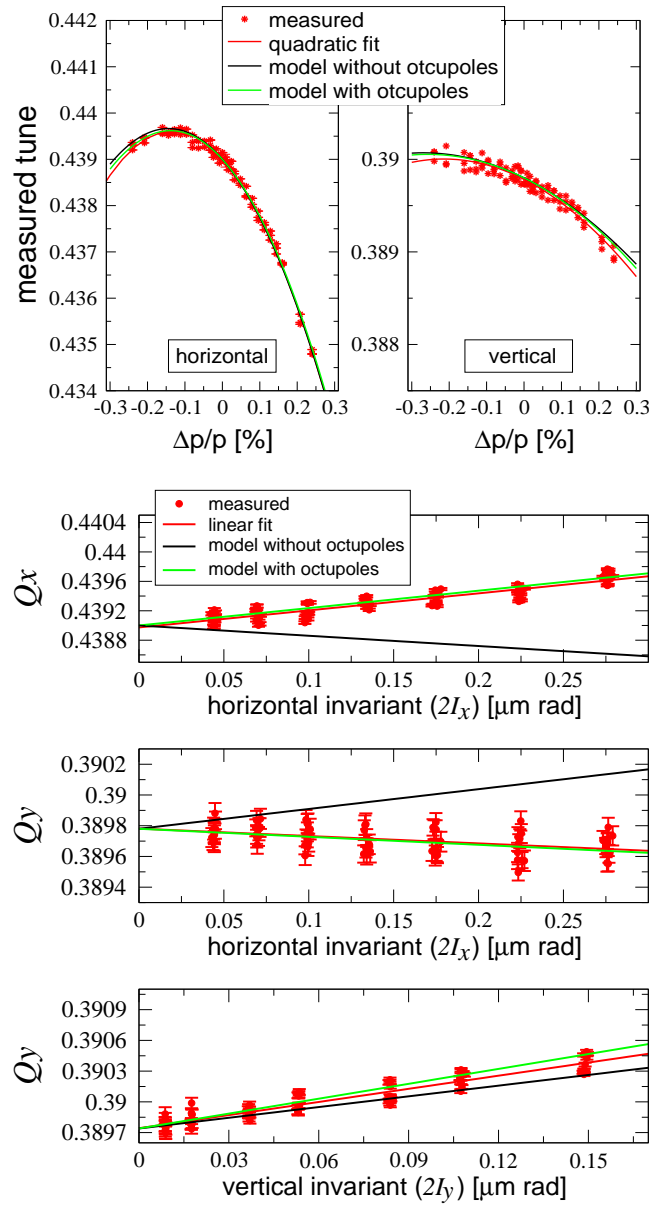

\rule{0mm}{0mm}
\centerline{
  \includegraphics[width=8.5cm,height=6cm,angle=0]
  {fig18A.eps}}
\vskip0.4cm
\centerline{
  \includegraphics[width=8.5cm,angle=0]
  {fig18B.eps}}
\rule{0mm}{-0mm}
  \caption{\label{fig_ChromDetuning} (Color) Nonlinear chromaticity 
           (top two graphs) and linear amplitude dependent detuning (bottom 
           three plots). The measured curves (red) are compared to the ones 
           computed from lattice model including all errors up to the 
           sextupolar components (black), and after including octupolar 
           components in the main quadrupoles (green). The corresponding 
           coefficients are listed in the Tk-FF columns (thick sextupoles 
           and fringe fields in all magnets) of Table~\ref{tab-NonLinParam}.}
\rule{0mm}{0mm}
\end{figure}

Even though no physical octupole is installed in the 
ESRF storage ring, the nonlinear model including 
sextupoles (with errors and tilts) alone does not 
suffice to reproduce global nonlinear parameters  
such as amplitude dependent detuning 
$\nu'=\partial Q$/$\partial(2I)$, as reported in 
Table~\ref{tab-NonLinParam} and 
Fig.~\ref{fig_ChromDetuning}.  The agreement between 
this model and measurements is however already good 
as far  as the second-order chromatic terms 
$Q''=\partial^2Q$/$\partial\delta^2$ are concerned.
As stressed in Sec.~\ref{sec_TheoIntro}, 
no significant difference has been observed in the 
RDTs when representing sextupoles either as thin 
or thick elements and when including fringe fields. 
This is not the case for the global nonlinear 
parameters and in Table~\ref{tab-NonLinParam} 
values for the three different models are reported. 
Plots and fit (of global parameters and octupolar 
CRDTs) however refer to the thick 
model and including fringe fields.

\begin{table*}[!t]
\caption{List of lines in the spectra of $\tilde{x}(N)$ and 
	$\tilde{y}(N)$ excited by normal octupole CRDTs $F=|F|e^{iq_F}$
        and excited resonances. For each line (defined as in 
        Table~\ref{tab:line-selection}), expressions for its amplitude 
	and phase are given. The tune lines $H(1,0)$ and $V(0,1)$ 
        are also affected by octupolar terms that are however not 
        observable. Quadrupole errors are to be included in the model when 
	computing the Courant-Snyder (C-S) parameters used to evaluate 
        $\tilde{x}(N)$, $\tilde{y}(N)$ and the CRDTs $g_{jklm}$, which are 
        defined in Table~\ref{tab:g_jklm_NO}.
        \vspace{-0.2 cm}}
\vskip 0.0cm
\centering 
{
\begin{tabular}{c}
\end{tabular} \\ 
\begin{tabular}{c c l l c c}
\hline\hline\vspace{-0.25cm} \\ \vspace{0.10 cm}
                 spectral line                &
{\hskip 0.4 cm}\ amplitude{\hskip 1.2 cm}\    & 
{\hskip 0.0 cm}  phase $\phi$  {\hskip 2.4 cm}\     &
{\hskip 0.0 cm}  Combined RDT{\hskip 0.4 cm}\ &
{\hskip 0.2 cm}\ resonances {\hskip 0.2 cm}\  &
 magnetic term     \\
\hline\hline\vspace{-0.30 cm}\\\vspace{0.07 cm}
$V(0, 3)$& $(2I_y)^{3/2}|F_{NO5}|$       & $q_{F_{NO3}}+\frac{\pi}{2}+3\psi_{y0}$         & $F_{NO5}=4g_{0040}^*-g_{0013}$  &(0,4),(0,2) & $y^4$   \\
\hline\vspace{-0.30 cm}\\\vspace{0.07 cm}
$H(1, 2)$& $(2I_x)^{1/2}(2I_y)|F_{NO4}|$ & $q_{F_{NO4}}+\frac{\pi}{2}+\psi_{x0}+2\psi_{y0}$& $F_{NO4}=2g_{2020,H}^*-g_{1102}$&(2,2),(0,2) & $x^2y^2$\\
\hline\vspace{-0.30 cm}\\\vspace{0.07 cm}
$H(3, 0)$& $(2I_x)^{3/2}|F_{NO3}|$       & $q_{F_{NO3}}+\frac{\pi}{2}+3\psi_{x0}$         & $F_{NO3}=4g_{4000}^*-g_{1300}$  &(4,0),(2,0) & $x^4$   \\
\hline\vspace{-0.30 cm}\\\vspace{0.07 cm}
$H(-1,2)$& $(2I_x)^{1/2}(2I_y)|F_{NO2}|$ & $q_{F_{NO2}}-\frac{\pi}{2}-\psi_{x0}+2\psi_{y0}$& $F_{NO2}=2g_{2002}-g_{1120}^*$ &(2,-2),(0,2)& $x^2y^2$\\
\hline\vspace{-0.30 cm}\\\vspace{0.07 cm}
$V(2,-1)$& $(2I_x)(2I_y)^{1/2}|F_{NO1}|$ & $q_{F_{NO1}}-\frac{\pi}{2}+2\psi_{x0}-\psi_{y0}$& $F_{NO1}=2g_{0220}-g_{2011}^*$  &(2,-2),(2,0)& $x^2y^2$\\
\hline\vspace{-0.30 cm}\\\vspace{0.07 cm}
$V(2, 1)$& $(2I_x)(2I_y)^{1/2}|F_{NO0}|$ & $q_{F_{NO0}}+\frac{\pi}{2}+2\psi_{x0}+\psi_{y0}$& $F_{NO0}=2g_{2020,V}^*-g_{0211}$&(2,2),(2,0) & $x^2y^2$\\
\hline
\end{tabular}}
\vskip 0.0cm
\label{tab:line-selection-octu}
\caption{Formulas to evaluate octupolar CRDTs from the 
	secondary lines in the spectra of $\tilde{x}(N)$ 
	and $\tilde{y}(N)$ assuming properly calibrated 
	BPMs, turn-by-turn oscillations without 
	decoherence and quadrupole errors included  
	in the C-S parameters.\vspace{-0.2 cm}}
\vskip 0.0cm
\centering 
{
\begin{tabular}{c}
\end{tabular} \\ 
\begin{tabular}{l l l}
\hline\hline\vspace{-0.25cm} \\ \vspace{0.10 cm}
   Combined RDT{\hskip 1.4 cm}\    &{\hskip 1.6 cm}\ amplitude {\hskip 3.2 cm}\  & {\hskip 0.7 cm}  phase $q_F$ {\hskip 0.0 cm}             \\
\hline\hline\vspace{-0.30 cm}\\\vspace{0.07 cm}
$F_{NO5}=|F_{NO5}|e^{iq_{F_{NO5}}}$& $|F_{NO5}|=|V( 0,3)|/[8|V(0,1)|^3]$       & $q_{F_{NO5}}=\phi_{V(0,3)}-3\phi_{V(0,1)}-\frac{\pi}{2}$ \\
\hline\vspace{-0.30 cm}\\\vspace{0.07 cm}
$F_{NO4}=|F_{NO4}|e^{iq_{F_{NO4}}}$& $|F_{NO4}|=|H( 1,2)|/[8|H(1,0)||V(0,1)|^2]$& $q_{F_{NO4}}=\phi_{H(1,2)}-\phi_{H(1,0)}-2\phi_{V(0,1)}-\frac{\pi}{2}$ \\
\hline\vspace{-0.30 cm}\\\vspace{0.07 cm}
$F_{NO3}=|F_{NO3}|e^{iq_{F_{NO3}}}$& $|F_{NO3}|=|H( 3,0)|/[8|H(1,0)|^3]$       & $q_{F_{NO3}}=\phi_{H(3,0)}-3\phi_{H(1,0)}-\frac{\pi}{2}$ \\
\hline\vspace{-0.30 cm}\\\vspace{0.07 cm}
$F_{NO2}=|F_{NO2}|e^{iq_{F_{NO2}}}$& $|F_{NO2}|=|H(-1,2)|/[8|H(1,0)||V(0,1)|^2]$& $q_{F_{NO2}}=\phi_{H-1,2)}+\phi_{H(1,0)}-2\phi_{V(0,1)}+\frac{\pi}{2}$ \\
\hline\vspace{-0.30 cm}\\\vspace{0.07 cm}
$F_{NO1}=|F_{NO1}|e^{iq_{F_{NO1}}}$& $|F_{NO1}|=|V(2,-1)|/[8|H(1,0)|^2|V(0,1)|]$& $q_{F_{NO1}}=\phi_{V( 2,-1)}-2\phi_{H(1,0)}+\phi_{V(0,1)}+\frac{\pi}{2}$ \\
\hline\vspace{-0.30 cm}\\\vspace{0.07 cm}
$F_{NO0}=|F_{NO0}|e^{iq_{F_{NO0}}}$& $|F_{NO0}|=|V(2, 1)|/[8|H(1,0)|^2|V(0,1)|]$& $q_{F_{NO0}}=\phi_{V( 2, 1)}-2\phi_{H(1,0)}-\phi_{V(0,1)}-\frac{\pi}{2}$ \\
\hline
\end{tabular}}
\vskip -0.0cm
\label{tab:line-rdt-octu}
\end{table*}
\begin{figure*}
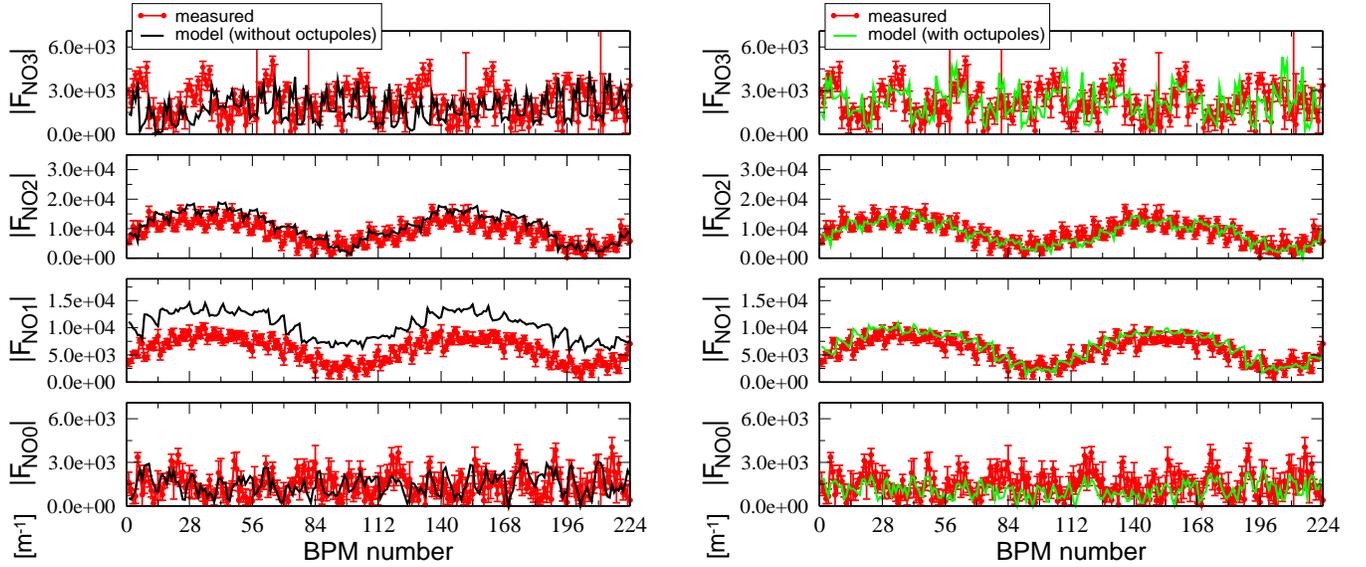

\rule{0mm}{-0mm}
\centerline{
  \includegraphics[width=8.5cm,height=7.45cm,angle=0]
  {fig19A.eps} \hspace{0.6cm}
  \includegraphics[width=8.5cm,height=7.45cm,angle=0]
  {fig19B.eps}}
  \caption{\label{fig_OctupoleCRDTAmpliMeas}(Color) Amplitudes of normal 
           octupole CRDTs (red curves) 
           measured at the ESRF storage ring (with a special dedicated 
           optics). Left: comparison with the model curves obtained from 
           a lattice with no octupolar source (CRDTs are  
           excited by sextupoles to the second order only). 
           Right:  comparison with the same model after introducing 
           octupole components in quadrupoles so to fit the nonlinear 
           parameters of Table~\ref{tab-NonLinParam} and the measured 
           octupolar CRDTs. 
           \vspace{-1.3cm}}
\rule{0mm}{0mm}
\end{figure*}
\clearpage

\begin{table}[t]
\caption{ Detuning coefficients and second-order 
          chromaticity for the special lattice optics studied 
          throughout this paper. Measured values are listed 
          in the first column, whereas the corresponding numbers 
          for the lattice model (all comprising the  
          errors inferred in the previous sections) are reported 
          in the second column. The model is presented in three 
          flavors: thin (Tn) and thick (Tk) sextupoles, and with 
          additional fringe fields in all magnets (Tk-FF).
          Octupolar fields are introduced in the 256 main normal 
          quadrupoles to reproduce the measured coefficients 
          (first) and the octupolar CRTDs (then), yielding to the 
          values in the last column.
          \vspace{0.2 cm}}
%
\vskip 0.0cm
\begin{tabular}{l|r|r|r}
\hline \hline
             &                 & model    \hspace{1.1cm} & model\\
parameter\ \ &\ \ measured\ \  & without  \hspace{9.5mm} & with\\
             &                 & octupoles\hspace{9.5mm} & octupoles\\
    &            &Tk-FF\hspace{0.6cm}Tk\hspace{0.9cm}Tn\ \ \ & Tk-FF\\
\hline\hline\vspace{-0.25cm} & & & \\\vspace{0.0 cm}
 $\nu_{xx}'$ \ [m$^{-1}$]& 2430$\pm$30 \ &-1396\ \ \    (-2515)\ \ \    (4185)\ & 2364 \\
 $\nu_{xy}'$ \ [m$^{-1}$]& -470$\pm$15 \ & 1289\ \ \ \  ( -115)\ \ \    (4962)\ & -521 \\
 $\nu_{yy}'$ \ [m$^{-1}$]& 4750$\pm$50 \ & 3485\ \ \    ( 2518)\ \ \    (8206)\ & 4846 \\
 $Q_x''\hspace{7.2mm}$[ ]& $-683\pm24$  \ & -649\ \ \ \ (-657)\ \ \ \    (-522)\ & -660 \\
 $Q_y''\hspace{7.2mm}$[ ]& $-96\pm17$ \   &  -67\ \ \ \ \ ( -74)\ \ \ \  ( -30)\ &  -82 \\
\hline
\end{tabular}\vspace{0.5cm}
\label{tab-NonLinParam}
\end{table}

To the first 
order, sextupoles do not drive any detuning with amplitude, 
which is a second-order effect (i.e. induced by cross-product 
between sextupole terms). Octupolar CRDTs too are excited by 
second-order sextupolar terms even in absence of physical 
octupoles. This can be seen when comparing the four 
measured octupolar CRDTs with those evaluated from the 
lattice model (including all errors), as shown in the left 
plot of Fig.~\ref{fig_OctupoleCRDTAmpliMeas}. Even though 
the global behavior is rather well reproduced by the model, 
a clear ingredient looks missing. Model CRDTs are evaluated 
by using the same procedure developed for the skew sextupole 
terms and described in Appendix~\ref{app:1}.

The first natural source of octupolar fields in a storage ring 
is represented by normal quadrupoles, because of the shape of 
their poles, fringe fields and their geometry in two halves 
(any variation of the vertical distance between the two 
yokes generates octupolar fields). The linear lattice of the ESRF 
storage ring is based on eight quadrupoles per cell (256 
magnets in total), with 
two pairs of identical magnets in the achromat fed by common 
power supplies, hence leaving six degrees of freedom. The first 
step in building the octupolar lattice model was to introduce 
six octupolar fields in the corresponding quadrupole families 
to best reproduce the five nonlinear parameters (three detuning 
coefficients and two second-order chromatic terms). The 
256 octupole fields have been then adjusted independently so to 
best reproduce the measured CRDTs of 
Fig.~\ref{fig_OctupoleCRDTAmpliMeas}. As for the skew sextupole 
analysis, even though model and measured CRDTs are excited 
mainly by second-order contributions stemming from sextupoles,
their difference will 

\hskip -0.3cm depend linearly on the octupolar field, 
and the same SVD pseudo-inversion carried out in the previous 
sections may be generalized to octupoles. The results of such 
a fit are shown in the right plot of 
Fig.~\ref{fig_OctupoleCRDTAmpliMeas} as far as the CRDTs are 
concerned, whereas the values of the global nonlinear parameters 
are reported in the last column of Table~\ref{tab-NonLinParam} 
and in Fig.~\ref{fig_ChromDetuning}. 

It shall be mentioned that measurements of octupolar CRDTs may 
be partially affected by BPM nonlinearities, these scaling with 
$x^3$ ($y^3$) to the first order~\cite{soleil} and the largest 
BPM data used for this analysis reaching about 6~mm. 
On the other hand, the software driving the 
ESRF BPMs computes the beam position using 
a nonlinear calibration curve, based on finite-element 
simulations of the BPM block, that accounts already 
for such nonlinearities.

\section{Conclusion}
This paper has shown how the harmonic analysis of 
turn-by-turn data from beam position monitors 
may be exploited for the reconstruction and correction 
of machine nonlinearities up to octupolar terms.  
An important peculiarity of this approach is that 
the systems to be solved are always linear, even though 
referring to nonlinear magnets. This represents a 
considerable step forward compared to precedent works 
on the same subject. Handy formulas have been derived 
and tested with real data for the quantitative 
evaluation of realistic magnetic model, for the 
calibration of individual nonlinear magnets and for 
the correction of the resonance driving terms. The analysis 
of the latter could predict the sextupolar 
field component measured in bending magnets, as well 
some inconsistencies with the calibration curves of 
sextupole magnets.~Correcting sextupolar 
resonance driving terms in the 
ESRF storage ring with a low-chromaticity optics (for 
low-intensity multi-bunch modes) resulted in increased 
lifetime, though the gain was minimal when applied 
to optics with larger chromaticity (for high-intensity 
few-bunch filling patterns).\vskip -0.3cm

\section{Acknowledgments}
We are grateful to the ESRF Operation Group for its support 
in preparing and carrying out measurements, to F. Epaud for 
adjusting the BPM device server to our needs, to M. Dubrulle for 
setting up the kickers, and to F. Taoutaou for helping with 
the magnetic measurements. We are also indebted with 
R. Tom\'as, for proofreading the original manuscript and making 
valuable suggestions.

\appendix

\section{Nonresonant normal forms and RDTs up to the 2nd order}
\label{app:1}
For a reliable analysis of the ESRF nonlinear 
lattice model, the normal form and RDT description 
of Refs.~\cite{Bartolini1,Rogelio1,Andrea-thesis,prstab_strength} 
needs to be extended to the second order. 
The theoretical background has been already 
developed in previous works, among which 
the fundamental Ref.~\cite{Bengtsson}. The 
first part of this Appendix does not provide 
any further development to that work: 
Results of main interest in the context of 
this article are reported for sake of 
consistency and nomenclature only. In order to 
tackle the heavy mathematics of second-order 
computations, it may be desirable to briefly 
review the concept of nonresonant normal 
forms of Refs.~\cite{Turchetti1,Forest1}.

\begin{figure}[t]
\rule{0mm}{0mm}
\centerline{
  \includegraphics[width=8.5cm,angle=0]
  {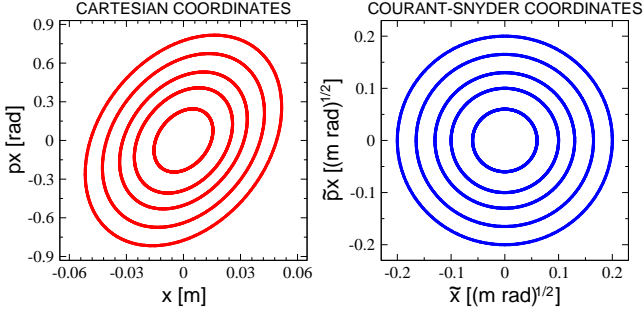}}
  \caption{\label{fig_ph0} Horizontal phase space portrait corresponding 
	to the linear, uncoupled, betatron motion induced by normal 
	quadrupoles in Cartesian coordinates (left) and in 
	Courant-Snyder's (right).}
\rule{0mm}{3mm}
\end{figure}
 
In the linear uncoupled case the turn-by-turn 
dynamics is described by the one-turn   
matrix $M(s)$ according to 
\begin{equation}
\vec{x}_{s+C}=M(s)\vec{x}_{s}\ ,
\end{equation}
where $C$ is the ring circumference and $s$ a
generic position along the ring. The 
corresponding phase space portrait 
($x,\ p_x$) of the betatronic motion inside a circular 
accelerator is typically a tilted ellipse (identical 
considerations apply of course to the vertical plane). 
Tilt  and main axis vary along the ring according 
to the focusing lattice (normal quadrupoles). Any 
regular (i.e. noncahotic) closed curve in phase-space 
may be transformed into a circle. The Courant-Snyder 
change of coordinates (Fig.~\ref{fig_ph0})
\begin{equation}
\left(\begin{array}{c} \tilde{x} \\ \ \\\tilde{p}_x\end{array}\right)=
\left(\begin{array}{c c}(\beta_x)^{-1/2}& 0 \\ \ \\
                        \alpha_x(\beta_x)^{-1/2} & (\beta_x)^{1/2} \end{array}\right)
\left(\begin{array}{c} x \\ \ \\ p_x\end{array}\right) \nonumber
\end{equation}
is one among the possible transformations preserving 
the phase-space. The Courant-Snyder (C-S) 
parameters in the above matrix contain 
all the information regarding the linear lattice. 
The circle is invariant along the ring in amplitude 
(i.e. radius and surface) and rotation frequency (i.e. 
the tune), which is independent on the initial 
conditions. The system is equivalent to a harmonic 
oscillator and the motion to a rotation.

\begin{figure}[h]
  \centerline{
  \includegraphics[height=5.5cm,angle=0]
	{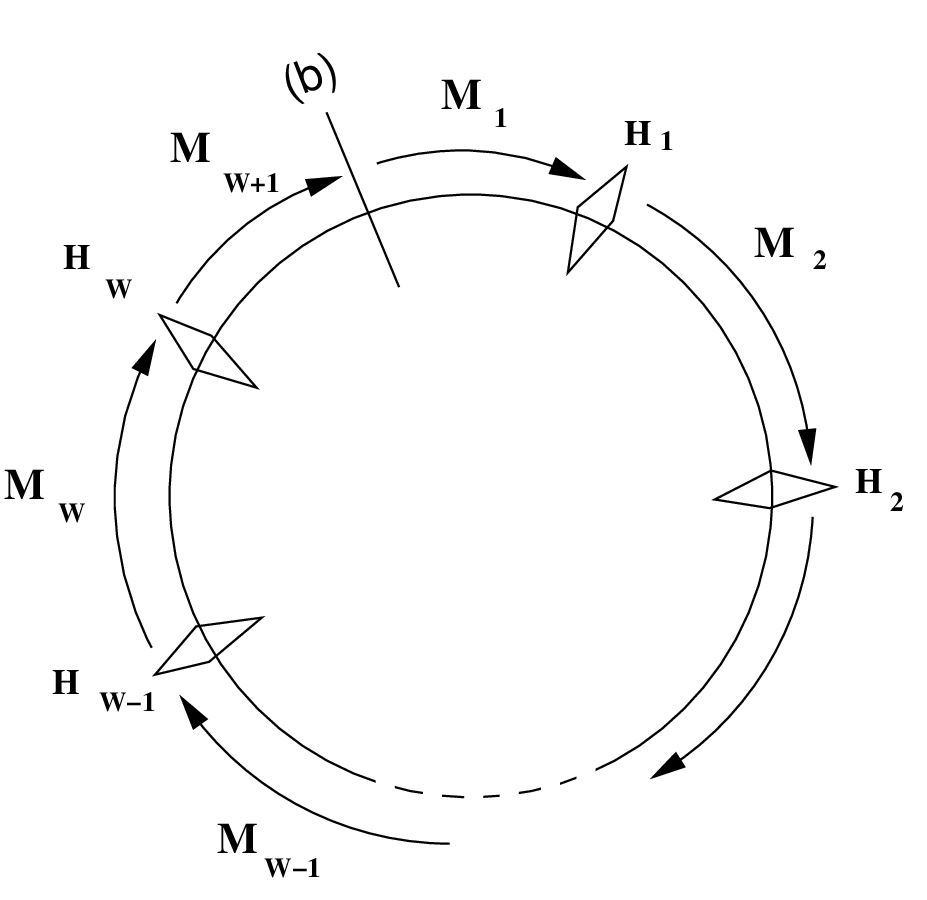}}
  \caption{\label{fig:ring1} Schematic view of a ring and its
           transfer maps, $M_w$ refer to sections free of nonlinearities,
           $H_w$ represent the nonlinear kicks.}
  \rule{0mm}{0mm}
\end{figure}

In the nonlinear case the one-turn map 
$\mathcal{M}(s)$ replaces the linear matrix $M(s)$ 
\begin{equation}
\vec{x}_{s+C}=\mathcal{M}(s)\vec{x}_{s}\ ,
\end{equation}
where the map is the composition of linear 
transfer matrices $M_w$ and Lie operators  
corresponding to localized nonlinear magnets, such as 
sextupoles, see Fig.~\ref{fig:ring1}:
\begin{equation}\label{eq:first}
\mathcal{M}=M_{W+1}\prod_{w=1}^W{e^{:H_{w}:}M_w},
\end{equation}
where $W$ is the total number of nonlinear elements 
in the machine.
Besides Ref.~\cite{Bengtsson}, detailed derivation 
within the RDT formalism may be found in  
Refs.~\cite{Bartolini1,Rogelio1,Andrea-thesis}. 
In the complex Courant-Snyder coordinates, 
$h_{q,\pm}=\tilde{q}\pm i\tilde{p}_q
          =\sqrt{2J_q}e^{\mp i(\phi_q+\phi_{q,0})}$, 
the system reduces to 
\begin{equation} \label{eq:OTM1}
\vec{h}_{s+C}=\tilde{\mathcal{M}}(s)\vec{h}_{s}\ ,\qquad
\tilde{\mathcal{M}}=\prod_{w=1}^W{e^{:\tilde{H}_{w}(s):}}R\ \ ,
\end{equation}
where $\vec{h}=(h_{x,+},h_{x,-},h_{y,+},h_{y,-})$, 
$R$ denotes the phase-space rotation, whose s
frequencies are the betatron tunes. The one-turn Lie 
operator then reads
\begin{equation}\label{eq:Hbw3A}
e^{:\tilde{H}:}=\prod_{w=1}^W{e^{:\tilde{H}_{w}:}}\quad ,\quad
\tilde{\mathcal{M}}(s)=e^{:\tilde{H}:}R\ \ ,
\end{equation}
and the Hamiltonian term of the generic nonlinear magnet 
$w$ reads 
\begin{eqnarray}
\tilde{H}_{w}(s)&=&\sum_{n\ge2}\hskip-0.2cm
		   \sum_{jklm}^{\ \ n=j+k+l+m}\hskip-0.4cm
  {h_{w,jklm}e^{i[(j-k)\Delta\phi_{w,x}^{(s)}+(l-m)\Delta\phi_{w,y}^{(s)}]}}
	\times \nonumber \\\label{eq:Hbw3B}
         &&    h_{w,x,+}^jh_{w,x,-}^kh_{w,y,+}^lh_{w,y,-}^m\ \ ,
\end{eqnarray}
where $h_{w,q,\pm}$ is the coordinate at the 
generic magnet $w$, while $\Delta\phi_{w,q}^{(s)}$ 
is the phase advance between the latter and the 
observation point $s$. $n$ denotes the magnet order: 
$n=2,3,4$ for quadrupoles, sextupoles and octupoles, 
respectively. The coefficients within the sums read
\begin{eqnarray}
h_{w,jklm}&=&-\displaystyle
\frac{\bigl[K_{w,n-1}\Omega(l+m)+
           iJ_{w,n-1}\Omega(l+m+1)\bigr]}
     {j!\quad k!\quad l!\quad m!\quad 2^{j+k+l+m}}\nonumber \\
     &&\times\  
     \ i^{l+m} \bigl(\beta_{w,x}\bigr)^{\frac{j+k}{2}}
     \bigl(\beta_{w,y}\bigr)^{\frac{l+m}{2}},\label{eq:h_Vs_KJ} \nonumber\\
     \nonumber \\
     \Omega&(i)&=1 \hbox{ if } i \hbox{ is even},\quad
                      \Omega(i)=0 \hbox{ if } i \hbox{ is odd}
     \qquad \ .
\end{eqnarray}
$\Omega(i)$ is introduced to select either the 
normal or the skew multipoles. $K_{w,n-1}$ and 
$J_{w,n-1}$ are the integrated magnet strengths 
of the multipole expansion (MADX definition)
\begin{equation}
-\Re\left[\sum_{n\geq2}{(K_{w,n-1}+iJ_{w,n-1})
            \frac{(x_w+iy_w)^n}{n!}}\right]
\end{equation}
from which Eqs.~\eqref{eq:Hbw3B} and ~\eqref{eq:h_Vs_KJ}  
are derived when moving from the Cartesian coordinates 
to the complex Courant-Snyder's: 
$x_w=\sqrt{\beta_{w,x}}(h_{w,x,-}+h_{w,x,+})/2$ and 
$y_w=\sqrt{\beta_{w,y}}(h_{w,y,-}+h_{w,y,+})/2$.

\begin{figure}[t]
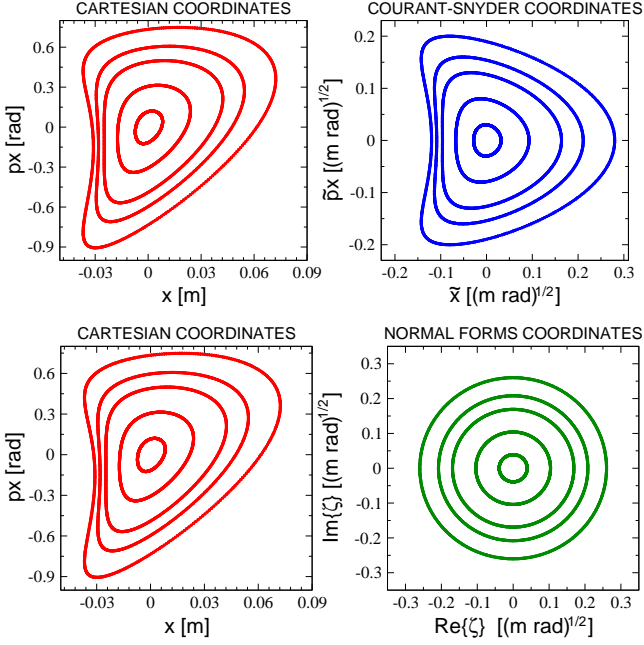

\rule{0mm}{0mm}
\centerline{
  \includegraphics[width=8.5cm,angle=0]
  {fig22A.eps}}
\vskip0.2cm
\centerline{
  \includegraphics[width=8.5cm,angle=0]
  {fig22B.eps}}
  \caption{\label{fig_ph1} Horizontal phase space portrait corresponding 
	to the linear betatron motion with normal sextupoles in Cartesian 
	coordinates (left), in Courant-Snyder's (upper right) and normal 
	forms (lower right).}
\rule{0mm}{3mm}
\end{figure}

If nonlinear terms are of relevance, at 
sufficiently large amplitudes the 
phase space is deformed from ellipses to more complex 
curves. In the case of normal sextupoles, the 
horizontal phase space ($x,\ p_x$) assumes a typical 
triangular shape, as a result of additional harmonics 
excited by these magnets (see Table~\ref{tab:line-selection}).
The Courant-Snyder transformation, being linear, will 
just remove the dependence on the linear parameters,  
rotating the triangle in an upright position and 
smoothing out its distortion. However, the 
invariant circle may not retrieved, since 
the linear transformation may not {\sl include} 
nonlinearities, see upper right portrait of 
Fig.~\ref{fig_ph1}. The phase-space trajectory  
being still closed and regular, it must still 
exist a transformation to retrieve a circle. Such
transformation indeed exists and is a polynomial 
function $F$
\begin{eqnarray}\label{eq:Fdef}
F=\sum_{n\ge2}\sum_{jklm}^{n=j+k+l+m}{f_{jklm}
             \zeta_{x,+}^j\zeta_{x,-}^k\zeta_{y,+}^l\zeta_{y,-}^m}\ ,
\end{eqnarray}
where $n$ denotes the multipole order, $f_{jklm}$ are 
the RDTs and $\zeta_{q,\pm}=\sqrt{2I_q}e^{\mp i(\psi_q+\psi_{q,0})}$ 
are the new complex normal form coordinates, which are 
the nonlinear generalization of the complex 
Courant-Snyder complex variable of $h_{q,\pm}$. 
The normal form approach for the turn-by-turn betatron 
motion is usually depicted in the scheme of 
Ref.~\cite{Turchetti1}:

\begin{table}[!h]
\vskip 0.2cm
\centering 
\begin{tabular}{r c l}
                  & \hspace{0.3 cm}$\tilde{\mathcal{M}}=e^{:\tilde{H}:}R$ \hspace{0.3 cm}\      &                \\ \vspace{0.1 cm}
$\vec{h}(N)$      &$\xrightarrow[]{\hspace{1.0 cm}\ \hspace{1.0 cm} }$ & $\vec{h}(N+1)$ \\ \vspace{0.1 cm}
$F^{-1}\Bigg\downarrow$ &                                               & $\Bigg\uparrow\quad F$ \\
$\vec{\zeta}(N)$  &$\xrightarrow[]{\hspace{1.0 cm}\ \hspace{1.0 cm} }$ & $\vec{\zeta}(N+1)$ \\ \vspace{0.1 cm}
                  & \hspace{0.3 cm}$\mathsf{R}=e^{:\mathsf{H}:}R$ \hspace{0.3 cm}\                &                \\ \vspace{0.1 cm}
\end{tabular} 
\rule{0mm}{0.1cm}
\end{table}
\hskip -0.3cm $R$ is the same rotation of Eq.~\eqref{eq:OTM1}, 
$\tilde{H}=\tilde{H}(J_x,J_y,\phi_x,\phi_y)$ is the 
phase-dependent Hamiltonian  of Eq.~\eqref{eq:Hbw3A} 
($J$ and $\phi$ are  the linear invariant and 
betatron phase), whereas $\mathsf{H}(I_x,I_y)$ 
is the phase-independent Hamiltonian in 
normal forms,  with $I$ the nonlinear 
invariant. Any dependence on the angles would 
indeed be related to the existence of fixed points 
different from the origin, around which stable orbits 
may exist, hence creating  discontinuous phase space 
trajectories, as the ones of Ref.~\cite{MTE}: In this 
(resonant) case, no regular transformation may convert 
all separated trajectories in continuous circles.

There are two ways to predict the evolution of 
the betatron coordinate $\vec{h}=(h_{x,+},h_{x,-},h_{y,+},h_{y,-})$ 
after one turn: Either solving the problem in 
the Courant-Snyder coordinates 
$\vec{h}(N+1)=e^{:\tilde{H}:}R\ \vec{h}(N)$, 
or finding the generating function $F$, moving 
into normal forms $\vec{\zeta}(N)$, applying the 
amplitude-dependent rotation $\mathsf{R}$ and 
coming back to the Courant-Snyder coordinates 
$\vec{h}(N+1)=e^{:F:}\vec{\zeta}(N+1)$. The 
second strategy, even if longer, turns out to 
be indeed the most effective. In fact, 
analytical expressions for the RDTs $f_{jklm}$ 
(and hence for $F$) may be derived up to a 
certain degree of precision (second order in this 
paper), while the evolution in normal form is 
already known, being an amplitude-dependent 
rotation, whose operator $\mathsf{R}$ is defined
by the tunes ($Q_{x,y}$) and the invariants 
($2I_{x,y}$) only, all measurable quantities, as shown 
later.  In normal forms, the circle surface 
$\pi|\zeta|^2$ is the new (nonlinear) 
invariant, see lower right portrait of Fig.~\ref{fig_ph1}. 
The nonlinearity is also transferred in 
the frequency, which exhibits a dependence on the 
amplitude $|\zeta|=\sqrt{2I}$ (amplitude-dependent tune). 
The system in these coordinates is equivalent to an 
anharmonic oscillator and the motion is a rotation 
whose frequency depends on the initial conditions. 
As the Courant-Snyder matrix contains all the 
information regarding the linear lattice, the 
RDTs of $F$ carry its nonlinear content. 

The equation establishing the change of coordinates 
in normal form of the above scheme may be written 
in terms of the Lie operators 
\begin{eqnarray}
\underbrace{\vec{\zeta}(N+1)}&=&
e^{:-F:}\underbrace{\vec{h}(N+1)}
\nonumber\\
\left(e^{:\mathsf{H}:}R\vec{\zeta}(N)\right)&=&
e^{:-F:}\left(e^{:\tilde{H}:}R\underbrace{\vec{h}(N)}\right)
\nonumber
\end{eqnarray}
\begin{eqnarray}
e^{:\mathsf{H}:}R\vec{\zeta}(N)&=&
e^{:-F:}e^{:\tilde{H}:}R\left(e^{:F:}\vec{\zeta}(N)\right)
\nonumber \\
\left[e^{:\mathsf{H}:}R\right]\vec{\zeta}(N)&=&
\left[e^{:-F:}e^{:\tilde{H}:}Re^{:F:}\right]\vec{\zeta}(N)
\nonumber\ .
\end{eqnarray}
The above relation holding for any $\vec{\zeta}(N)$, 
the following homological equation is obtained
\begin{eqnarray}\label{eq:omol_start}
e^{:\mathsf{H}:}R&=&e^{:-F:}e^{:\tilde{H}:}Re^{:F:} 
\end{eqnarray}
Following the same procedure of Ref.~\cite{Bengtsson}, 
all operators may be decomposed in first- and 
second-order terms as follows
\begin{eqnarray}\label{eq:deco1}
        F &=&         F^{(1)}+         F^{(2)} \\
\mathsf{H}&=&\mathsf{H}^{(1)}+\mathsf{H}^{(2)} \\
\tilde{H} &=& \tilde{H}^{(1)}+ \tilde{H}^{(2)} 
           =\sum\limits_{w=1}^W{\tilde{H}_w}+\frac{1}{2}
		\sum\limits_{w=1}^W{\sum\limits_{u=1}^{w-1}
		{\left[\tilde{H}_u,\tilde{H}_w\right]}}
		\nonumber \\ \label{eq:deco2}
\end{eqnarray}
Equation~\eqref{eq:deco2} results from Eq.~\eqref{eq:Hbw3A}: 
The composition of Lie Operators indeed obeys to the 
Campbell-Backer-Hausdorff (CBH) theorem
\begin{eqnarray}\label{eq:CBH}
e^{:A:}e^{:B:}e^{:C:}=
e^{:A+B+C +\frac{1}{2}\left\{[A,B]+[A,C]+[B,C]\right\}:}
	+O(3^{{\tiny\hbox{rd}}})\ , \nonumber \\
\end{eqnarray}
where $[,]$ denotes the Poisson bracket, the first 
term (A+B+C) is the first-order 
truncation while the rest is the second-order 
contribution. Higher-order terms are here included
in the remainder $O(3^{{\tiny\hbox{rd}}})$.
Being these operators polynomial functions,
Eq.~\eqref{eq:omol_start} apply to all 
coefficients $h_{jklm}$ (for $\tilde{H}$), 
$f_{jklm}$ (for $F$), $\mathsf{h}_{jjll}$ 
(for $\mathsf{H}$). The rotation $R$ then 
will also act to each  term 
$(\zeta_{x,+}^j\zeta_{x,-}^k\zeta_{y,+}^l\zeta_{y,-}^m)$, 
i.e. 
\begin{eqnarray} \label{eq:R-def}
R = e^{2\pi i\left[(j-k)Q_x+(l-m)Q_y)\right]}\ ,\qquad
	\forall\quad j,k,l,m\ .\qquad
\end{eqnarray}
Since the Hamiltonian in normal forms must be phase 
independent (only $\mathsf{h}_{jjll}$ coefficients 
are nonzero), its first-order term must 
contain the detuning part of the Courant-Snyder 
Hamiltonian, which corresponds in turn to its
average over the phases, i.e. 
\begin{eqnarray}\label{eq:mathsfH1}
\mathsf{H}^{(1)}=<\tilde{H}^{(1)}>_{\phi}\ .
\end{eqnarray}
Detuning terms are those in Eq.~\eqref{eq:Hbw3B} 
having $j=k$ and $l=m$ (the octupolar-like 
$h_{2200}$, $h_{1111}$ and $h_{0022}$), since only when 
these two conditions are met, the corresponding 
phase is zero ($h_{x,+}^j h_{x,-}^j h_{y,+}^l h_{y,-}^j=
|h_x|^{2j}|h_y|^{2l}$, phase independent). 
In normal forms, hence, only 
detuning terms are kept, while those phase-dependent 
are {\sl absorbed} by the generating function $F$: This is 
the reason why RDTs will always be phase-dependent. In 
general, at each order, it is always possible in the 
Courant-Snyder Hamiltonian to decompose a phase-independent 
term  $<\tilde{H}^{(1,2)}>_{\phi}$ from a  
phase-dependent remainder $\tilde{H}^{\ddagger}$, 
\begin{eqnarray}\label{eq:tildeH2}
\tilde{H}^{(1,2)}=<\tilde{H}^{(1,2)}>_{\phi}+\tilde{H}^{(1,2)\ddagger}\ .
\end{eqnarray}
Since $I=R^{-1}R=e^{:-A:}e^{:A:}$, where $I$ is 
the identity operator and $A$ a generic function, 
Eq.~\eqref{eq:omol_start} may be rewritten as
\begin{eqnarray}
e^{:\mathsf{H}:}R&=&e^{:-F:}e^{:\tilde{H}:}Re^{:F:}
	\nonumber\\
e^{:F:}e^{:\mathsf{H}:}R&=&e^{:\tilde{H}:}Re^{:F:}
	\nonumber\\
e^{:F:}e^{:\mathsf{H}:}Re^{:-F:}&=&e^{:\tilde{H}:}R
	\nonumber\\
e^{:F:}e^{:\mathsf{H}:}Re^{:-F:}(R^{-1}R)&=&e^{:\tilde{H}:}R
	\nonumber\\
e^{:F:}e^{:\mathsf{H}:}(Re^{:-F:}R^{-1})R&=&e^{:\tilde{H}:}R
	\nonumber\\ 
e^{:F:}e^{:\mathsf{H}:}e^{:-RF:}R&=&e^{:\tilde{H}:}R
	\nonumber\\ \label{eq:omol_1}
\left[e^{:F:}e^{:\mathsf{H}:}e^{:-RF:}\right]R&=&
\left[e^{:\tilde{H}:}\right]R\ .
\end{eqnarray}
The two operators  next to the rotation $R$ 
must then be equal,
\begin{eqnarray}\label{eq:omol_2}
e^{:F:}e^{:\mathsf{H}:}e^{:-RF:}&=&e^{:\tilde{H}:}\ .
\end{eqnarray}
According to the CBH theorem then, up to second order 
Eq.~\eqref{eq:omol_2} may be then rewritten as 
\begin{eqnarray}
e^{:\tilde{H}:}&=&e^{:F+\mathsf{H}-RF
		+\frac{1}{2}[F,\mathsf{H}]-\frac{1}{2}[F,RF]
		-\frac{1}{2}[\mathsf{H},RF]:}
         	+O(3^{{\tiny\hbox{rd}}})\ ,
\nonumber \\\label{eq:omol_3}
\end{eqnarray}
which is equivalent to 
\begin{eqnarray}\label{eq:omol_4}
\tilde{H}&=&(I-R)F +\mathsf{H}
		+\frac{1}{2}[F,\mathsf{H}]-\frac{1}{2}[F,RF]
		-\frac{1}{2}[\mathsf{H},RF]\ . \nonumber
\end{eqnarray}
After substituting the first- and second-order 
decomposition of Eqs.~\eqref{eq:deco1} and 
\eqref{eq:deco2}, the following relations hold
\begin{eqnarray}\label{eq:omol_5}
\tilde{H}^{(1)}+\tilde{H}^{(2)}&=&(I-R)(F^{(1)}+F^{(2)})+
		\mathsf{H}^{(1)}+\mathsf{H}^{(2)}
		\nonumber \\
	&&+\frac{1}{2}[F^{(1)}+F^{(2)},\mathsf{H}^{(1)}+\mathsf{H}^{(2)}]
		\nonumber \\
        &&-\frac{1}{2}[F^{(1)}+F^{(2)},R(F^{(1)}+F^{(2)})]
		\nonumber \\
  	&&-\frac{1}{2}[\mathsf{H}^{(1)}+\mathsf{H}^{(2)},R(F^{(1)}+F^{(2)})]
	\ .\qquad
\end{eqnarray}
Up to the second order the above equation 
reduces to
\begin{eqnarray}\label{eq:omol_6}
\tilde{H}^{(1)}+\tilde{H}^{(2)}&=&(I-R)(F^{(1)}+F^{(2)})+
		\mathsf{H}^{(1)}+\mathsf{H}^{(2)}
		\nonumber \\
	     &&+\frac{1}{2}[F^{(1)},\mathsf{H}^{(1)}]
		-\frac{1}{2}[F^{(1)},RF^{(1)}]\nonumber \\
             &&
               -\frac{1}{2}[\mathsf{H}^{(1)},RF^{(1)}]
	       +O(3^{{\tiny\hbox{rd}}})
	\ .\qquad
\end{eqnarray}
By separating first- and second-order terms we obtain
\begin{eqnarray}\label{eq:omol_7}
\tilde{H}^{(1)}&=&(I-R)F^{(1)}+\mathsf{H}^{(1)} \\
\tilde{H}^{(2)}&=&(I-R)F^{(2)}+\mathsf{H}^{(2)}\label{eq:omol_8}
		  -\frac{1}{2}[\mathsf{H}^{(1)},(I+R)F^{(1)}]
		\nonumber \\
	        &&-\frac{1}{2}[F^{(1)},RF^{(1)}]
\end{eqnarray}

{\sl\bf First-order RDTs } By substituting 
Eqs.~\eqref{eq:mathsfH1} and~\eqref{eq:tildeH2}  
in Eq.~\eqref{eq:omol_7} the normal form 
transformation up to the first order 
reads

\begin{eqnarray}\label{eq:omol_1st_1}
\tilde{H}^{(1)\ddagger}=(I-R)F^{(1)}\quad\Rightarrow
F^{(1)}=\frac{\tilde{H}^{(1)\ddagger}}{I-R}\ .
\end{eqnarray}
Recalling that both $F$ and $\tilde{H}^{(1)}$ are 
polynomial functions (see Eqs.~\eqref{eq:Fdef} and 
\eqref{eq:Hbw3B}), and the definition of $R$ given in  
Eq.~\eqref{eq:R-def}, the above relation 
must hold for any index set $j,k,l,m$, i.e. 
\begin{equation}\label{eq:RDT-1st}
f^{(1)}_{jklm}(s) =\frac{\sum\limits_w 
h_{w,jklm}e^{i[(j-k)\Delta\phi_{w,x}^{(s)}+(l-m)\Delta\phi_{w,y}^{(s)}]}}
                     {1-e^{2\pi i[(j-k)Q_x+(l-m)Q_y]}}\ \ .
\end{equation}
The explicit dependence of the RDTs on $s$ is 
due to the presence of the phase advances between the 
location where $f_{jklm}$ is either evaluated or 
measured ($s$) and the one of the magnet $w$: RDTs 
measured at different places will contain hence 
different phase advances, making $f_{jklm}$ 
to vary. Formulas in Table~\ref{tab:lattice-rdt} 
are derived from Eqs.~\eqref{eq:RDT-1st} 
and~\eqref{eq:h_Vs_KJ}.

Equation~\eqref{eq:RDT-1st} shows a fundamental 
feature of the first-order RDTS: they depend linearly on 
$h_{w,jklm}$ and hence on the magnet strengths 
$K_{w,n-1}$ and $J_{w,n-1}$, through 
Eq.~\eqref{eq:h_Vs_KJ}. By measuring sextupole RDTs, 
for instance, it is possible to infer a sextupole 
error model and to correct undesired deviation 
from the ideal setting. This is done by first rewriting 
Eq.~\eqref{eq:RDT-1st} as linear system
\begin{eqnarray}\label{eq:RDT_linsys}
\vec{f}_{\tiny\hbox{norm. sext}}=
{\bf B_{N_k}}(\beta_x,\beta_y,\phi_x,\phi_y,Q_x,Q_y)\vec{K}_2\ ,\qquad
\end{eqnarray}
where $\vec{f}_{\tiny\hbox{norm. sext}}$ is a 
vector containing the sextupolar RDTs evaluated 
or measured at $N_f$ {\sl observation points} along the ring, 
$\vec{K}_2$ is a vector with the strengths 
of the $N_k$ normal sextupoles, and ${\bf B_{N_k}}$ is a 
$N_f\times N_k$ response matrix which depends on the C-S 
parameters only (betatron functions and phases at the 
sextupoles, betatron phases at the observation points, 
and the tunes). Ideal sextupoles would induce a given 
RDT vector, 
\begin{eqnarray}
\vec{f}_{\tiny\hbox{norm. sext}}^{\tiny\hbox{mod}}=
{\bf B_{N_k}}\vec{K}_2^{\tiny\hbox{mod}}\ .\qquad
\end{eqnarray}
If sextupolar RDTs are measurable, a 
difference vector may be defined as 
$\Delta\vec{f}=\vec{f}_{\tiny\hbox{norm. sext}}^{\tiny\hbox{mod}}- 
\vec{f}_{\tiny\hbox{norm. sext}}^{\tiny\hbox{meas}}$ 
and the above system may be rewritten and 
pseudo-inverted, via single-value decomposition 
(SVD) for instance, to infer sextupole errors
\begin{eqnarray}
\Delta\vec{f}={\bf B_{N_k}}\Delta\vec{K}_2\ .\qquad
\end{eqnarray}
If corrector sextupoles are available, a second 
system may be defined as 
\begin{eqnarray}\label{eq:RDT_lincor}
\Delta\vec{f}={\bf B_{N_c}}\vec{K}_2^{\tiny\hbox{cor}}\ ,\qquad
\end{eqnarray}
where $\vec{K}_2^{\tiny\hbox{cor}}$ contains 
the strengths of the $N_c$ corrector magnets 
and ${\bf B_{N_c}}$ is the same matrix of 
Eq.~\eqref{eq:RDT_linsys}, whose dimensions are 
$N_f\times N_c$ and whose C-S parameters correspond 
to the corrector sextupoles in place of the main 
magnets. $\vec{K}_2^{\tiny\hbox{cor}}$ may be then 
evaluated from the difference between the measured 
and expected RDT vector ($\Delta\vec{f}$) after 
pseudo-inverting Eq.~\eqref{eq:RDT_lincor}, providing 
the strengths of the corrector sextupoles to be set 
in order to minimize the deviation from the ideal 
non-linear lattice model. The system may be extended 
in order to include  in the RDT vector the linear 
chromaticity (which is also linear in the sextupole 
strengths).

To complete the normal form approach, explicit formulas 
for the change of coordinates are missing. They are 
readily derived from 
\begin{eqnarray}\label{eq:NFtransform}
\vec{h}(s,N)=e^{:F^{(s)}:}\vec{\zeta}(s,N)\ .
\end{eqnarray}
$\vec{h}(s,N)$ is the turn-by-turn evolution of the 
complex Courant-Snyder coordinates at a given $s$, 
and $\vec{\zeta}(s,N)$ is the equivalent in normal 
forms, where the circular orbit in phase space 
may be written as  
\begin{eqnarray}\label{eq:zetadef}
\zeta_{x,\pm}(s,N)=\sqrt{2I_x}e^{\mp i(2\pi Q_xN+\psi_{s,x,0})}\ .
\end{eqnarray}
$2I_x$ is the nonlinear invariant, whereas 
$\psi_{s,x,0}$ denotes the initial phase (in 
normal form). Even if not explicitly indicated,   
$Q_x$ depends on $2I_x$ (amplitude-dependent 
detuning). The Lie operator may be expanded 
as a Taylor series 
\begin{eqnarray}\label{eq:NFtransform2}
e^{:F:}\vec{\zeta}=\vec{\zeta}+[F,\vec{\zeta}\ ]
	+\frac{1}{2}[F,[F,\vec{\zeta}\ ]] 
	+O(3^{{\tiny\hbox{rd}}})\ ,
\end{eqnarray}
where the remainder contains higher-order Poisson 
brackets. Up to the first order the inverse normal 
form transformation then reads
\begin{eqnarray}\label{eq:tbt-RDT}
\vec{h}(s,N)\simeq\vec{\zeta}(s,N)+[F^{(s)},\vec{\zeta}(s,N)]\ .
\end{eqnarray}
The Poisson bracket may be evaluated after 
substituting $F$ of Eq.~\eqref{eq:Fdef} into 
Eq~\eqref{eq:tbt-RDT} and applying the rule  
\begin{eqnarray}\label{eq:Poisson-bracket}
[\zeta_{r,+}^j,\zeta_{r,-}^k]=-2i(jk)\zeta_{r,+}^{j-1}\zeta_{r,-}^{k-1}
			     =-[\zeta_{r,-}^k,\zeta_{r,+}^j]\ ,\qquad
\end{eqnarray}
where $r$ stands for either $x$ or $y$ and all 
other combinations yield zero Poisson brackets.
The result is
\begin{eqnarray}\label{e:horizontal}
h_{x,-}(s,N)&=&\sqrt{2I_x}e^{i(2\pi Q_xN+\psi_{s,x,0})}\ - 
\\\nonumber \\
&&\hskip -0.5cm 2i\sum_{jklm}{jf^{(s)}_{jklm}(2I_x)^{\frac{j+k-1}{2}}
                  (2I_y)^{\frac{l+m}{2}}}\ \times \nonumber\\
&&\hskip -0.5cm
 e^{i[(1-j+k)(2\pi Q_xN+\psi_{s,x,0})+
                  (m-l)(2\pi Q_yN+\psi_{s,y,0})]}\nonumber
\end{eqnarray}
and
\begin{eqnarray}\label{e:vertical}
h_{y,-}(s,N)&=&\sqrt{2I_y}e^{i(2\pi Q_yN+\psi_{s,y,0})}\ -
\\\nonumber \\
&&\hskip -0.5cm 2i\sum_{jklm}{lf^{(s)}_{jklm}(2I_x)^{\frac{j+k}{2}}
                 (2I_y)^{\frac{l+m-1}{2}}} \times \nonumber\\
&&\hskip -0.5cm
e^{i[(k-j)(2\pi Q_xN+\psi_{s,x,0})+
                 (1-l+m)(2\pi Q_yN+\psi_{s,y,0})]}\ \ ,\nonumber
\end{eqnarray}
from which 
Table IV of Ref.~\cite{prstab_strength} is 
derived. 

In conclusion, up to the first order, 
Eq.~\eqref{eq:RDT-1st} is the formula to 
evaluate the RDTs from the lattice parameters, 
while Eqs.~\eqref{e:horizontal} and~\eqref{e:vertical} 
may be used to compute (or measure) them  from 
the FFT of turn-by-turn position data, either simulated 
via tracking or measured by BPMs. Both approaches provide 
the same equivalent RDTs. In the following part it will 
be shown how this equivalence is no longer true when 
second-order terms are to be taken into account.

{\sl\bf Second-order RDTs }
The evaluation up to the second order of the RDTs 
passes through Eq.~\eqref{eq:omol_8} (from the lattice) 
and the Poisson bracket of Eq.~\eqref{eq:NFtransform2} (from 
turn-by-turn data). Equation~\eqref{eq:omol_8} may be 
rewritten as 
\begin{eqnarray}
(I-R)F^{(2)}+\mathsf{H}^{(2)}&=&\tilde{H}^{(2)}
		  +\frac{1}{2}[\mathsf{H}^{(1)},(I+R)F^{(1)}]
		\nonumber \\ \label{eq:omol_9}
	        &&+\frac{1}{2}[F^{(1)},RF^{(1)}]
\end{eqnarray}
By inserting Eq.~\eqref{eq:omol_1st_1} into the Poisson 
brackets, the following relations apply
\begin{eqnarray}
[\mathsf{H}^{(1)},(I+R)F^{(1)}]&=&
    [\mathsf{H}^{(1)},\frac{I+R}{I-R}\tilde{H}^{(1)\ddagger}]\ ,
\end{eqnarray}
\begin{eqnarray}
[F^{(1)},RF^{(1)}]
&=&[\frac{\tilde{H}^{(1)\ddagger}}{I-R},\frac{R}{I-R}\tilde{H}^{(1)\ddagger}]
\nonumber \\
&=&[\frac{\tilde{H}^{(1)\ddagger}}{I-R},\frac{I+R-I}{I-R}\tilde{H}^{(1)\ddagger}]
\nonumber \\
&=&[\frac{\tilde{H}^{(1)\ddagger}}{I-R},\frac{\tilde{H}^{(1)\ddagger}}{I-R}]-
   [\frac{\tilde{H}^{(1)\ddagger}}{I-R},\tilde{H}^{(1)\ddagger}]
\nonumber \\
&=&[\tilde{H}^{(1)\ddagger},\frac{\tilde{H}^{(1)\ddagger}}{I-R}]\ ,
\end{eqnarray}
providing
\begin{eqnarray}
(I-R)F^{(2)}+\mathsf{H}^{(2)}&=&\tilde{H}^{(2)}
      +\frac{1}{2}[\mathsf{H}^{(1)},\frac{I+R}{I-R}\tilde{H}^{(1)\ddagger}]
		\nonumber \\ \label{eq:omol_10}
&&+\frac{1}{2}[\tilde{H}^{(1)\ddagger},\frac{\tilde{H}^{(1)\ddagger}}{I-R}]\ .
\end{eqnarray}
The average over the phases of Eq.~\eqref{eq:omol_10} 
reads
\begin{eqnarray}
<(I-R)F^{(2)}>_{\phi}+<\mathsf{H}^{(2)}>_{\phi}&=&<\tilde{H}^{(2)}>_{\phi}+
		\label{eq:omol_11} \\\nonumber
&&\hskip -5cm
\frac{1}{2}<[\mathsf{H}^{(1)},\frac{I+R}{I-R}\tilde{H}^{(1)\ddagger}]>_{\phi}
+\frac{1}{2}<[\tilde{H}^{(1)\ddagger},\frac{\tilde{H}^{(1)\ddagger}}{I-R}]>_{\phi}\ .
\end{eqnarray}
It is worthwhile recalling that, according to  
Eq.~\eqref{eq:RDT-1st}, $F^{(1)}$ is phase-dependent, 
while $\mathsf{H}^{(2)}$, by definition, must be 
phase-independent, i.e. $\mathsf{H}^{(2)}=<\mathsf{H}^{(2)}>_{\phi}$. 
Hence 
\begin{eqnarray}
<(I-R)F^{(2)}>_{\phi}&=&0  \\
<\mathsf{H}^{(2)}>_{\phi}&=&\mathsf{H}^{(2)} \\
<[\mathsf{H}^{(1)},\frac{I+R}{I-R}\tilde{H}^{(1)\ddagger}]>_{\phi}&=&0\ ,
\end{eqnarray}
where the later relation results from the fact that 
$\mathsf{H}^{(1)}$ is phase-independent, conversely to 
$\tilde{H}^{(1)\ddagger}$: the resulting Poisson 
bracket is then necessarily phase-dependent, hence with 
zero average. Equation~\eqref{eq:omol_11} thus reads
\begin{eqnarray} \label{eq:Ham_NF_2}
\mathsf{H}^{(2)}=<\tilde{H}^{(2)}>_{\phi}+
\frac{1}{2}<[\tilde{H}^{(1)\ddagger},\frac{\tilde{H}^{(1)\ddagger}}{I-R}]>_{\phi}
\ .\qquad
\end{eqnarray}
Substituting the above expression in 
Eq.~\eqref{eq:omol_10}, the latter is equivalent to 
\begin{eqnarray}
&&(I-R)F^{(2)}+<\tilde{H}^{(2)}>_{\phi}+
\frac{1}{2}<[\tilde{H}^{(1)\ddagger},\frac{\tilde{H}^{(1)\ddagger}}{I-R}]>_{\phi}
\quad =\nonumber \\
&&\tilde{H}^{(2)}
      +\frac{1}{2}[\mathsf{H}^{(1)},\frac{I+R}{I-R}\tilde{H}^{(1)\ddagger}]
\label{eq:omol_12}
+\frac{1}{2}[\tilde{H}^{(1)\ddagger},\frac{\tilde{H}^{(1)\ddagger}}{I-R}]\ .
\end{eqnarray}
According to Eq.~\eqref{eq:tildeH2}
\begin{eqnarray}\nonumber
\tilde{H}^{(2)}-<\tilde{H}^{(2)}>_{\phi}&=&\tilde{H}^{(2)\ddagger} \ ,
\end{eqnarray}
\begin{eqnarray}\nonumber
[\tilde{H}^{(1)\ddagger},\frac{\tilde{H}^{(1)\ddagger}}{I-R}]-
<[\tilde{H}^{(1)\ddagger},\frac{\tilde{H}^{(1)\ddagger}}{I-R}]>_{\phi}&=&
[\tilde{H}^{(1)\ddagger},\frac{\tilde{H}^{(1)\ddagger}}{I-R}]^{\ddagger}\ ,
\end{eqnarray}
where $[,]^{\ddagger}$ denotes the phase-dependent 
part of the Poisson bracket. Equation~\eqref{eq:omol_11} 
then reads
\begin{eqnarray}
(I-R)F^{(2)}&=&\tilde{H}^{(2)\ddagger}
      +\frac{1}{2}[\mathsf{H}^{(1)},\frac{I+R}{I-R}\tilde{H}^{(1)\ddagger}]
		\nonumber \\ \nonumber 
&&+\frac{1}{2}[\tilde{H}^{(1)\ddagger},\frac{\tilde{H}^{(1)\ddagger}}{I-R}]^{\ddagger}\ ,
\end{eqnarray}
providing the general equation to evaluate the 
second-order contribution to the RDTs 
(from the lattice)
\begin{eqnarray}
F^{(2)}&=&\left(\tilde{H}^{(2)\ddagger}
      +\frac{1}{2}[\mathsf{H}^{(1)},\frac{I+R}{I-R}\tilde{H}^{(1)\ddagger}]
		\right.\nonumber \\ \label{eq:omol_13} &&\left.
+\frac{1}{2}[\tilde{H}^{(1)\ddagger},\frac{\tilde{H}^{(1)\ddagger}}{I-R}]^{\ddagger}
\right)/(I-R)\ .
\end{eqnarray}
The situation is less desperate than it could appear. 
$\mathsf{H}^{(1)}$ is the first-order Hamiltonian 
in normal forms that contains first-order detuning 
terms only (phase independent). These may be
generated firstly by normal quadrupole errors 
$\delta K_1$, through the Hamiltonian terms 
$\sum\limits_w{h_{w,1100}}=\sum\limits_w\beta_{x,w}\delta K_{1,w}$ 
and 
$\sum\limits_w{h_{w,0011}}=\sum\limits_w\beta_{y,w}\delta K_{1,w}$. 
Nevertheless, assuming that the machine is tuned to 
the desired ideal tune, those sums are equal to zero. 
The next nonzero first-order detuning terms 
may then be excited by octupoles (through the Hamiltonian 
terms $\sum\limits_w{h_{w,2200}}=\sum\limits_w\beta_{x,w}^2K_{2,w}$ 
and the like), which are assumed to be a negligible source 
of detuning compared to sextupoles, which act only in the 
second-order Hamiltonian $\mathsf{H}^{(2)}$ of 
Eq.~\eqref{eq:Ham_NF_2}. 
Under thse assumptions the first-order 
Hamiltonian in the complex Courant-Snyder coordinates 
will be always phase-dependent, i.e. 
$\tilde{H}^{(1)\ddagger}\equiv\tilde{H}^{(1)}$. The same is  
true for second-order RDTs: only phase-dependent terms of 
$[\tilde{H}^{(1)\ddagger},\frac{\tilde{H}^{(1)\ddagger}}{I-R}]$ 
will then be used.
Therefore, in the context of this paper, 
$\mathsf{H}^{(1)}=0$ and 
$[\tilde{H}^{(1)\ddagger},\frac{\tilde{H}^{(1)\ddagger}}{I-R}]^\ddagger=
[\tilde{H}^{(1)},\frac{\tilde{H}^{(1)}}{I-R}]$, 
simplifying the above equation
\begin{eqnarray}\label{eq:omol_14}
F^{(2)}&=&\left(\tilde{H}^{(2)\ddagger}
+\frac{1}{2}[\tilde{H}^{(1)},\frac{\tilde{H}^{(1)}}{I-R}]
\right)/(I-R)\ .\qquad
\end{eqnarray} 
The assumption of having no detuning term in the 
first-order Hamiltonian ($\mathsf{H}^{(1)}=0$) shall 
not be confused with the unnecessary assumption of having 
zero linear detuning, that may come from sextupolar 
terms in the second-order Hamiltonian ($\mathsf{H}^{(2)}\ne0$).
If first-order octupolar terms may not be neglected, 
the more general Eq.~\eqref{eq:omol_13} shall be used. 

The first step in making Eq.~\eqref{eq:omol_14} 
more explicit is the evaluation of 
$\tilde{H}^{(2)}$, whose phase-independent terms 
will contribute to the detuning with amplitude of 
$\mathsf{H}^{(2)}$, through Eq.~\eqref{eq:Ham_NF_2}, 
while those phase-dependent will define the second-order 
RDTs via Eq.~\eqref{eq:omol_14}. $\tilde{H}^{(2)}$ 
results from the CBH theorem and, according to 
Eq.~\eqref{eq:deco2},
\begin{eqnarray}
\tilde{H}^{(2)}(s)=\frac{1}{2}
		\sum\limits_{w=1}^W{\sum\limits_{u=1}^{w-1}
		{\left[\tilde{H}_u(s),\tilde{H}_w(s)\right]}}
		\ ,
\end{eqnarray}
where the dependence on $s$ is kept explicit 
to prevent incorrect simplification. Equation~\eqref{eq:Hbw3B} 
provides the explicit expressions for both 
$\tilde{H}_u(s)$ and $\tilde{H}_u(s)$
\begin{widetext}
\begin{eqnarray}
\tilde{H}_{w}(s)&=&\sum_{jklm}
  {h_{w,jklm}e^{i[(j-k)\Delta\phi_{w,x}^{(s)}+(l-m)\Delta\phi_{w,y}^{(s)}]}}
h_{x,+}^jh_{x,-}^kh_{y,+}^lh_{y,-}^m\ \ , \\
\tilde{H}_{u}(s)&=&\sum_{pqrt}
  {h_{u,pqrt}e^{i[(p-q)\Delta\phi_{u,x}^{(s)}+(r-t)\Delta\phi_{u,y}^{(s)}]}}
h_{x,+}^ph_{x,-}^qh_{y,+}^rh_{y,-}^t\ \ ,
\end{eqnarray}
yielding
\begin{eqnarray}
\tilde{H}^{(2)}(s)&=&\frac{1}{2}
		\sum\limits_{w=1}^W{\sum\limits_{u=1}^{w-1}\Bigg\{
		{\sum_{jklm}\sum_{pqrt}h_{w,jklm}h_{u,pqrt} 	
		 e^{i[(p-q)\Delta\phi_{u,x}^{(s)}+(r-t)\Delta\phi_{u,y}^{(s)}
		     +(j-k)\Delta\phi_{w,x}^{(s)}+(l-m)\Delta\phi_{w,y}^{(s)}]}
\times}}\nonumber \\
&&\hskip 3 cm	[h_{x,+}^ph_{x,-}^qh_{y,+}^rh_{y,-}^t,
		 h_{x,+}^jh_{x,-}^kh_{y,+}^lh_{y,-}^m]\Bigg\}
		\ .
\end{eqnarray}
Applying the distributive property of the Poisson 
bracket, $[AB,CD]=[A,C]BD+C[A,D]B+A[B,C]D+AC[B,D]$, 
and Eq.~\eqref{eq:Poisson-bracket}, the above definition 
reads
\begin{eqnarray}\label{eq:H2tilde}
\tilde{H}^{(2)}(s)&=&i
		\sum\limits_{w=1}^W{\sum\limits_{u=1}^{w-1}\Bigg\{
		{\sum_{jklm}\sum_{pqrt}h_{w,jklm}h_{u,pqrt} 	
		 e^{i[(p-q)\Delta\phi_{u,x}^{(s)}+(r-t)\Delta\phi_{u,y}^{(s)}
		     +(j-k)\Delta\phi_{w,x}^{(s)}+(l-m)\Delta\phi_{w,y}^{(s)}]}
\times}}     \\
&&\hskip 2 cm	
\bigg[(jq-kp)h_{x,+}^{j+p-1}h_{x,-}^{k+q-1}h_{y,+}^{l+r}  h_{y,-}^{m+t}+
      (lt-mr)h_{x,+}^{j+p}  h_{x,-}^{k+q}  h_{y,+}^{l+r-1}h_{y,-}^{m+t-1}
\bigg]\Bigg\}\ .\nonumber
\end{eqnarray}
This equation shows how $\tilde{H}^{(2)}$ depends 
explicitly on $s$ via the phases advances 
$\Delta\phi_{u}^{(s)}$ and $\Delta\phi_{w}^{(s)}$
in the exponential term. Such a dependence disappears only
when $p-q=-(j-k)$ and $r-t=-(l-m)$, i.e. whenever 
$j+p=q+k$ and $l+r=m+t$: a condition, however, that 
generates detuning terms, since 
$h_{x,+}^{j+p-1}h_{x,-}^{k+q-1}h_{y,+}^{l+r}h_{y,-}^{m+t}=
|h_x|^{2(j+p-1)}|h_y|^{2(l+r)}$ and 
$h_{x,+}^{j+p}h_{x,-}^{k+q}h_{y,+}^{l+r-1}h_{y,-}^{m+t-1}=
|h_x|^{2(j+p}|h_y|^{2(l+r-1)}$, both phase-independent. 
$\tilde{H}^{(2),\ddagger}$, and hence $F^{(2)}$, are 
then $s$-dependent, while the detuning 
(phase-independent) term $<\tilde{H}^{(2)}>_\phi$ 
is invariant along the ring, like the tunes. 

In general, to evaluate the second-order Hamiltonian 
terms, only those sets of index $jklm$ and $pqrt$ 
in Eq.~\eqref{eq:H2tilde} satisfying the following 
relations shall be evaluated and summed up
\begin{eqnarray}\label{eq:select0}
\tilde{H}^{(2)}_{abcd}(s)=i\left\{\tilde{h}^{(2)}_{abcd}(s)\right\}
                   h_{x,+}^ah_{x,-}^bh_{y,+}^ch_{y,-}^d
\quad \Rightarrow\hspace{1.0cm}
\left\{
\begin{array}{l}
j+p-1=a \\
k+q-1=b \\
l+r=c   \\
m+t=d   
\end{array}
\right.
\qquad\hbox{or}\qquad
\left\{
\begin{array}{l}
j+p=a   \\
k+q=b   \\
l+r-1=c \\
m+t-1=d 
\end{array}
\right. \quad , 
\end{eqnarray}
An example may be of help understanding this 
procedure. Let us evaluate the skew sextupole-like
Hamiltonian term $\tilde{H}^{(2)}_{0030}$. According to 
Eqs.~\eqref{eq:H2tilde} and~\eqref{eq:select0}, the 
possible combinations of indexes $jklm$ and $pqrt$ 
generating a term proportional to $h_{y,+}^3$ must
satisfy the following conditions
\begin{eqnarray}\label{eq:select}
\tilde{H}^{(2)}_{0030}(s)=i\left\{\tilde{h}^{(2)}_{0030}(s)\right\}
                   h_{y,+}^3
\quad \Rightarrow\hspace{1.0cm}
\left\{
\begin{array}{l}
j+p-1=0 \\
k+q-1=0 \\
l+r=3   \\
m+t=0   
\end{array}
\right.
\qquad\hbox{or}\qquad
\left\{
\begin{array}{l}
j+p=0   \\
k+q=0   \\
l+r-1=3 \\
m+t-1=0 
\end{array}
\right. \quad , 
\end{eqnarray}
resulting in 
\begin{eqnarray}\label{eq:tildeh2_0030}
\tilde{H}^{(2)}_{0030}(s)&=&i\left\{\tilde{h}^{(2)}_{0030}(s)\right\}h_{y,+}^3
  \\&&\hskip-1cm=
	         i\Bigg\{\sum\limits_{w=1}^W\sum\limits_{u=1}^{w-1}
	         -h_{w,1010}\ h_{u,0120}\  	
		 e^{i[ \Delta\phi_{u,x}^{(s)}+ \Delta\phi_{u,y}^{(s)}
		      -\Delta\phi_{w,x}^{(s)}+2\Delta\phi_{w,y}^{(s)}]}
	         +h_{w,0110}\ h_{u,1020}\  	
		 e^{i[-\Delta\phi_{u,x}^{(s)}+ \Delta\phi_{u,y}^{(s)}
		      +\Delta\phi_{w,x}^{(s)}+2\Delta\phi_{w,y}^{(s)}]}
 \nonumber \\&&\hskip 0.1cm
	         -h_{w,1020}\ h_{u,0110}\ 
		 e^{i[ \Delta\phi_{u,x}^{(s)}+2\Delta\phi_{u,y}^{(s)}
		      -\Delta\phi_{w,x}^{(s)}+ \Delta\phi_{w,y}^{(s)}]}
	         +h_{w,0120}\ h_{u,1010}\ 
		 e^{i[-\Delta\phi_{u,x}^{(s)}+2\Delta\phi_{u,y}^{(s)}
		      +\Delta\phi_{w,x}^{(s)}+ \Delta\phi_{w,y}^{(s)}]}
	   \Bigg\}h_{y,+}^3 \ .\nonumber
\end{eqnarray}
This shows how skew-sextupole-like Hamiltonian 
terms, and hence second-order RDTs, may be 
generated by cross-products between skew 
quadrupoles (via $h_{1010}$ and $h_{0110}$) 
and normal sextupoles (via $h_{1020}$ and 
$h_{0120}$). They are also generated by quadrupole 
errors ($h_{2000}$, $h_{1100}$ and the like) together 
with skew sextupoles ($h_{0030}$ and $h_{0012}$). 
Nevertheless, it is assumed that no physical strong 
skew sextupole is powered and that skew sextupole 
Hamiltonian coefficients are generated by small 
sextupole tilts. Being quadrupole errors also small, 
their products is generally negligible compared to the 
products between the larger normal sextupole coefficients 
and those generated by coupling. Throughout the following 
calculations, this simplification will be always applied 
and all products between either beta-beat or coupling terms 
and those generated by skew sextupoles will be put in the 
third-order remainder and ignored.

The second step in evaluating Eq.~\eqref{eq:omol_14} 
is the computation of $[\tilde{H}^{(1)},\frac{\tilde{H}^{(1)}}{I-R}]$. 
The two arguments within the bracket read 
\begin{eqnarray}
\tilde{H}^{(1)}&=&\sum\limits_{w=1}^W{\tilde{H}_{w}(s)}=
  \sum\limits_{w=1}^W\sum_{jklm}
	{h_{w,jklm}e^{i[(j-k)\Delta\phi_{w,x}^{(s)}+(l-m)\Delta\phi_{w,y}^{(s)}]}}
h_{x,+}^jh_{x,-}^kh_{y,+}^lh_{y,-}^m\ \ , \\
\frac{\tilde{H}^{(1)}}{I-R}&=&
  \sum\limits_{u=1}^W\sum_{pqrt}
	{h_{u,pqrt}\frac
	{e^{i[(p-q)\Delta\phi_{u,x}^{(s)}+(r-t)\Delta\phi_{u,y}^{(s)}]}}
	{1-e^{2\pi i[(p-q)Q_x+(r-t)Q_y]}}}
h_{x,+}^ph_{x,-}^qh_{y,+}^rh_{y,-}^t\ \ .
\end{eqnarray}
The Poisson bracket may be then split as follows
\begin{eqnarray}
[\tilde{H}^{(1)},\frac{\tilde{H}^{(1)}}{I-R}]&=&
		\sum\limits_{w=1}^W{\sum\limits_{u=1}^{W}\Bigg\{
		{\sum_{jklm}\sum_{pqrt}h_{w,jklm}h_{u,pqrt} 
		 \frac	
		 {e^{i[(p-q)\Delta\phi_{u,x}^{(s)}+(r-t)\Delta\phi_{u,y}^{(s)}
		     +(j-k)\Delta\phi_{w,x}^{(s)}+(l-m)\Delta\phi_{w,y}^{(s)}]}}
		 {1-e^{2\pi i[(p-q)Q_x+(r-t)Q_y]}}}}
\times\nonumber \\
&&\hskip 3 cm	\left[h_{x,+}^ph_{x,-}^qh_{y,+}^rh_{y,-}^t\ ,\ 
		 h_{x,+}^jh_{x,-}^kh_{y,+}^lh_{y,-}^m\right]\Bigg\}
		\ .
\end{eqnarray}
The argument within the Poisson bracket in the r.h.s. 
is the same of Eq.~\eqref{eq:H2tilde}, resulting in 
\begin{eqnarray}
\frac{1}{2}[\tilde{H}^{(1)},\frac{\tilde{H}^{(1)}}{I-R}]&=&i
		\sum\limits_{w=1}^W{\sum\limits_{u=1}^{W}\Bigg\{
		{\sum_{jklm}\sum_{pqrt}h_{w,jklm}h_{u,pqrt} 
		 \frac	
		 {e^{i[(p-q)\Delta\phi_{u,x}^{(s)}+(r-t)\Delta\phi_{u,y}^{(s)}
		     +(j-k)\Delta\phi_{w,x}^{(s)}+(l-m)\Delta\phi_{w,y}^{(s)}]}}
		 {1-e^{2\pi i[(p-q)Q_x+(r-t)Q_y]}}}}
\times \nonumber\\
&&\hskip 2 cm	
\bigg[(jq-kp)h_{x,+}^{j+p-1}h_{x,-}^{k+q-1}h_{y,+}^{l+r}  h_{y,-}^{m+t}+
      (lt-mr)h_{x,+}^{j+p}  h_{x,-}^{k+q}  h_{y,+}^{l+r-1}h_{y,-}^{m+t-1}
\bigg]\Bigg\}\ ,\nonumber \\
\label{eq:PBH1}
\frac{1}{2}[\tilde{H}^{(1)},\frac{\tilde{H}^{(1)}}{I-R}]_{abcd}&=&
              i\left\{\hat{h}^{(2)}_{abcd}(s)\right\}
                   h_{x,+}^ah_{x,-}^bh_{y,+}^ch_{y,-}^d
\quad \Rightarrow\hspace{0.4cm}
\left\{
\begin{array}{l}
j+p-1=a \\
k+q-1=b \\
l+r=c   \\
m+t=d   
\end{array}
\right.
\quad\hbox{or}\quad
\left\{
\begin{array}{l}
j+p=a   \\
k+q=b   \\
l+r-1=c \\
m+t-1=d 
\end{array}
\right. \quad , 
\end{eqnarray}

The above expression is similar to the one of 
Eq.~\eqref{eq:H2tilde} with two notable differences.
Both summations here extend over the total number of 
magnets $W$, whereas the two are nested in 
Eq.~\eqref{eq:H2tilde}. The exponential term is 
now scaled by the rotational term $I-R$. However, 
the same selection rules of Eq.~\eqref{eq:select0} 
apply. It is worthwhile mentioning that the more 
general case 
$[\tilde{H}^{(1)\ddagger},\frac{\tilde{H}^{(1)\ddagger}}{I-R}]^\ddagger$ 
is retrieved by requesting that only those terms 
with $j\ne k$, $l\ne m$, $p\ne q$ and $r\ne t$ 
are included in the summations.

The explicit expression for the 0030 example, 
after applying the same selection rules of 
Eq.~\eqref{eq:select}, reads
\begin{eqnarray}\label{eq:hath2_0030}
\frac{1}{2}[\tilde{H}^{(1)},\frac{\tilde{H}^{(1)}}{I-R}]_{0030}(s)&=&i
        \left\{\hat{h}_{0030}(s)\right\}h_{y,+}^3
  \\&&\hskip-2.7cm=
	         i\Bigg\{\sum\limits_{w=1}^W\sum\limits_{u=1}^{W}
	         -h_{w,1010}\ h_{u,0120}\  	
	\frac{e^{i[ \Delta\phi_{u,x}^{(s)}+ \Delta\phi_{u,y}^{(s)}
		      -\Delta\phi_{w,x}^{(s)}+2\Delta\phi_{w,y}^{(s)}]}}
	     {1-e^{2\pi i[Q_x+Q_y]}}
	         +h_{w,0110}\ h_{u,1020}\  	
	\frac{e^{i[-\Delta\phi_{u,x}^{(s)}+ \Delta\phi_{u,y}^{(s)}
		      +\Delta\phi_{w,x}^{(s)}+2\Delta\phi_{w,y}^{(s)}]}}
	     {1-e^{2\pi i[-Q_x+Q_y]}}
 \nonumber \\&&\hskip -1.7cm
	         -h_{w,1020}\ h_{u,0110}\ 
	\frac{e^{i[ \Delta\phi_{u,x}^{(s)}+2\Delta\phi_{u,y}^{(s)}
		      -\Delta\phi_{w,x}^{(s)}+ \Delta\phi_{w,y}^{(s)}]}}
	     {1-e^{2\pi i[Q_x+2Q_y]}}
	         +h_{w,0120}\ h_{u,1010}\ 
	\frac{e^{i[-\Delta\phi_{u,x}^{(s)}+2\Delta\phi_{u,y}^{(s)}
		      +\Delta\phi_{w,x}^{(s)}+ \Delta\phi_{w,y}^{(s)}]}}
	     {1-e^{2\pi i[-Q_x+2Q_y]}}
	   \Bigg\}h_{y,+}^3 \ .\nonumber
\end{eqnarray}
The example is completed by writing the analytic 
expression for $f_{0030}$ up to the second order
\begin{eqnarray}\label{eq:f0030}
f_{0030}&=&f_{0030}^{(1)}+f_{0030}^{(2)}=\frac{1}{1-e^{2\pi i(3Q_y)}}
	\Bigg\{\sum\limits_{w=1}^W{h_{w,0030}\ e^{i3\Delta\phi_{w,y}^{(s)}}}+
	       i\tilde{h}^{(2)}_{0030}(s)+i\hat{h}_{0030}(s)\Bigg\}
           \ \ , 
\end{eqnarray}
where $\tilde{h}^{(2)}_{0030}(s)$ and 
$\hat{h}_{0030}(s)$ are excited by skew 
quadrupoles (via $h_{1010}$ and $h_{0110}$) 
and normal sextupoles (via $h_{1020}$ and 
$h_{0120}$), see Eqs.~\eqref{eq:tildeh2_0030} 
and~\eqref{eq:hath2_0030}, while the first term 
in the r.h.s. is excited by either skew sextupoles, if 
any, or tilted normal sextupoles through 
$h_{0030}$ and Eq.~\eqref{eq:RDT-1st}.
The Hamiltonian coefficients ($h_{0030}$, 
$h_{1010}$ and the like) may be computed from 
the magnet strengths and the C-S parameters 
via Eq.~\eqref{eq:h_Vs_KJ}. 
The same computation may be carried out for any 
other RDTs in the same way. 

\begin{table*}
\caption{List of Hamiltonian coefficients 
$h_{jklm}$ and the corresponding magnetic element 
contributing to the second-order RDTs (up to sextupoles). Note that 
skew sextupole coefficients are not in the list. It is indeed 
assumed that no physical strong skew sextupole is powered and 
that first-order skew sextupole Hamiltonian coefficients are 
generated by small sextupole tilts. In the second-order terms 
$\tilde{h}^{(2)}_{jklm}$ and $\hat{h}_{jklm}$ of 
Eqs.~\eqref{eq:select0}, and~\eqref{eq:PBH1} skew sextupole 
coefficients enter always multiplied by either beta-beat or 
coupling already-small coefficients, hence making these products 
negligible. See the 0030 example in the main text. 
\vspace{-0.0 cm}}
\vskip 0.0cm
\centering 
{
\begin{tabular}{c}
\end{tabular} \\ 
\begin{tabular}{l l l r}
\hline\hline\vspace{-0.25cm} \\ \vspace{0.10 cm}
   RDT{\hskip 0.7 cm}\    &{\hskip 0.0 cm}magnet-like {\hskip 1.7 cm}\  & 
{\hskip 0.0 cm}second-order contribution from {\hskip 0.3 cm} & (magnet)      \\
\hline\hline\vspace{-0.30 cm}\\ & & & \\ 
\hline\vspace{-0.30 cm}\\\vspace{0.07 cm}
$f_{2000}^{(2)}$&   normal quadrupole $x^2$       & $h_{2000}$, $h_{1100}$             &(normal quadrupole)\\
                &                                 & $h_{1010}$, $h_{1001}$             &(skew quadrupole)\\
\hline\vspace{-0.30 cm}\\\vspace{0.07 cm}
$f_{0020}^{(2)}$&   normal quadrupole $y^2$       & $h_{0020}$, $h_{0011}$             &(normal quadrupole)\\
                &                                 & $h_{1010}$, $h_{0110}$             &(skew quadrupole)\\
\hline \\ & &  \vspace{-0.33cm}\\
\hline\vspace{-0.30 cm}\\\vspace{0.07 cm}
$f_{1010}^{(2)}$&   skew quadrupole $xy$       & $h_{1100}$, $h_{2000}$, $h_{0011}$, $h_{0020}$ &(normal quadrupole)\\
                &                              & $h_{1010}$, $h_{1001}$, $h_{0110}$             &(skew quadrupole) \\
\hline\vspace{-0.30 cm}\\\vspace{0.07 cm}
$f_{1001}^{(2)}$&   skew quadrupole $xy$       & $h_{1100}$, $h_{2000}$, $h_{0011}$, $h_{0002}$ &(normal quadrupole)\\
                &                              & $h_{1010}$, $h_{0101}$, $h_{1001}$             &(skew quadrupole) \\
\hline \\ & &  \vspace{-0.33cm}\\
\hline\vspace{-0.30 cm}\\\vspace{0.07 cm}
$f_{3000}^{(2)}$&   normal sextupole $x^3$        & $h_{1100}$, $h_{2000}$              &(normal quadrupole) \\
                &                                 & $h_{2100}$,$h_{3000}$               &(normal sextupole)\\
\hline\vspace{-0.30 cm}\\\vspace{0.07 cm}
$f_{1200}^{(2)}$&   normal sextupole $x^3$        & $h_{1100}$, $h_{2000}$, $h_{0200}$  &(normal quadrupole) \\
                &                                 & $h_{1200}$, $h_{2100}$, $h_{0300}$  &(normal sextupole)\\
\hline\vspace{-0.30 cm}\\\vspace{0.07 cm}
$f_{0111}^{(2)}$&   normal sextupole $xy^2$       & $h_{1100}$, $h_{0200}$, $h_{0020}$, $h_{0002}$  &(normal quadrupole) \\
                &                                 & $h_{0111}$, $h_{1011}$, $h_{0120}$, $h_{0102}$  &(normal sextupole)\\
\hline\vspace{-0.30 cm}\\\vspace{0.07 cm}
$f_{0120}^{(2)}$&   normal sextupole $xy^2$       & $h_{1100}$, $h_{0200}$, $h_{0011}$, $h_{0020}$  &(normal quadrupole) \\
                &                                 & $h_{0111}$, $h_{0120}$, $h_{1020}$              &(normal sextupole)\\
\hline\vspace{-0.30 cm}\\\vspace{0.07 cm}
$f_{1020}^{(2)}$&   normal sextupole $xy^2$       & $h_{1100}$, $h_{2000}$, $h_{0011}$, $h_{0020}$  &(normal quadrupole) \\
                &                                 & $h_{1011}$, $h_{0120}$, $h_{1020}$              &(normal sextupole)\\
\hline \\ & &  \vspace{-0.33cm}\\
\hline\vspace{-0.30 cm}\\\vspace{0.07 cm}
$f_{0030}^{(2)}$&   skew sextupole $y^3$          & $h_{1010}$, $h_{0110}$             &(skew quadrupole) \\
                &                                 & $h_{0120}$, $h_{1020}$             &(normal sextupole)\\
\hline\vspace{-0.30 cm}\\\vspace{0.07 cm}
$f_{0012}^{(2)}$&   skew sextupole $y^3$          & $h_{0101}$, $h_{1010}$, $h_{0110}$,$h_{1001}$   &(skew quadrupole)\\
                &                                 & $h_{1011}$, $h_{0111}$, $h_{0102}$, $h_{1002}$  &(normal sextupole)\\
\hline\vspace{-0.30 cm}\\\vspace{0.07 cm}
$f_{1110}^{(2)}$&   skew sextupole $x^2y$         & $h_{0101}$, $h_{1010}$, $h_{0110}$, $h_{1001}$   &(skew quadrupole)\\
                &                                 & $h_{1011}$, $h_{0111}$, $h_{0120}$, $h_{1020}$, $h_{1200}$, $h_{2100}$ &(normal sextupole)\\
\hline\vspace{-0.30 cm}\\\vspace{0.07 cm}
$f_{2001}^{(2)}$&   skew sextupole $x^2y$         & $h_{1010}$, $h_{1001}$, $h_{0101}$             &(skew quadrupole) \\
                &                                 & $h_{1011}$, $h_{1002}$, $h_{3000}$, $h_{2100}$ &(normal sextupole)\\
\hline\vspace{-0.30 cm}\\\vspace{0.07 cm}
$f_{2010}^{(2)}$&   skew sextupole $x^2y$         & $h_{1010}$, $h_{1001}$, $h_{0110}$             &(skew quadrupole) \\
                &                                 & $h_{1011}$, $h_{1020}$, $h_{3000}$, $h_{2100}$ &(normal sextupole)\\
\hline\vspace{0.2 cm}
\end{tabular}}
\rule{0mm}{0mm}
\label{tab:2nd_rdt_lat} 
\end{table*}

Table~\ref{tab:2nd_rdt_lat} 
lists for each RDT (up to the sextupole terms) 
the Hamiltonian coefficients 
$h_{jklm}$ and the corresponding magnetic element 
contributing to the second order $f_{jklm}^{(2)}$, 
as derived from Eqs.~\eqref{eq:select0}
and~\eqref{eq:PBH1}. The result is manifold: 
to the second order, $(i)$ skew quadrupoles (coupling) 
induce focusing errors; $(ii)$ focusing errors modify 
coupling and sextupole RDTs; $(iii)$ coupling and 
normal sextupoles induce skew sextupole RDTs. 
Similar list for the second-order octupolar RDTs 
may be computed, but are not reported here for 
sake of space. 

The last missing step is to include second-order 
terms in the change of coordinates of 
Eqs.~\eqref{eq:NFtransform} and~\eqref{eq:NFtransform2}. 
In analogy with the previous apporach, the function 
$F$ is split in its first- an second-order parts
\begin{eqnarray}\label{eq:NFtransform3}
\vec{h}(s,N)&=&e^{:F:}\vec{\zeta}=\vec{\zeta}+[F^{(1)}+F^{(2)},\vec{\zeta}\ ]
	+\frac{1}{2}[F^{(1)}+F^{(2)},[F^{(1)}+F^{(2)},\vec{\zeta}\ ]] 
	+O(3^{{\tiny\hbox{rd}}})\nonumber \\
	&=&\vec{\zeta}+[F^{(1)}+F^{(2)},\vec{\zeta}\ ]
	+\frac{1}{2}[F^{(1)},[F^{(1)},\vec{\zeta}\ ]] 
	+O(3^{{\tiny\hbox{rd}}})\ ,
\end{eqnarray}
where the second Poisson bracket was reduced, as 
any term proportional to $[F^{(2)},F^{(1)}]$ or 
$[F^{(2)},F^{(2)}]$ goes into the third-order remainder. 
$F^{(1)}$ and $F^{(2)}$ may be evaluated from 
Eqs.~\eqref{eq:omol_1st_1} and~\eqref{eq:omol_14}. 
Hence the only quantity to be evaluated is the 
double Poisson bracket $[F^{(1)},[F^{(1)},\vec{\zeta}\ ]]$. 
The inner bracket has been already evaluated in 
Eq.~\eqref{e:horizontal} (horizontal plane) and 
Eq.~\eqref{e:vertical} (vertical plane), yielding
\begin{eqnarray}\label{eq:FLie1}
F^{(1)}&=&\sum_{jklm}{f_{jklm}^{(1)}
             \zeta_{x,+}^j\zeta_{x,-}^k\zeta_{y,+}^l\zeta_{y,-}^m}  \ ,  \\
{[F^{(1)}+F^{(2)},\zeta_{x,-}]}&=& -2i\sum_{pqrt}
	     {p\ \left[f_{pqrt}^{(1)}+f_{pqrt}^{(2)}\right]
             \zeta_{x,+}^{p-1}\zeta_{x,-}^q\zeta_{y,+}^r\zeta_{y,-}^t}\ , \\
{[F^{(1)}+F^{(2)},\zeta_{y,-}]}&=& -2i\sum_{pqrt}
	     {r\ \left[f_{pqrt}^{(1)}+f_{pqrt}^{(2)}\right]
             \zeta_{x,+}^p\zeta_{x,-}^q\zeta_{y,+}^{r-1}\zeta_{y,-}^t}\ .
\end{eqnarray}
The double Poisson bracket then reads
\begin{eqnarray}
\frac{1}{2}{[F^{(1)},[F^{(1)},\zeta_{x,-}]]}&=& i\sum_{jklm}\sum_{pqrt}
	{p\ f_{jklm}^{(1)}f_{pqrt}^{(1)}
	[\zeta_{x,+}^j\zeta_{x,-}^k\zeta_{y,+}^l\zeta_{y,-}^m,\ 
             \zeta_{x,+}^{p-1}\zeta_{x,-}^q\zeta_{y,+}^r\zeta_{y,-}^t}]\ , \\
\frac{1}{2}{[F^{(1)},[F^{(1)},\zeta_{y,-}]]}&=& i\sum_{jklm}\sum_{pqrt}
	{r\ f_{jklm}^{(1)}f_{pqrt}^{(1)}
	[\zeta_{x,+}^j\zeta_{x,-}^k\zeta_{y,+}^l\zeta_{y,-}^m,\ 
             \zeta_{x,+}^p\zeta_{x,-}^q\zeta_{y,+}^{r-1}\zeta_{y,-}^t]}\ .
\end{eqnarray}
By applying the distributive property and 
Eq.~\eqref{eq:Poisson-bracket} the following 
relations hold
\begin{eqnarray}
\frac{1}{2}{[F^{(1)},[F^{(1)},\zeta_{x,-}]]}&=& -2\sum_{jklm}\sum_{pqrt}
	p\ f_{jklm}^{(1)}f_{pqrt}^{(1)}
	\bigg\{[jq-k(p-1)]\zeta_{x,+}^{j+p-2}\zeta_{x,-}^{k+q-1}\zeta_{y,+}^{l+r}\zeta_{y,-}^{m+t}+
\\ &&\hskip 4cm
      (lt-mr)\zeta_{x,+}^{j+p-1}\zeta_{x,-}^{k+q}\zeta_{y,+}^{l+r-1}\zeta_{y,-}^{m+t-1}
	\bigg\}\ , \nonumber \\
\frac{1}{2}{[F^{(1)},[F^{(1)},\zeta_{y,-}]]}&=& -2\sum_{jklm}\sum_{pqrt}
	r\ f_{jklm}^{(1)}f_{pqrt}^{(1)}
	\bigg\{(jq-kp)\zeta_{x,+}^{j+p-1}\zeta_{x,-}^{k+q-1}\zeta_{y,+}^{l+r-1}\zeta_{y,-}^{m+t}+
\\ &&\hskip 4cm
      [lt-m(r-1)]\zeta_{x,+}^{j+p}\zeta_{x,-}^{k+q}\zeta_{y,+}^{l+r-2}\zeta_{y,-}^{m+t-1}
	\bigg\}\nonumber\  .
\end{eqnarray}
Equation~\eqref{eq:NFtransform3}, up to the second order, 
may be eventually rewritten as
\begin{eqnarray}\label{eq:NFtransform4x}
{h}_{x,-}(s,N)&=&\zeta_{x,-}
	   -2i\sum_{abcd}{j\ \left[f_{abcd}^{(1)}+f_{abcd}^{(2)}\right]
             \zeta_{x,+}^{a-1}\zeta_{x,-}^b\zeta_{y,+}^c\zeta_{y,-}^d} \\
	&&-2\sum_{jklm}\sum_{pqrt}p\ f_{jklm}^{(1)}f_{pqrt}^{(1)}
	\bigg\{[jq-k(p-1)]\zeta_{x,+}^{j+p-2}\zeta_{x,-}^{k+q-1}\zeta_{y,+}^{l+r}\zeta_{y,-}^{m+t}+
      (lt-mr)\zeta_{x,+}^{j+p-1}\zeta_{x,-}^{k+q}\zeta_{y,+}^{l+r-1}\zeta_{y,-}^{m+t-1}
	\bigg\}\ , \nonumber  \\
{h}_{y,-}(s,N)&=&\zeta_{y,-}\label{eq:NFtransform4y}
	   -2i\sum_{abcd}{l\ \left[f_{abcd}^{(1)}+f_{abcd}^{(2)}\right]
             \zeta_{x,+}^a\zeta_{x,-}^b\zeta_{y,+}^{c-1}\zeta_{y,-}^d} \\
	&&-2\sum_{jklm}\sum_{jklm}r\ f_{jklm}^{(1)}f_{pqrt}^{(1)}
	\bigg\{(jq-kp)\zeta_{x,+}^{j+p-1}\zeta_{x,-}^{k+q-1}\zeta_{y,+}^{l+r-1}\zeta_{y,-}^{m+t}+
      [lt-m(r-1)]\zeta_{x,+}^{j+p}\zeta_{x,-}^{k+q}\zeta_{y,+}^{l+r-2}\zeta_{y,-}^{m+t-1}
	\bigg\}\nonumber\  .
\end{eqnarray}

\begin{table*}
\caption{List and definition of second-order 
	ORDTs $g_{jklm}$ from the skew quadrupole-like 
	secondary spectral lines of the complex C-S signals 
        $h_{x,-}=\tilde{x}-i\tilde{p}_x$ ($h_{y,-}=\tilde{y}-i\tilde{p}_y$). 
        First-order coupling
	RDTs $f_{1001}^{(1)}$ and $f_{1010}^{(1)}$ are those of 
	Table~\ref{tab:lattice-rdt}, derived from 
	Eqs.~\eqref{eq:RDT-1st} and~\eqref{eq:h_Vs_KJ}. 
	The second-order Hamiltonian terms $\tilde{h}^{(2)}_{jklm}$
	and $\hat{h}_{jklm}$ are to be computed from 
	Eqs.~\eqref{eq:select0}
	and~\eqref{eq:PBH1}. Beta-beat RDTs $f_{2000}^{(1)}$ 
	and $f_{0020}^{(1)}$ are defined in Eqs~\eqref{eq:def_f2000} 
	and~\eqref{eq:def_f0020}. If C-S parameters are evaluated 
	from the linear lattice model with quadrupole errors  
	included the second-order contributions vanish, i.e. 
	$g_{jklm}=f_{jklm}^{(1)}$, as discussed in Appendix~\ref{app:3}. 
	T.B.T. YET!\vspace{-0.0 cm}}
\vskip 0.0cm
\centering 
{
\begin{tabular}{c}
\end{tabular} \\ 
\begin{tabular}{l l r}
\hline\hline\vspace{-0.25cm} \\ \vspace{0.10 cm}
$h$ spectral line {\hskip 3.5 cm} & 
{\hskip 4.2 cm}$\left\{g_{jklm}\right\}${\hskip 4.5 cm}  & 
{\hskip -0.4 cm}magnetic term   \\
\hline\hline\vspace{-0.30 cm}\\ & & \\ 
\hline\vspace{-0.30 cm}\\\vspace{0.07 cm}
$H_h(0,1)=-2i\left\{g_{1001}\right\}\zeta_{y,-}$ &
$\displaystyle\left\{
f_{1001}^{(1)}+i\frac{\tilde{h}^{(2)}_{1001}+\hat{h}_{1001}}
{1-e^{2\pi i(Q_x-Q_y)}}+i\left[2f_{0020}^{(1)*}f_{1010}^{(1)}
		              +2f_{2000}^{(1)} f_{1010}^{(1)*}
\right]\right\}$ &
$xy$\\ 
\hline\vspace{-0.30 cm}\\\vspace{0.07 cm}
$H_h(0,-1)=-2i\left\{g_{1010,H}\right\}\zeta_{y,+}$ &
$\displaystyle\left\{
f_{1010}^{(1)}+i\frac{\tilde{h}^{(2)}_{1010}+\hat{h}_{1010}}
{1-e^{2\pi i(Q_x+Q_y)}}-i\left[2f_{0020}^{(1)}f_{1001}^{(1)}
		              -2f_{2000}^{(1)}f_{1001}^{(1)*}
\right]\right\}$ &
$xy$\\ 
\hline\vspace{-0.30 cm}\\\vspace{0.07 cm}
$V_h(1,0)=-2i\left\{g_{0110}\right\}\zeta_{x,-}$ &
$\displaystyle\left\{
f_{0110}^{(1)}+i\frac{\tilde{h}^{(2)}_{0110}+\hat{h}_{0110}}
{1-e^{-2\pi i(Q_x-Q_y)}}+i\left[2f_{0020}^{(1)}f_{1010}^{(1)*}
		              +2f_{2000}^{(1)*}f_{1010}^{(1)}
\right]\right\}$ &
$xy$\\ 
\hline\vspace{-0.30 cm}\\\vspace{0.07 cm}
$V_h(-1,0)=-2i\left\{g_{1010,V}\right\}\zeta_{x,+}$ &
$\displaystyle\left\{
f_{1010}^{(1)}+i\frac{\tilde{h}^{(2)}_{1010}+\hat{h}_{1010}}
{1-e^{2\pi i(Q_x+Q_y)}}+i\left[2f_{0020}^{(1)}f_{1001}^{(1)}
		              -2f_{2000}^{(1)}f_{1001}^{(1)*}
\right]\right\}$ &
$xy$\\ 
\hline\vspace{0.2 cm}
\end{tabular}}
\rule{0mm}{0mm}
\label{tab:g_jklm_SQ} 
\caption{List and definition of second-order
	ORDTs $g_{jklm}$ from the normal sextupole-like 
	secondary spectral lines of the complex C-S signals 
        $h_{x,-}=\tilde{x}-i\tilde{p}_x$ ($h_{y,-}=\tilde{y}-i\tilde{p}_y$). 
        First-order sextupolar
	RDTs ($f_{3000}^{(1)}$, $f_{1200}^{(1)}$ and the like) are 
	those of Table~\ref{tab:lattice-rdt}, derived from 
	Eqs.~\eqref{eq:RDT-1st} and~\eqref{eq:h_Vs_KJ}. 
	The second-order Hamiltonian terms $\tilde{h}^{(2)}_{jklm}$
	and $\hat{h}_{jklm}$ are to be computed from 
	Eqs.~\eqref{eq:select0}
	and~\eqref{eq:PBH1}. Beta-beat RDTs $f_{2000}^{(1)}$ 
	and $f_{0020}^{(1)}$ are defined in Eqs~\eqref{eq:def_f2000} 
	and~\eqref{eq:def_f0020}. If C-S parameters are evaluated 
	from the linear lattice model with quadrupole errors  
	included, the second-order contributions vanish, i.e. 
	$g_{jklm}=f_{jklm}^{(1)}$, as discussed in Appendix~\ref{app:3}.
	Products between skew sextupole and  coupling RDTs have been 
	excluded. As for Table~\ref{tab:2nd_rdt_lat}, it is indeed 
	assumed that no physical strong skew sextupole is powered and 
	that first-order skew sextupole RDTs are generated 
	by small sextupole tilts. In Eqs.~\eqref{eq:NFtransform4x} 
	and~\eqref{eq:NFtransform4y} skew sextupole 
        first-order RDTs enter always multiplied by the already-small 
	first-order coupling RDTs, hence making these products 
	negligible. \vspace{-0.0 cm}}
\vskip 0.0cm
\centering 
{
\begin{tabular}{c}
\end{tabular} \\ 
\begin{tabular}{l l r}
\hline\hline\vspace{-0.25cm} \\ \vspace{0.10 cm}
$h$ spectral line {\hskip 3.5 cm} & 
{\hskip 4.2 cm}$\left\{g_{jklm}\right\}${\hskip 4.5 cm}  & 
{\hskip -0.4 cm}magnetic term   \\
\hline\hline\vspace{-0.30 cm}\\ & & \\ 
\hline\vspace{-0.30 cm}\\\vspace{0.07 cm}
$H_h(-2,0)=-6i\left\{g_{3000}\right\}\zeta_{x,+}^2$ &
$\displaystyle\left\{
f_{3000}^{(1)}+i\frac{\tilde{h}^{(2)}_{3000}+\hat{h}_{3000}}
{1-e^{2\pi i(3Q_x)}}-\frac{i}{3}\left[2f_{2000}^{(1)}f_{1200}^{(1)*}
\right]\right\}$ &
$x^3$\\ 
\hline\vspace{-0.30 cm}\\\vspace{0.07 cm}
$H_h(+2,0)=-2i\left\{g_{1200}\right\}\zeta_{x,-}^2$ &
$\displaystyle\left\{
f_{1200}^{(1)}+i\frac{\tilde{h}^{(2)}_{1200}+\hat{h}_{1200}}
{1-e^{2\pi i(-Q_x)}}+i\left[4f_{2000}^{(1)*}f_{1200}^{(1)*}
		           +6f_{2000}^{(1) }f_{3000}^{(1)*}
			   \right]\right\} $ &
$x^3$\\
\hline\vspace{-0.30 cm}\\\vspace{0.07 cm}
$H_h(0,-2)=-2i\left\{g_{1020,H}\right\}\zeta_{y,+}^2$ &
$\displaystyle\left\{
f_{1020}^{(1)}+i\frac{\tilde{h}^{(2)}_{1020}+\hat{h}_{1020}}
{1-e^{2\pi i(Q_x+2Q_y)}}+i\left[2f_{2000}^{(1)}f_{0120}^{(1)}
			       -2f_{0020}^{(1)}f_{0111}^{(1)*}
			   \right]\right\}$ &
$xy^2$\\
\hline\vspace{-0.30 cm}\\\vspace{0.07 cm}
$H_h(0,+2)=-2i\left\{g_{1002}\right\}\zeta_{y,-}^2$ &
$\displaystyle\left\{
f_{1002}^{(1)}+i\frac{\tilde{h}^{(2)}_{1002}+\hat{h}_{1002}}
{1-e^{2\pi i(Q_x-2Q_y)}}+i\left[2f_{0020}^{(1)*}f_{0111}^{(1)*}
			       +2f_{2000}^{(1) }f_{1020}^{(1)*}
			   \right]\right\}$ &
$xy^2$\\
\hline\vspace{-0.30 cm}\\\vspace{0.07 cm}
$V_h(+1,+1)=-2i\left\{g_{0111}\right\}\zeta_{x,-}\zeta_{y,-}$ &
$\displaystyle\left\{
f_{0111}^{(1)}+i\frac{\tilde{h}^{(2)}_{0111}+\hat{h}_{0111}}
{1-e^{2\pi i(-Q_x)}}+i\left[4f_{0020}^{(1)*}f_{0120}^{(1)}
			   +4f_{0020}^{(1) }f_{1020}^{(1)*}
		           +2f_{2000}^{(1)}f_{0111}^{(1)*}
			   \right]\right\}$ &
$xy^2$\\ 
\hline\vspace{-0.30 cm}\\\vspace{0.07 cm}
$V_h(-1,-1)=-4i\left\{g_{1020,V}\right\}\zeta_{x,+}\zeta_{y,+}$ &
$\displaystyle\left\{
f_{1020}^{(1)}+i\frac{\tilde{h}^{(2)}_{1020}+\hat{h}_{1020}}
{1-e^{2\pi i(Q_x+2Q_y)}}-i\left[2f_{2000}^{(1)}f_{0120}^{(1)}
			\right]\right\}$ &
$xy^2$ \\
\hline\vspace{-0.30 cm}\\\vspace{0.07 cm}
$V_h(+1,-1)=-4i\left\{g_{0120}\right\}\zeta_{x,-}\zeta_{y,+}$ &
$\displaystyle\left\{
f_{0120}^{(1)}+i\frac{\tilde{h}^{(2)}_{0120}+\hat{h}_{0120}}
{1-e^{2\pi i(-Q_x+2Q_y)}}+i\left[2f_{2000}^{(1)*}f_{1020}^{(1)}
			    \right]\right\} $ &
$xy^2$ \\
\hline\vspace{-0.30 cm}\\\vspace{0.07 cm} 
$V_h(-1,+1)=-2i\left\{g_{1011}\right\}\zeta_{x,+}\zeta_{y,-}$ &
$\displaystyle\left\{
f_{1011}^{(1)}+i\frac{\tilde{h}^{(2)}_{1011}+\hat{h}_{1011}}
{1-e^{2\pi i(Q_x)}}+i\left[4f_{0020}^{(1)*}f_{1020}^{(1) }
		          +4f_{0020}^{(1) }f_{0120}^{(1)*}
		          -2f_{2000}^{(1) }f_{0111}^{(1)}\right]\right\}$ &
$xy^2$\\
\hline\vspace{0.2 cm}
\end{tabular}}
\rule{0mm}{0mm}
\label{tab:g_jklm_NS} 
\end{table*}

\begin{table*}
\caption{List and definition of second-order
	ORDTs $g_{jklm}$ from the skew sextupole-like 
	secondary spectral lines of the complex C-S signals 
        $h_{x,-}=\tilde{x}-i\tilde{p}_x$ ($h_{y,-}=\tilde{y}-i\tilde{p}_y$). 
        First-order sextupolar
	RDTs ($f_{3000}^{(1)}$, $f_{1200}^{(1)}$ and the like) are 
	those of Table~\ref{tab:lattice-rdt}, derived from 
	Eqs.~\eqref{eq:RDT-1st} and~\eqref{eq:h_Vs_KJ}. 
	The second-order Hamiltonian terms $\tilde{h}^{(2)}_{jklm}$
	and $\hat{h}_{jklm}$ are to be computed from 
	Eqs.~\eqref{eq:select0}
	and~\eqref{eq:PBH1}. ORDTs have no longer 
	two RDTs properties: $g_{jklm}\ne g_{kjml}^*$ and 
	$g_{jklm,H}\ne g_{jklm,V}$. Products between skew sextupole 
	and  beta-beat RDTs have been excluded. As for 
	Table~\ref{tab:2nd_rdt_lat}, it is indeed assumed 
	that no physical strong skew sextupole is powered and 
	that first-order skew sextupole RDTs are generated 
	by small sextupole tilts. In Eqs.~\eqref{eq:NFtransform4x} 
	and~\eqref{eq:NFtransform4y} skew sextupole 
        first-order RDTs enter always multiplied by the already-small 
	beta-beat first-order RDTs, hence making these products 
	negligible. See the 0030 example in the main text. \vspace{-0.0 cm}}
\vskip 0.0cm
\centering 
{
\begin{tabular}{c}
\end{tabular} \\ 
\begin{tabular}{l l r}
\hline\hline\vspace{-0.25cm} \\ \vspace{0.10 cm}
$h$ spectral line {\hskip 3.5 cm} & 
{\hskip 4.2 cm}$\left\{g_{jklm}\right\}${\hskip 4.5 cm}  & 
{\hskip -0.4 cm}magnetic term   \\
\hline\hline\vspace{-0.30 cm}\\ & & \\ 
\hline\vspace{-0.30 cm}\\\vspace{0.07 cm}
$V_h(0,-2)=-6i\left\{g_{0030}\right\}\zeta_{y,+}^2$ &
$\displaystyle\left\{
f_{0030}^{(1)}+i\frac{\tilde{h}^{(2)}_{0030}+\hat{h}_{0030}}
{1-e^{2\pi i(3Q_y)}}-\frac{i}{3}\left[ f_{1010}^{(1)}f_{0120}^{(1)}
		                      -f_{1001}^{(1)*}f_{1020}^{(1)}
\right]\right\}$ &
$y^3$\\
\hline\vspace{-0.30 cm}\\\vspace{0.07 cm}
$V_h(0,+2)=-2i\left\{g_{0012}\right\}\zeta_{y,-}^2$ &
$\displaystyle\left\{
f_{0012}^{(1)}+i\frac{\tilde{h}^{(2)}_{0012}+\hat{h}_{0012}}
{1-e^{2\pi i(-Q_y)}}-i\left[ f_{1001}^{(1)}f_{0111}^{(1)}
		            -f_{1010}^{(1)*}f_{0111}^{(1)*}
			    +f_{1001}^{(1)*}f_{0120}^{(1)*}\right.\right. $ &
$y^3$\\
	& 		    $\hskip 4cm\left.
			    -f_{1010}^{(1)}f_{1020}^{(1)*}\right]\Bigg\}$ &\\
\hline\vspace{-0.30 cm}\\\vspace{0.07 cm}
$V_h(+2,0)=-2i\left\{g_{0210}\right\}\zeta_{x,-}^2$ &
$\displaystyle\left\{
f_{0210}^{(1)}+i\frac{\tilde{h}^{(2)}_{0210}+\hat{h}_{0210}}
{1-e^{2\pi i(-2Q_x+Q_y)}}-i\left[ f_{1001}^{(1)*}\left(f_{0111}^{(1)}+f_{1200}^{(1)}\right)
		            \right.\right. $ &
$x^2y$\\
	& 		    $\hskip 4.5cm\left.
   			    -2f_{1010}^{(1)*}f_{0120}^{(1)}
			    -3f_{1010}^{(1)}f_{3000}^{(1)*}\right]\Bigg\}$ &\\
\hline\vspace{-0.30 cm}\\\vspace{0.07 cm}
$V_h(-2,0)=-2i\left\{g_{2010,V}\right\}\zeta_{x,+}^2$ &
$\displaystyle\left\{
f_{2010}^{(1)}+i\frac{\tilde{h}^{(2)}_{2010}+\hat{h}_{2010}}
{1-e^{2\pi i(2Q_x+Q_y)}}-i\left[3f_{1001}^{(1)*}f_{3000}^{(1)}
			       -2f_{1001}^{(1)}f_{1020}^{(1)}
		            \right.\right. $ &
$x^2y$\\
	& 		    $\hskip 4.5cm\left.
   			    +f_{1010}^{(1)}\left(f_{0111}^{(1)*}-f_{1200}^{(1)*}
			    \right)\right]\Bigg\}$ &\\
\hline\vspace{-0.30 cm}\\\vspace{0.07 cm}
$H_h(+1,+1)=-2i\left\{g_{1101}\right\}\zeta_{x,-}\zeta_{y,-}$ &
$\displaystyle\left\{
f_{1101}^{(1)}+i\frac{\tilde{h}^{(2)}_{1101}+\hat{h}_{1101}}
{1-e^{2\pi i(-Q_y)}}-i\left[2f_{1001}^{(1)*}f_{0120}^{(1)*}
			   +f_{1001}^{(1)}\left(f_{0111}^{(1)}+2f_{1200}^{(1)}\right)\right.\right. $ &
$x^2y$\\
	& 		    $\hskip 4cm\left.
		           -2f_{1010}^{(1)}f_{1020}^{(1)*}
   		           +f_{1010}^{(1)*}\left(-f_{0111}^{(1)*}-2f_{1200}^{(1)*}\right)
			   \right]\Bigg\}$ &\\
\hline\vspace{-0.30 cm}\\\vspace{0.07 cm}
$H_h(-1,-1)=-4i\left\{g_{2010,H}\right\}\zeta_{x,+}\zeta_{y,+}$ &
$\displaystyle\left\{
f_{2010}^{(1)}+i\frac{\tilde{h}^{(2)}_{2010}+\hat{h}_{2010}}
{1-e^{2\pi i(2Q_x+Q_y)}}-\frac{i}{2}\left[2f_{1010}^{(1)}f_{1200}^{(1)*}
		            -6f_{1001}^{(1)*}f_{3000}^{(1)}\right]\right\}$ &
$x^2y$\\
\hline\vspace{-0.30 cm}\\\vspace{0.07 cm}
$H_h(+1,-1)=-2i\left\{g_{1110}\right\}\zeta_{x,-}\zeta_{y,+}$ &
$\displaystyle\left\{
f_{1110}^{(1)}+i\frac{\tilde{h}^{(2)}_{1110}+\hat{h}_{1110}}
{1-e^{2\pi i(Q_y)}} -i\left[2f_{1001}^{(1)}f_{0120}^{(1)}
			    +f_{1010}^{(1)}\left(-f_{0111}^{(1)}+2f_{1200}^{(1)}\right)
			    \right.\right. $ &
$x^2y$\\
	& 		    $\hskip 4cm\left.
			    -2f_{1010}^{(1)*}f_{1020}^{(1)}
			    +f_{1001}^{(1)*}\left(f_{0111}^{(1)*}-2f_{1200}^{(1)*}\right)\right]\Bigg\}$ &\\
\hline\vspace{-0.30 cm}\\\vspace{0.07 cm} 
$H_h(-1,+1)=-4i\left\{g_{2001}\right\}\zeta_{x,+}\zeta_{y,-}$ &
$\displaystyle\left\{
f_{2001}^{(1)}+i\frac{\tilde{h}^{(2)}_{2001}+\hat{h}_{2001}}
{1-e^{2\pi i(2Q_x-Q_y)}}-\frac{i}{2}\left[2f_{1001}^{(1)}f_{1200}^{(1)*}
		                -6f_{1010}^{(1)*}f_{3000}^{(1)}\right]\right\}$ &
$x^2y$\\
\hline\vspace{0.2 cm}
\end{tabular}}
\rule{0mm}{0mm}
\vskip 0.6cm
\label{tab:g_jklm_SS} 
\end{table*}

\vskip 0.0cm
\begin{table*}
\caption{List and definition of second-order
	ORDTs $g_{jklm}$ from the normal octupole-like 
	secondary spectral lines of the complex C-S signals 
        $h_{x,-}=\tilde{x}-i\tilde{p}_x$ ($h_{y,-}=\tilde{y}-i\tilde{p}_y$). 
        First-order sextupolar terms 
        ($f_{3000}^{(1)}$, $f_{1200}^{(1)}$ and the like)  are those of
	Table~\ref{tab:lattice-rdt}, whereas first-order octupolar
	RDTs ($f_{4000}^{(1)}$, $f_{1300}^{(1)}$ and the like) may be
        computed from Eqs.~\eqref{eq:RDT-1st} and~\eqref{eq:h_Vs_KJ}. 
	The second-order Hamiltonian terms $\tilde{h}^{(2)}_{jklm}$
	and $\hat{h}_{jklm}$ are to be computed from 
	Eqs.~\eqref{eq:select0}
	and~\eqref{eq:PBH1}. ORDTs have no longer 
	two RDTs properties: $g_{jklm}\ne g_{kjml}^*$ and 
	$g_{jklm,H}\ne g_{jklm,V}$. If quadrupole errors are 
        included in the model when computing the C-S  
        parameters, first-order quadrupolar RDTs vanish, i.e.  
        $f_{2000}^{(1)}=f_{0020}^{(1)}=0$.\vspace{-0.0 cm}}
\vskip 0.0cm
\centering 
{
\begin{tabular}{c}
\end{tabular} \\ 
\begin{tabular}{l l r}
\hline\hline\vspace{-0.25cm} \\ \vspace{0.10 cm}
$h$ spectral line {\hskip 2.9 cm} & 
{\hskip 4.2 cm}$\left\{g_{jklm}\right\}${\hskip 5.3 cm}  & 
{\hskip -0.4 cm}magnetic term   \\
\hline\hline\vspace{-0.30 cm}\\ & & \\ 
\hline\vspace{-0.30 cm}\\\vspace{0.07 cm}
$H_h(-3,0)=-8i\left\{g_{4000}\right\}\zeta_{x,+}^3$ &
$\displaystyle\left\{
f_{4000}^{(1)}+i\frac{\tilde{h}^{(2)}_{4000}+\hat{h}_{4000}}
{1-e^{2\pi i(4Q_x)}}-i\left[ f_{2000}^{(1)}f_{3100}^{(1)}
\right]\right\}$ &
$x^4$\\
\hline\vspace{-0.30 cm}\\\vspace{0.07 cm}
$H_h(3,0)=-2i\left\{g_{1300}\right\}\zeta_{x,-}^3$ &
$\displaystyle\left\{
f_{1300}^{(1)}+i\frac{\tilde{h}^{(2)}_{1300}+\hat{h}_{1300}}
{1-e^{2\pi i(-2Q_x)}}+i\left[6f_{3000}^{(1)*}f_{1200}^{(1)*}-2f_{1200}^{(1)\ 2}
                              +8f_{2000}^{(1)}f_{4000}^{(1)}
\right]\right\}$ &$x^4$ \\
\hline\vspace{-0.30 cm}\\\vspace{0.07 cm}
$H_h(-1,2)=-4i\left\{g_{2002}\right\}\zeta_{x,+}\zeta_{y,-}^2$ &
$\displaystyle\left\{
f_{2002}^{(1)}+i\frac{\tilde{h}^{(2)}_{2002}+\hat{h}_{2002}}
{1-e^{2\pi i(2Q_x-2Q_y)}}+i\left[3f_{1020}^{(1)*}f_{3000}^{(1)}
                                 -f_{1002}^{(1)}f_{1200}^{(1)*}
                                +2f_{0020}^{(1)*}f_{2011}^{(1)}
\right]\right\}$ & $x^2y^2$\\
\hline\vspace{-0.30 cm}\\\vspace{0.07 cm}
$H_h(1,-2)=-2i\left\{g_{1120}\right\}\zeta_{x,-}\zeta_{y,+}^2$ &
$\displaystyle\left\{
f_{1120}^{(1)}+i\frac{\tilde{h}^{(2)}_{1120}+\hat{h}_{1120}}
{1-e^{2\pi i(2Q_y)}}-2i\left[f_{1020}^{(1)}\left(f_{1200}^{(1) }-f_{0111}^{(1) }\right)
                            +f_{0120}^{(1)}\left(f_{0111}^{(1)*}-f_{1200}^{(1)*}\right)
\right.\right.$ & $x^2y^2$\\ &   $\hskip 4cm
               -2\left(f_{2000}^{(1)*}f_{2020}^{(1)}+f_{2000}^{(1)}f_{0220}^{(1)}\right)
\Big]\Bigg\}$ & \\

\hline\vspace{-0.30 cm}\\\vspace{0.07 cm}
$H_h(1,2)=-2i\left\{g_{1102}\right\}\zeta_{x,-}\zeta_{y,-}^2$ &
$\displaystyle\hspace{-2mm}\left\{
f_{1102}^{(1)}+i\frac{\tilde{h}^{(2)}_{1102}+\hat{h}_{1102}}
{1-e^{2\pi i(-2Q_y)}}-2i\left[f_{0120}^{(1)*}\left(f_{1200}^{(1) }+f_{0111}^{(1) }\right)
                             -f_{1020}^{(1)*}\left(f_{0111}^{(1)*}+f_{1200}^{(1)*}\right)
\right.\right.$ & $x^2y^2$\\ &   $\hskip 4cm
               -2\left(f_{2000}^{(1)*}f_{2002}^{(1)}+f_{2000}^{(1)}f_{2020}^{(1)}\right)
\Big]\Bigg\}$ & \\

\hline\vspace{-0.30 cm}\\\vspace{0.07 cm}
$H_h(-1,-2)=-4i\left\{g_{2020,H}\right\}\zeta_{x,+}\zeta_{y,+}^2$ &
$\displaystyle\left\{
f_{2020}^{(1)}+i\frac{\tilde{h}^{(2)}_{2020}+\hat{h}_{2020}}
{1-e^{2\pi i(2Q_x+2Q_y)}}+i\left[3f_{0120}^{(1)}f_{3000}^{(1)}
                                 -f_{1020}^{(1)}f_{1200}^{(1)*}
                                +2f_{0020}^{(1)}f_{2011}^{(1)}
\right]\right\}$ & $x^2y^2$\\
\hline\vspace{-0.30 cm}\\\vspace{0.07 cm}
$V_h(-2,-1)=-4i\left\{g_{2020,V}\right\}\zeta_{x,+}^2\zeta_{y,+}$ &
$\displaystyle\left\{
f_{2020}^{(1)}+i\frac{\tilde{h}^{(2)}_{2020}+\hat{h}_{2020}}
{1-e^{2\pi i(2Q_x+2Q_y)}}-i\left[3f_{0120}^{(1)}f_{3000}^{(1)}
                                 -f_{1020}^{(1)}f_{1200}^{(1)*}
                                +2f_{2000}^{(1)}f_{1120}^{(1)}
\right]\right\}$ & $x^2y^2$\\
\hline\vspace{-0.30 cm}\\\vspace{0.07 cm}
$V_h(-2,1)=-2i\left\{g_{2011}\right\}\zeta_{x,+}^2\zeta_{y,-}$ &
$\displaystyle\left\{
f_{2011}^{(1)}+i\frac{\tilde{h}^{(2)}_{2011}+\hat{h}_{2011}}
{1-e^{2\pi i(2Q_x)}}-i\left[3f_{3000}^{(1)}f_{0111}^{(1)}-f_{1200}^{(1)*}f_{0111}^{(1)*}
                            -4f_{0120}^{(1)*}f_{1020}^{(1)}
\right.\right.$ & $x^2y^2$\\ &   $\hskip 3.9cm
                +f_{0111}^{(1)*\ 2}\ 
               -4\left(f_{0020}^{(1)*}f_{2020}^{(1)}+f_{0020}^{(1)}f_{2002}^{(1)}\right)
\Big]\Bigg\}$ & \\
\hline\vspace{-0.30 cm}\\\vspace{0.07 cm}
$V_h(2,-1)=-4i\left\{g_{0220}\right\}\zeta_{x,-}^2\zeta_{y,+}$ &
$\displaystyle\hspace{-1mm}\left\{
f_{0220}^{(1)}+i\frac{\tilde{h}^{(2)}_{0220}+\hat{h}_{0220}}
{1-e^{2\pi i(-2Q_x+2Q_y)}}+i\left[3f_{3000}^{(1)*}f_{1020}^{(1) }
                                  -f_{1200}^{(1) }f_{0120}^{(1) }
                                 +2f_{2000}^{(1)*}f_{1120}^{(1) }
\right]\right\}$ & $x^2y^2$\\
\hline\vspace{-0.30 cm}\\\vspace{0.07 cm}
$V_h(2,1)=-2i\left\{g_{0211}\right\}\zeta_{x,-}^2\zeta_{y,-}$ &
$\displaystyle\hspace{-1mm}\left\{
f_{0211}^{(1)}+i\frac{\tilde{h}^{(2)}_{0211}+\hat{h}_{0211}}
{1-e^{2\pi i(-2Q_x)}}-i\left[f_{0111}^{(1)}\left(f_{1200}^{(1)}+f_{0111}^{(1)}\right)
                            -3f_{3000}^{(1)*}f_{0111}^{(1)*}
\right.\right.$ & $x^2y^2$\\ &   $\hskip 3.9cm
                            -4f_{1020}^{(1)*}f_{0120}^{(1)}
               -4\left(f_{0020}^{(1)}f_{0202}^{(1)}+f_{0020}^{(1)*}f_{0220}^{(1)}\right)
\Big]\Bigg\}$ & \\
\hline\vspace{-0.30 cm}\\\vspace{0.07 cm}
$v_h(0,-3)=-8i\left\{g_{0040}\right\}\zeta_{y,+}^3$ &
$\displaystyle\hspace{-1mm}\left\{
f_{0040}^{(1)}+i\frac{\tilde{h}^{(2)}_{0040}+\hat{h}_{0040}}
{1-e^{2\pi i(2Q_y)}}-i\left[f_{0020}^{(1)}f_{0013}^{(1)*}
\right]\right\}$ &$y^4$ \\
\hline\vspace{-0.30 cm}\\\vspace{0.07 cm}
$v_h(0,3)=-2i\left\{g_{0013}\right\}\zeta_{y,-}^3$ &
$\displaystyle\hspace{-1mm}\left\{
f_{0013}^{(1)}+i\frac{\tilde{h}^{(2)}_{0013}+\hat{h}_{0013}}
{1-e^{2\pi i(-2Q_y)}}+i\left[f_{1020}^{(1)*}f_{0111}^{(1)*}
                            +f_{0120}^{(1)*}f_{0111}^{(1)}
                           -8f_{0020}^{(1)}f_{0040}^{(1)*}
\right]\right\}$ &$y^4$ \\
\hline\vspace{0.2 cm}
\end{tabular}}
\rule{0mm}{0mm}
\vskip 0.6cm
\label{tab:g_jklm_NO} 
\end{table*}
\clearpage

If second-order terms may be neglected, 
Eqs.~\eqref{e:horizontal} and~\eqref{e:vertical} 
are retrieved. The above equations show how 
second-order terms prevent the direct measurement of 
the RDTs $f_{jklm}=f_{jklm}^{(1)}+f_{jklm}^{(2)}$ from 
the secondary spectral lines (harmonics in the first 
summation). Among the terms in the second double 
summation, in fact, there will be always some harmonics 
overlapping those of the first summation. An example 
may again help clarifying this point. Let us select 
in Eq.~\eqref{eq:NFtransform4y} those terms proportional 
to $\zeta_{y,+}^2$, i.e. those exciting the spectral line 
$V_h(0,-2)$ (of the complex Courant-Snyder variable 
$h_{y,-}=\tilde{y}-i\tilde{p}_y$). The first sum selects 
the index 0030, while in the second double summation 
only those index satisfying the following conditions shall 
be kept
\begin{eqnarray}\label{eq:select2}
\left\{
\begin{array}{l}
j+p-1=0 \\
k+q-1=0 \\
l+r-1=2 \\
m+t=0   
\end{array}
\right.
\qquad\hbox{or}\qquad
\left\{
\begin{array}{l}
j+p=0   \\
k+q=0   \\
l+r-2=2 \\
m+t-1=0 
\end{array}
\right. \quad , 
\end{eqnarray}
resulting in 
\begin{eqnarray}
V_h(0,-2)=-6i\left\{f_{0030}^{(1)}+f_{0030}^{(2)} 
		   -\frac{i}{3}\left[f_{1010}^{(1)}f_{0120}^{(1)}-
				     f_{1001}^{(1)*}f_{1020}^{(1)}+
				    2f_{0020}^{(1)}f_{0021}^{(1)}-
				    6f_{0030}^{(1)}f_{0011}^{(1)}\right]
	\right\}\zeta_{y,+}^2 \ .
\end{eqnarray}
Assuming than no physical skew sextupoles 
exists, first-order skew sextupole RDTs 
$f_{0030}^{(1)}$ and $f_{0021}^{(1)}$ are generated 
by small tilts of normal sextupoles. Since 
focusing errors are assumed to be already small the products 
$f_{0020}^{(1)}f_{0021}^{(1)}$ and 
$f_{0030}^{(1)}f_{0011}^{(1)}$ are negligible 
compared to the products $f_{1010}^{(1)}f_{0120}^{(1)}$ and 
$f_{1001}^{(1)*}f_{1020}^{(1)}$, both generated by the low 
coupling and the large sextupole RDTs. By 
making use of Eq.~\eqref{eq:f0030}, the observable 
quantity $g_{0030}$ is 
\begin{eqnarray}\label{eq:vh02}
V_h(0,-2)=-6i\left\{g_{0030}\right\}\zeta_{y,+}^2
	 =-6i\left\{f_{0030}^{(1)}+i\frac{\tilde{h}^{(2)}_{0030}+\hat{h}_{0030}}
			  {1-e^{2\pi i(3Q_y)}}
		   -\frac{i}{3}\left[f_{1010}^{(1)}f_{0120}^{(1)}-
				     f_{1001}^{(1)*}f_{1020}^{(1)}\right]
	\right\}\zeta_{y,+}^2 \ .
\end{eqnarray}
This example shows how the equivalence between the RDTs 
computed from the lattice and those either measured or 
evaluated from turn-by-turn data breaks down when 
second-order terms are to be included. In Table IV of 
Ref.~\cite{prstab_strength} by using the spectral line 
$V_h(0,-2)$ to infer $f_{0030}$ and hence  the 
sextupole tilts may under- or over-estimate the reality, 
since part of $V_h(0,-2)$ is also excited by second-order 
terms uncorrelated to the sextupole tilts.  

Again, the situation is less desperate than it could appear 
and a way out exists to build a reliable non-linear 
lattice error model, since second-order terms may be 
computed beforehand and substracted to the measured 
RDTs. \\

\end{widetext}

\section{Resonance driving terms from turn-by-turn data of single 
            (dual-plane) BPM}
\label{app:2}
RDTs of Refs.~\cite{Bartolini1,Rogelio1,Andrea-thesis,prstab_strength} 
may be evaluated either via analytical formulas 
or from turn-by-turn BPM data. More precisely, 
BPM data need first to be normalized by the 
beta function (Courant-Snyder coordinates) and 
then to be composed in order to obtain the 
complex turn-by-turn variable 
$h_{r}(N)=\tilde{r}(N)-i\tilde{p}_r(N)$, where 
$r$ stands for either $x$ or $y$ and $N$ is the 
turn number. The harmonic analysis of $h_{r}(N)$, 
via the Fast Fourier Transform (FFT), allows the 
direct measurement of the RDTs.  As discussed 
in Ref.~\cite{Rogelio2} the reconstruction of 
the momentum $\tilde{p}_r$ from the positions 
at two consecutive BPMs is reliable as long as 
the region between the two monitors is free of 
nonlinearities (such as sextupole magnets). 
This condition is almost never met in electron 
storage rings like the one at the ESRF, where 
there is always one sextupole between two BPMs 
within each cell. A way out was found in 
Ref.~\cite{Rogelio2} to combine the signal over 
three consecutive BPMs. The resulting variable 
is no longer affected by errors in the momentum 
reconstruction and its harmonic analysis may be 
used to measure local resonance terms. The 
method was successfully applied to the CERN Super 
Proton Synchrotron~\cite{prstab_strength}, as 
well as to the Relativistic Heavy Ion 
Collider~\cite{Rogelio2}. On the other hand, the 
applicability of the 3-BPM signal is hampered 
whenever the phase advance between the monitors 
is close to either zero or $\pi$, which is 
unfortunately the case of many BPMs in the ESRF 
storage ring. Moreover, both 2-BPM and 3-BPM signals 
require a perfect synchronization among the BPMs 
(i.e. triggering on the same turn and on the same 
position along the bunch train).

\begin{figure}[h]
\rule{0mm}{0mm}
\centerline{
  \includegraphics[width=8.5cm,angle=0]
  {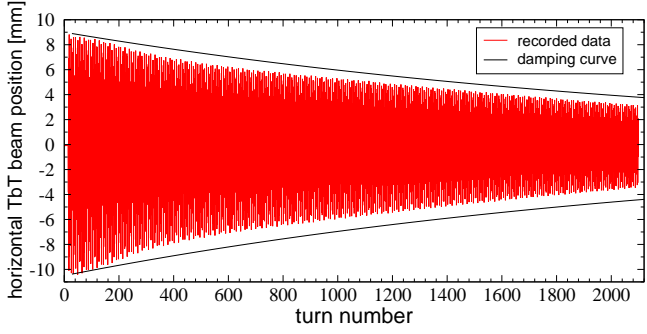}}
  \caption{\label{fig_TbT} (Color) Example of horizontal turn-by-turn 
	beam oscillation after the pulse of an horizontal kicker 
	magnet measured at the ESRF electron storage ring. A special 
	sextupole setting was used to minimize amplitude-dependent 
	detuning and chromaticity. The decreasing amplitude is believed 
	to be the result of radiation damping (black curve, 
        damping time of 7 ms, corresponding to about 2500 turns).}
\rule{0mm}{3mm}
\end{figure}
 
\begin{figure*}[t]
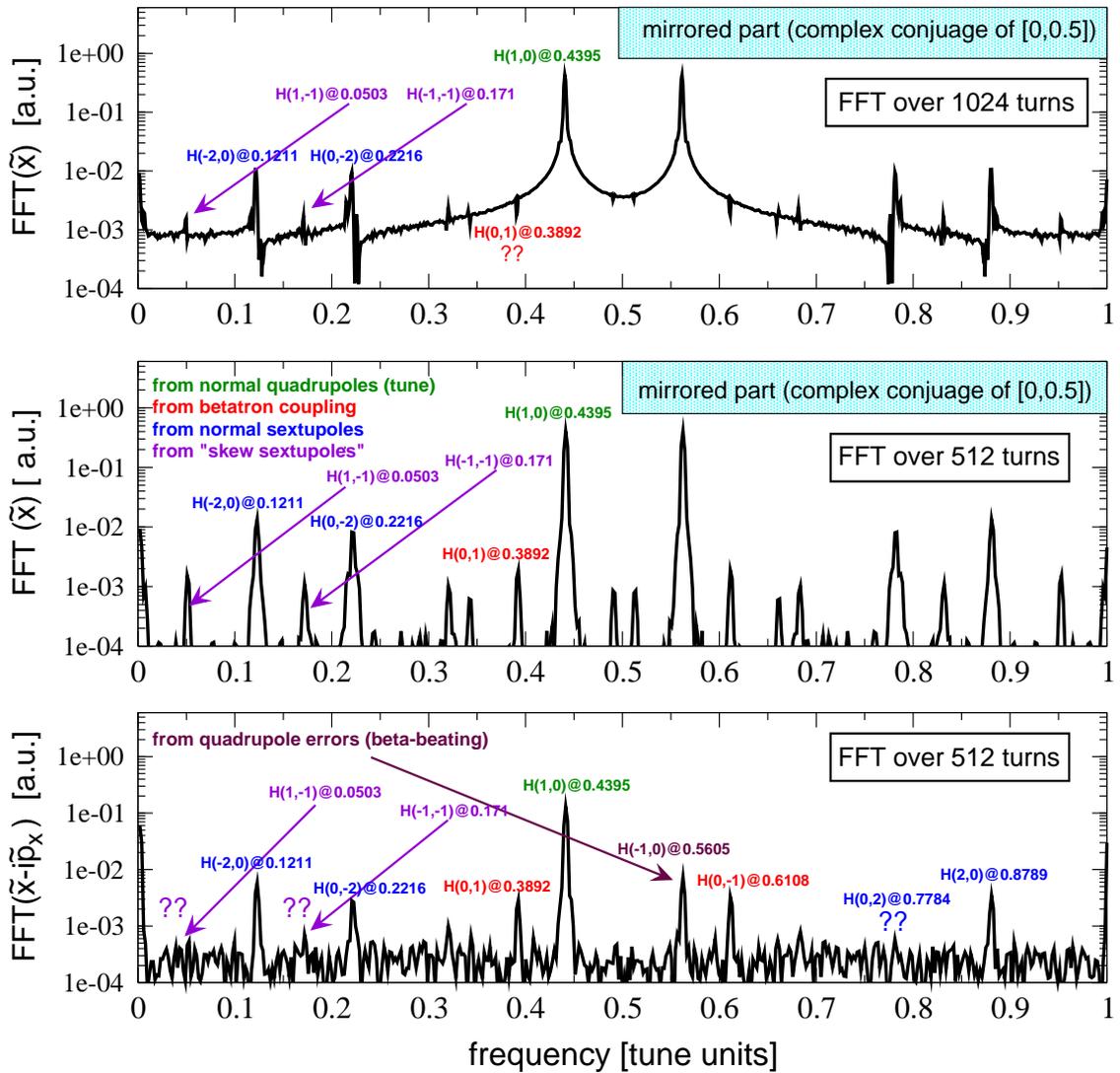

\rule{0mm}{0mm}
\centerline{
  \includegraphics*[width=15.0cm,angle=0]
  {fig24A.eps}
  \vspace{-1cm}} 
\centerline{
  \includegraphics*[width=15.0cm,angle=0]
  {fig24B.eps}}
  \caption{\label{fig_FFT} (Color) Example of horizontal turn-by-turn 
	spectra measured 
	at the ESRF electron storage ring. The upper plots are the FFT of 
	the real signal $\tilde{x}(N)=x(N)/\sqrt{\beta_x}$ computed over 
	1024 and 512 turns, respectively. Despite the longer signal, the 
	spectrum with 1024 turns shows higher background noise, which 
	is attributed to radiation damping. In both cases, the region 
	[0.5,1] is the mirrored copy (i.e. complex conjugate) of 
	the region [0,0.5]. The 
	bottom plot shows the FFT over 512 turns of the complex signal 
	$h_x(N)=\tilde{x}(N)-i\tilde{p}_x(N)$: There is no mirroring around 
	0.5, but the background noise is higher than in the simple FFT 
	of $\tilde{x}$ (center plot) and some spectral lines are no longer 
	visible. For a complete (up to the first order) classification of 
	all spectral lines, see Table IV of Ref.~\cite{prstab_strength}.}
\rule{0mm}{3mm}
\end{figure*}

Another aspect limiting the applicability 
of the multi-BPM signals in electron machine 
is represented by radiation damping. Ideally, 
the greater the number of turns with exploitable 
data (i.e. before decoherence smooths out the 
signal to zero), the higher the spectral 
resolution and hence the quality of the RDT 
measurement. At the ESRF a special sextupole 
setting was found to provide low 
amplitude-dependent detuning and linear 
chromaticity in both planes, so to minimize 
decoherence. However, the damping time being of 
about 2500 turns, the signal is sufficiently 
depressed to compromise the whole measurement 
already after 1024 turns (see Fig.~\ref{fig_TbT}). 
It has been observed that the spectral background 
noise increases with the number of turns used to 
perform the FFT of the turn-by-turn signal (see 
upper plots of Fig.~\ref{fig_FFT}): An effect 
which is believed to be related to radiation damping. 
Moreover, when analyzing the complex signal $h_{x}$ 
constructed between two consecutive BPMs the 
background noise is enhanced 
(see bottom plot of Fig.~\ref{fig_FFT}). 
The sextupolar spectral lines being intrinsically 
of limited height, it is of capital interest to 
keep the spectral background noise as low as 
possible, so to enhance the signal-to-noise ratio. 
For this reason, even if 2048 turns may be stored 
by each BPM, only 256 or 512 are routinely used at the 
ESRF in order to carry out the harmonic analysis, 
a trade off between enhancement of spectral 
resolution and minimization background noise. 

A new approach has been then developed to measure 
(a different kind of) RDTs from single-BPM data, 
$\tilde{r}(N)=r(N)/\sqrt{\beta_r}$, in order to 
avoid the noise enhancement whenever 
data from two or three BPMs are used. By using single-BPM 
data the issue of synchronization between the monitors 
is no longer of concern, the harmonic analysis at each 
BPM being independent of the data acquired by the others.

The bottom plot of Fig.~\ref{fig_FFT} shows a typical 
horizontal spectrum of $h_x$ with the excited spectral 
lines: Besides the principal tune line generated by the main 
quadrupoles, quadrupole errors (beta-beat) excite a 
line at the frequency $1-Q_x$, $H_h(-1,0)$, while betatron 
coupling is visible from the presence of a secondary 
line at the frequency corresponding to the vertical 
tune, $H_h(0,1)$, and its reciprocal, $H_h(-1,0)$. The 
main sextupoles (chromatic and harmonic) generate several 
lines, among which $H_h(\pm2,0)$ and $H_h(0,\pm2)$. Skew 
sextupole lines appear also at the position $Q_x-Q_y$, 
$H_h(1,-1)$, and $1-Q_x-Q_y$, $H_h(-1,-1)$: their origin will 
be discussed later. Other lines generated by 2-nd order 
sextupolar terms are more visible in the spectrum 
of $\tilde{x}$ (center plot of Fig.~\ref{fig_FFT}).

The harmonic analysis of the real signal $\tilde{r}$ 
produces a spectrum that is mirrored around 0.5 
(in tune units). By looking at Table IV of 
Ref.~\cite{prstab_strength} and Fig.~\ref{fig_FFT} 
it is clear that secondary spectral lines of the complex 
signal $h_r$ and excited by either coupling or sextupoles, 
become indistinguishable.  For instance, the two sextupolar 
horizontal spectral lines $H_h(2,0)$, located at the 
frequency $2Q_x$, and $H_h(-2,0)$, located at $1-2Q_x$, 
are excited by the two different 
RDTs,  $f_{1200}$ and $f_{3000}$, 
respectively. The latter are distinguishable and 
measurable from the FFT of the complex signal $h_x$, 
while the mirroring around 0.5 prevents such 
distinction if the FFT is performed on the real signal 
$\tilde{x}$.  An example may help understanding this 
merging. A complex signal $h_x$ contains two harmonics 
in the following combination
\begin{eqnarray}
h_x(N)=A_+e^{i(4\pi QxN+\psi)} +A_-e^{i(-4\pi QxN+\theta)}\ . \nonumber
\end{eqnarray}
The first term would correspond ideally at the spectral 
line $H_h(2,0)$, while the second to $H_h(-2,0)$. The FFT 
of $h_x(N)$ would yield two spectral lines located 
at $2Q_x$ and $1-2Q_x$, respectively, with measurable 
amplitudes ($A_+$ and $A_-$) and phases ($\psi$ and 
$\theta$). The real part of $h_x(N)$, corresponding to 
the single-BPM turn-by-turn data $\tilde{x}$, reads
\begin{eqnarray*}
\tilde{x}&=&\Re\{h_x\}(N) \nonumber \\
        &=&A_+\frac{e^{i(4\pi QxN+\psi   )}+e^{-i(4\pi QxN+\psi  )}}{2}+\\
        && A_-\frac{e^{i(-4\pi QxN+\theta)}+e^{i(4\pi QxN-\theta)}}{2} \\
        &=&\frac{\left[A_+e^{i\psi}+A_-e^{-i\theta}\right]}{2}e^{i4\pi QxN}
		+c.c. \qquad .
\end{eqnarray*}
The omitted term represents the complex conjugate 
of the first (i.e. its mirrored copy) and 
contains no additional information. 
The FFT of $\tilde{x}(N)$ would still 
generate two spectral lines $H(\pm2,0)$, which are 
however complex conjugates. This implies that 
the only (complex) observable left is 
$\left[A_+e^{i\psi}+A_-e^{-i\theta}\right]$,  
while the four quantities $A_{\pm}$, $\psi$ and 
$\theta$ are no longer measurable. Back to the RDT 
context, the consequence of analyzing the single-BPM 
signal instead of $h_r$ is that only linear 
combinations of RDTs are directly measurable. It 
will be shown that this is still sufficient to 
infer lattice errors and provide a correction scheme. 

Generally speaking, each line in the spectrum of 
$\tilde{x}(N)$ is the superposition of two lines 
in the spectrum of $h_x(N)$, according to
\begin{eqnarray}\label{eq:fft_Real}
H(n_x,n_y)=\frac{\left[H_h(n_x,n_y)+H_h(-n_x,-n_y)\right]}{2}\ . 
\end{eqnarray}
$H(n_x,n_y)$ without suffix denotes a line in 
spectrum of the real signal $\tilde{x}(N)$. 
$H_h(n_x,n_y)$ refers to a generic line in 
the spectrum of the complex signal $h_x(n)$. 
Identical relations hold  for the vertical 
spectrum.

\section{Quadrupole errors and tune-line amplitude}
\label{app:3}
In Table~\ref{tab:line-selection} (easier) 
first-order RDTs $f_{jklm}$ for skew quadrupole and 
normal sextupole spectral lines are used, while 
for skew sextupoles (more complex) second-order 
ORDTs $g_{jklm}$ are invoked. The latter are also 
used in Table~\ref{tab:line-selection-octu} for 
normal octupoles. In this section motivations and 
conditions for such a choice are provided. It will 
be shown how this is related to the removal of 
RDTs generated by focusing errors when using 
the modulated beta functions 
(i.e. computed from a lattice including quadrupole 
errors) rather than those of the ideal lattice. 

Parenthetically, a novel procedure for using single-BPM 
turn-by-turn data to measure the beta functions is 
described, which is complementary to the traditional approach
based on the direct measurement of the phase advance 
between two consecutive BPMs~\cite{Castro}. The latter 
is independent on calibration errors, but requires a 
perfect synchronization among the BPMs. The proposed 
scheme, on the contrary, necessitates well calibrated 
monitors but does not need any synchronization, as 
for the nonlinear analysis of 
Table~\ref{tab:line-selection}. 

In the Appendix of Ref.~\cite{prstab_coup} the 
complete Lie series of Eq.~\eqref{eq:NFtransform3} 
was computed in presence of betatron coupling, i.e. 
with the normal form transformation $F$ defined 
exclusively by coupling RDTs, namely
\begin{eqnarray}\label{eq:F-coup}
F=&&f_{1001}\zeta_{x,+}\zeta_{y,-} + f_{1001}^*\zeta_{x,-}\zeta_{y,+} 
   +f_{1010}\zeta_{x,+}\zeta_{y,+} \nonumber \\
  &&\hskip 0.5cm + f_{1010}^*\zeta_x^-\zeta_y^-\ .
\end{eqnarray}
Recursive expressions for the Poisson brackets 
at any order were derived leading to the 
following expression for the turn-by-turn 
evolution of the complex Courant-Snyder coordinates
\begin{eqnarray}
h_{x}&=&\hskip-0.05cm\cosh{(2\mathcal{P})}\ \zeta_{x,-} -\hskip-0.1cm
       i\sinh{(2\mathcal{P})}\left[\frac{f_{1001}}{\mathcal{P}}\zeta_{y,-} + 
                        \frac{f_{1010}}{\mathcal{P}}\zeta_{y,+} \right]
	\hskip-0.1cm\nonumber \\ \label{hx-coup}
&& \hskip 0.5cm \Uparrow\hskip4 cm\Uparrow   \hskip1.7 cm \Uparrow\\
&& \hskip 0.2cm H_h(1,0)\hskip2.8cm H_h(0,1)\hskip 0.5cm H_h(0,-1)
\nonumber \\
h_{y}&=&\hskip-0.05cm\cosh{(2\mathcal{P})}\ \zeta_{y,-} -\hskip-0.1cm
       i\sinh{(2\mathcal{P})}\left[\frac{f^*_{1001}}{\mathcal{P}}\zeta_{x,-} + 
                        \frac{f_{1010}}{\mathcal{P}}\zeta_{x,+} \right]
	\hskip-0.1cm\nonumber   \\ \label{hy-coup}
&& \hskip 0.5cm \Uparrow\hskip4 cm\Uparrow   \hskip1.7 cm \Uparrow\\
&& \hskip 0.2cm V_h(0,1)\hskip2.8cm V_h(1,0)\hskip 0.7cm V_h(-1,0)
\nonumber
\end{eqnarray}
where
\begin{eqnarray}
2\mathcal{P}&=&\sqrt{-|2f_{1001}|^2 +|2f_{1010}|^2}\ .
\end{eqnarray}
According to Eq.~\eqref{eq:fft_Real} the tune lines 
of the real Courant-Snyder coordinates read
\begin{eqnarray}
H(1,0)&=&\frac{\cosh{(2\mathcal{P})}}{2}\sqrt{2I_x}e^{i(2\pi NQ_x+\psi_{x,0})}
	+c.c.\quad ,\qquad\label{tunex-coup}\\
V(0,1)&=&\frac{\cosh{(2\mathcal{P})}}{2}\sqrt{2I_y}e^{i(2\pi NQ_y+\psi_{y,0})}
	+c.c.\quad ,\qquad\label{tuney-coup}
\end{eqnarray}
where $c.c.$ stands for complex conjugate. Note 
that in presence of pure coupling 
$H_h(-1,0)=V_h(0,-1)=0$.

The same algebra may be applied to another extreme 
case with normal quadrupole errors  only,  i.e. 
with the normal form transformation $F$ defined 
exclusively by focusing error RDTs
\begin{eqnarray}\label{eq:F-betabeat}
F=f_{2000}\zeta_{x,+}^2 +f_{2000}^*\zeta_{x,-}^2 +
  f_{0020}\zeta_{y,+}^2 +f_{0020}^*\zeta_{y,-}^2  \ .\qquad
\end{eqnarray}
The result is 
\begin{eqnarray}
h_{x}&=&\cosh{(4|f_{2000}|)}\ \zeta_{x,-} -
       i\sinh{(4|f_{2000}|)}\ e^{iq_{2000}}\ \zeta_{x,+} \ ,  
	\nonumber \\ \label{hx-betabeat}
&& \hskip 1.5cm \Uparrow  \hskip3.6 cm  \Uparrow \\
&& \hskip 1.2cm H_h(1,0)  \hskip2.2cm H_h(-1,0)\nonumber\\
h_{y}&=&\cosh{(4|f_{0020}|)}\ \zeta_{y,-} -
       i\sinh{(4|f_{0020}|)}\ e^{iq_{0020}}\ \zeta_{y,+} \ ,  
	\nonumber \\ \label{hy-betabeat}
&& \hskip 1.5cm \Uparrow  \hskip3.6 cm  \Uparrow \\
&& \hskip 1.2cm V_h(0,1)  \hskip2.2cm V_h(0,-1)\nonumber
\end{eqnarray}
where 
\begin{eqnarray}
q_{2000}&=&\hbox{arg}\left\{f_{2000}\right\} \quad \hbox{and} \quad
q_{0020}=\hbox{arg}\left\{f_{0020}\right\} \ , \qquad \\
\label{eq:def_f2000}
f_{2000}&=&\frac{\sum\limits_w^W \delta K_{w,1}\beta_{x}^w
        	e^{2i\Delta\phi_{w,x}}}
        {8(1-e^{4\pi iQ_x})}\ , \\
f_{0020}&=&\frac{\sum\limits_w^W \delta K_{w,1}\beta_{y}^w
	        e^{2i\Delta\phi_{w,y}}}
        {8(1-e^{4\pi iQ_y})}\ , \label{eq:def_f0020}
\end{eqnarray}
with $\beta_{x,y}$ and $\Delta\phi_{x,y}$ are the 
the C-S parameters of the ideal lattice (i.e. without 
focusing errors $\delta K_{w,1}$).  
Conversely to 
coupling, pure quadrupole errors excite directly the 
two lines $H_h(-1,0)$ and $V_h(0,-1)$. Applying 
again Eq.~\eqref{eq:fft_Real} the tune lines 
of the real Courant-Snyder coordinates read
\begin{eqnarray}
H(1,0)&=&\frac{1}{2}\Big[\cosh{(4|f_{2000}|)}
	            +i\sinh{(4|f_{2000}|)}e^{-iq_{2000}}
	       \Big] \nonumber \\
      &&\hskip 1cm\times\sqrt{2I_x}e^{i(2\pi NQ_x+\psi_{x,0})}
	+c.c.\quad ,\qquad\label{tunex-beat}\\
V(0,1)&=&\frac{1}{2}\Big[\cosh{(4|f_{0020}|)}
	            +i\sinh{(4|f_{0020}|)}e^{-iq_{0020}}
	       \Big] \nonumber \\
      &&\hskip 1cm\times\sqrt{2I_y}e^{i(2\pi NQ_y+\psi_{y,0})}
	+c.c.\quad ,\qquad\label{tuney-beat}
\end{eqnarray}

In the more general case with both quadrupole and 
coupling errors similar results may be derived. 
However, Eqs.~\eqref{tunex-coup}-\eqref{tuney-coup}
and~\eqref{tunex-beat}-\eqref{tuney-beat} already 
reveal an intrinsic feature of linear lattice errors: 
The amplitude of the tune lines $H(1,0)$ and 
$V(0,1)$ are modified by the RDTs. The implication 
of that is twofold. First, it becomes impossible to 
use single-BPM turn-by-turn data to disentangle and 
measure coupling and beta-beat RDTs, because the tune 
lines used to extract them from the secondary lines in 
Refs.~\cite{Rogelio1,Andrea-thesis,prstab_strength} 
depend on the RDTs too. For the same reason, measurement 
of sextupolar combined RDTs of Table~\ref{tab:line-rdt} 
will be corrupted. 

\begin{figure}
\rule{0mm}{0mm}
\centerline{
  \includegraphics[width=7.5cm,angle=0]
  {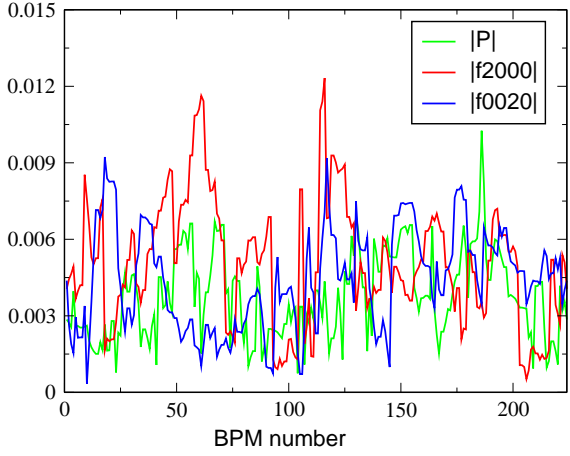}}
  \caption{\label{fig_latRDT} (Color) Comparison between the coupling 
	RDT factor $\mathcal{P}$ (green) and the focusing error RDTs 
	$f_{2000}$ (red), $f_{0020}$ (blue) inferred from ORM 
	measurement at the ESRF electron storage ring after 
	correction (peak beta-beating of about $\pm5\%$ and an 
	emittance ration lower than $1$\textperthousand). All three 
	quantities are of the same order of magnitude, though well 
	below $10^{-2}$ level.}
\rule{0mm}{3mm}
\end{figure}

A way out from this inconvenient situation is 
represented by measuring (and correcting) lattice 
errors before carrying out the turn-by-turn analysis 
with an independent technique, such as the orbit 
response matrix (ORM) and LOCO analysis of Ref.~\cite{ORM}. 
In Ref.~\cite{prstab_esr_coupling} the procedure 
to evaluate first-order focusing and coupling RDTs 
from ORM measurement is described in detail.  It can
be hence assumed that $f_{2000}$, $f_{0020}$, 
$f_{1001}$, and $f_{1010}$ be known prior to 
the turn-by-turn analysis. At the ESRF, with 
(peak) beta-beating usually corrected 
up to $\pm5\%$ and betatron coupling so to have 
an emittance ratio lower than $1$\textperthousand, 
typical values of $\mathcal{P}$, $|f_{2000}|$ and 
$|f_{0020}|$ are lower than $10^{-2}$, see 
Fig.~\ref{fig_latRDT}. Hence the perturbation 
introduced by coupling to the tune-line amplitudes of 
Eqs.~\eqref{tunex-coup} and \eqref{tuney-coup}
is negligible, $1-\cosh{(2\mathcal{P})}<10^{-4}$. The 
same is true only partially for quadrupole errors: 
while $1-\cosh{(4|f_{2000}|)}\simeq1-\cosh{(4|f_{0020}|)}<10^{-4}$, 
the residual modulation introduced by the hyperbolic 
sines in Eqs.~\eqref{tunex-beat} and 
\eqref{tuney-beat} may not be ignored, remaining of 
the level of percents. If such a modulation is not 
removed, it will be propagated as an error in the 
nonlinear error model.

\begin{figure}
\rule{0mm}{0mm}
\centerline{
  \includegraphics[width=7.5cm,angle=0]
  {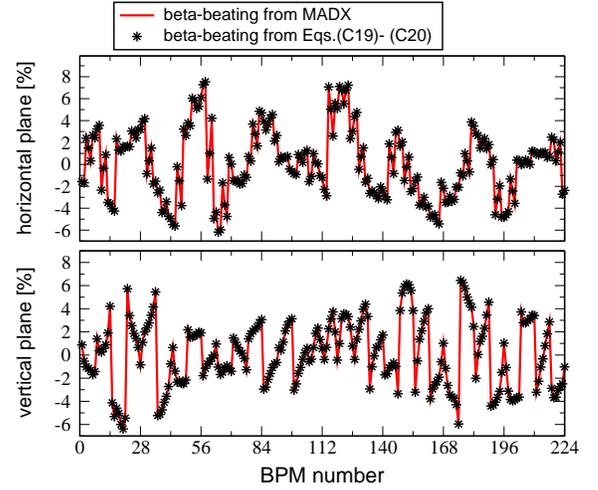}}
  \caption{\label{fig_betabeatRDT} (Color) Comparison between the 
	beta-beating computed by MADX after loading an error model 
	constructed from an ORM measurement at the ESRF storage 
	ring (red) and that evaluated from 
	Eqs~\eqref{eq:betabeat1} and~\eqref{eq:betabeat2}
        (black stars).}
\rule{0mm}{3mm}
\end{figure}

The task of making the tune-line amplitudes constant 
along the ring is actually much easier than it may 
appear. From Eq.~\eqref{tunex-beat} (the same applies 
to the vertical plane) the  tune-line amplitude reads
\begin{eqnarray}
|H(1,0)|&=&\frac{\sqrt{2I_x}}{2}\Bigg\{1 +2\sinh{(4|f_{2000}|)}\times 
		\\ &&\nonumber\hskip-0.5cm\Big[
		\sinh{(4|f_{2000}|)}+\cosh{(4|f_{2000}|)}\sin{q_{2000}}
		\Big]\Bigg\}^{1/2}\ .\label{tune-beat0}
\end{eqnarray}
In the ideal lattice, the Courant-Snyder coordinate
$\tilde{x}=x/\sqrt{\beta_{x}}$ would provide a 
a tune-line amplitude 
\begin{eqnarray}
|H(1,0)|_0&=&\frac{\sqrt{2I_x}}{2}\ .
\end{eqnarray}
It is straightforward to prove that by scaling the 
beta function by a factor $c$, $\beta_{x1}=c\beta_{x}$, 
the tune-line amplitude scales as 
\begin{eqnarray}
|H(1,0)|&=&\frac{1}{\sqrt{c}}\frac{\sqrt{2I_x}}{2}\ .
\end{eqnarray}
Hence, it is enough to replace the ideal beta 
function $\beta_{x}$ with 
\begin{eqnarray}
\beta_{x1}&=&\beta_{x}\Bigg\{1+2\sinh{(4|f_{2000}|)}\Big[
		\sinh{(4|f_{2000}|)}
		\nonumber\\ &&\hskip1.0cm \label{eq:betabeat0}
		+\cosh{(4|f_{2000}|)}\sin{q_{2000}}\Big]\Bigg\}
\end{eqnarray}
to retrieve a RDT-independent tune line, 
$|H(1,0)|=\sqrt{2I_x}/2$. Equation~\eqref{eq:betabeat0} 
provides an analytical expression for the 
horizontal beta-beating introduced by quadrupole 
errors:
\begin{eqnarray}
\frac{\Delta\beta_x}{\beta_{x}}&=&2\sinh{(4|f_{2000}|)}\Big[
		\sinh{(4|f_{2000}|)}
		 \label{eq:betabeat1}\\ \nonumber&&\hskip2.6cm
		+\cosh{(4|f_{2000}|)}\sin{q_{2000}}
		\Big]\ .
\end{eqnarray}
The same algebra applied to the vertical plane 
leads to 
\begin{eqnarray}
\frac{\Delta\beta_y}{\beta_{y}}&=&2\sinh{(4|f_{0020}|)}\Big[
		\sinh{(4|f_{0020}|)}
		 \label{eq:betabeat2}\\ \nonumber&&\hskip2.6cm
		+\cosh{(4|f_{0020}|)}\sin{q_{0020}}
		\Big]\ .
\end{eqnarray}
In Fig.\ref{fig_betabeatRDT} a comparison between the 
beta-beating as computed by MADX and by 
Eqs.~\eqref{eq:betabeat1} and~\eqref{eq:betabeat2} is 
shown. The lattice error model loaded in MADX 
is derived from an ORM measurement carried out at 
the ESRF storage ring. First-order RDTs $f_{2000}$ 
and $f_{0020}$ may be computed from the focusing 
errors via Eqs.~\eqref{eq:def_f2000} 
and~\eqref{eq:def_f0020}. For a more accurate 
comparison, the RDTs used in Fig.\ref{fig_betabeatRDT} 
are inferred from the harmonic analysis of 
single-particle tracking data, after isolating the 
$H_h(-1,0)$ and $V_h(0,-1)$ of 
Eqs.~\eqref{hx-betabeat} and~\eqref{hy-betabeat}. 
The agreement is remarkable. 

In practice then, the perturbations introduced by 
the focusing error RDTs $f_{2000}$ and $f_{0020}$ 
of Eqs.~\eqref{eq:def_f2000} and~\eqref{eq:def_f0020} 
may be {\sl absorbed} by replacing the ideal C-S parameters 
$\beta_{x,y}$ and $\Delta\phi_{x,y}$ with the modulated 
ones $\beta_{x1,y1}$ and $\Delta\phi_{x1,y1}$. 
By doing so, the new Courant-Snyder coordinates 
will no longer show quadrupole errors, and the 
beta-beat RDTs vanish, i.e. 
\begin{eqnarray}
\label{eq:def_f2000B}
f_{2000}(\beta_{x},\Delta\phi_{x})&\ne& 0
	\quad \hbox{but}\quad 
f_{2000}(\beta_{x1},\Delta\phi_{x1})\equiv 0\ , \qquad
\\
f_{0020}(\beta_{y},\Delta\phi_{y})&\ne& 0
	\quad \hbox{but}\quad 
f_{0020}(\beta_{x1},\Delta\phi_{x1})\equiv 0\ . \qquad  
\label{eq:def_f0020B}
\end{eqnarray}
Physically this corresponds to the fact that not 
including focusing errors in the computation 
of the C-S parameters results in {\sl mismatched} functions 
and a phase space topology (ellipses) dependent on the 
longitudinal position along the ring (through 
the $s$-dependent beta-beating RDTs $f_{2000}$ 
and $f_{0020}$). When, instead, the C-S parameters 
used to compute the C-S coordinates 
properly account for all normal quadrupole forces, 
the corresponding phase space topology becomes 
invariant along the ring (and circular).

Even though focusing errors and coupling may be 
efficiently measured and corrected from ORM 
measurement, the inferred model results usually 
from a best fit procedure which converges 
towards the measured matrix up to a certain degree 
of precision. Provided that coupling is well 
corrected (e.g. $\cosh{(2\mathcal{P})}=1$ up 
to a sufficient level), the same single-BPM 
turn-by-turn data may be used to validate and 
possibly to further correct the beta functions. 
The starting point are focusing RDTs that, despite 
the use of C-S parameters obtained from the 
error model, may still be nonzero because of  
residual fit errors, i.e. 
$0\ne f_{2000}(\beta_{x},\Delta\phi_{x}),\ 
 f_{0020}(\beta_{x},\Delta\phi_{x})\ll 1$. 
It is then reasonable to assume that 
$\cosh{(4|f_{2000}|)}\simeq\cosh{(4|f_{0020}|)}=1$ 
up to a sufficient level of tolerance. 
Equation~\eqref{tune-beat0} then simplifies to
\begin{eqnarray}
|H(1,0)|=\frac{\sqrt{2I_x}}{2}
	\Big\{1 +8|f_{2000}|\sin{q_{2000}}
		+O(|f_{2000}|^2)\Big\}^{1/2} .
	\nonumber \\\label{tune-beat1}
\end{eqnarray}
The above expression contains a constant term, 
$\sqrt{2I_x}/2$, and an oscillating one proportional to 
the residual focusing RDT, $|f_{2000}|\sin{q_{2000}}$. 
If the number of oscillations along the ring is 
sufficiently large and the number of BPM enough to 
cover uniformly the path, by averaging the tune-line 
amplitudes among all BPMs, the oscillating term cancels 
out and the invariant may be retrieved: 
\begin{eqnarray}
\sqrt{2I_x}\simeq2 \frac{1}{N}\sum_{n=1}^N{|H(1,0)|_n}
\ ,\label{tune-beat2} \\
\sqrt{2I_y}\simeq2 \frac{1}{N}\sum_{n=1}^N{|V(0,1)|_n}
\ ,
\end{eqnarray}
where $N$ is the number of available BPMs (not necessarily 
synchronized over the same turns and/or bunches) and 
$|H(1,0)|_n$ are the tune-line amplitudes measured at 
the different monitors. More intriguingly, the same 
single-BPM turn-by-turn data may be post-processed to 
extract the true beta function at its location. Once the 
average tune-line amplitude is computed, $<|H(1,0)|>$ 
($<|V(0,1)|>$), at each BPM the beta function may 
be scaled by a factor $c=[\ |H(1,0)|_n/<|H(1,0)|>]^2$ 
($c=[\ |V(0,1)|_n/<|V(0,1)|>]^2$) as done for 
Eq.~\eqref{eq:betabeat0}. The result is 
then straightforward:
\begin{eqnarray}
\beta_{x1,n}=\left(\frac{|H(1,0)|_n}{<|H(1,0)|>}\right)^2
		  \beta_{x,n}
\ ,\label{tune-beat3}\\
\beta_{y1,n}=\left(\frac{|V(0,1)|_n}{<|V(0,1)|>}\right)^2
		   \beta_{y,n}
\ ,
\end{eqnarray}
where $\beta_{x,n}$ denotes the initial horizontal beta 
function obtained from the lattice error model created, 
for instance, from the ORM measurement and fit, while 
$\beta_{x1,n}$ are those corrected by the harmonic 
turn-by-turn analysis. 

\begin{figure}
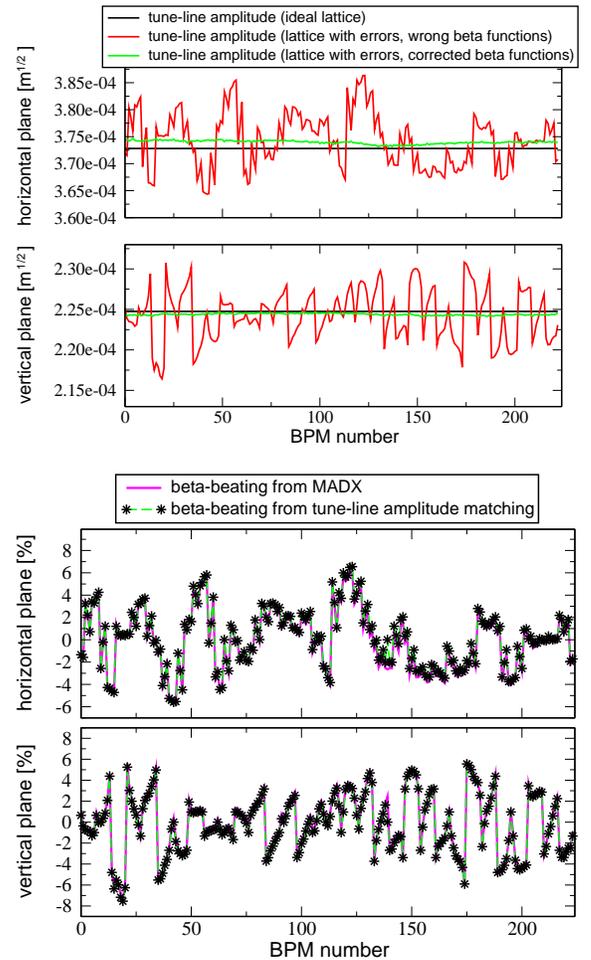

\rule{0mm}{0mm}
\centerline{
  \includegraphics[width=7.5cm,angle=0]
  {fig27A.eps}}
\vskip 0.3cm
\centerline{
  \includegraphics[width=7.5cm,angle=0]
  {fig27B.eps}}
  \caption{\label{fig_betabeatTuneAmpli} (Color) Top: Evolution of the 
	tune-line amplitudes along the SERF storage ring obtained 
	from single-particle tracking with the ideal lattice (black 
	curve), with lattice errors included in the tracking but 
	not in the beta functions (red), and with the beta functions 
	corrected after minimizing the spread between the tune-line 
	amplitudes among the BPMs (green). Bottom: Comparison between 
	the expected beta-beating as computed by MADX (magenta 
	curve) and the once inferred from the minimization of the 
	spread among the tune-line amplitudes (black stars). Note how 
	the tune-line amplitude pattern (red curve, upper plot) matches 
	that of the beta-beating (bottom plot).}
\rule{0mm}{3mm}
\end{figure}

In Fig.~\ref{fig_betabeatTuneAmpli} the results from a 
numerical test carried out from single-particle tracking 
with the same focusing errors and coupling of 
Fig.~\ref{fig_latRDT} is shown. Single-BPM turn-by-turn 
data are represented by the particle coordinates stored 
at the 224 BPMs of 
the ESRF storage ring for 256 turns. The Courant-Snyder 
variable is constructed from the ideal beta functions 
(hence without including the focusing errors). Once the 
average tune-line amplitude is computed, the new beta 
functions are computed from the ratio between the 
tune-line amplitude at the BPM and the mean value. 
The resulting beta-beating is compared with the one 
computed by MADX confirming the excellent agreement, while 
the spread between the tune-line amplitude evaluated 
at the 224 BPMs with the new beta functions drops 
dramatically. Note how the tune-line amplitude pattern 
(red curve in the upper plot) matches that of the 
beta-beating (bottom plot).

The use of the C-S parameters that already include 
focusing errors in evaluating the first-order RDTs 
of Table~\ref{tab:lattice-rdt} not only will correctly 
account for the modulation introduced by the beta 
functions $\beta_{x,y}$, but has a tremendous impact in 
simplifying the second-order analysis of coupling and 
normal sextupole RDTs of Tables~\ref{tab:g_jklm_SQ} 
and~\ref{tab:g_jklm_NS}, respectively: The 
first-order beta-beat RDTs being all zero (as well 
as the corresponding  Hamiltonian coefficients),   
second-order ORDTs $g_{jklm}$ are equivalent to the 
first-order RDTs $f_{jklm}^{(1)}$. For this reason 
the in Table~\ref{tab:line-selection} the latter are 
used in the entries corresponding to skew quadrupoles 
and normal sextupoles and no beta-beat RDT is reported 
in the tune lines. Second-order ORDTs $g_{jklm}$ are 
instead used for the skew sextupole-like entries 
because they are generated by nonzero coupling and 
normal sextupole RDTs.

Focusing errors hence, if not included in the lattice 
model, modulate the tune-line amplitudes and thus 
the beta functions. The modulation is independent on 
the initial conditions (action and phase). Another, more 
complex, amplitude-dependent modulation is introduced at 
the second order by normal sextupoles, requiring a  
careful preliminary analysis in order to make it negligible.  
In the example of Eq.~\eqref{eq:select2}, when computing 
the second-order contribution to the spectral line 
$V_h(0,-2)$, proportional to $\zeta_{y,+}^2$, the 
sets of indexes $jklm$ and $pqrt$ were selected 
by solving the following systems for $abcd=0020$: 
\begin{eqnarray}\label{eq:selectA1}
\left\{
\begin{array}{l}
j+p-1=a \\
k+q-1=b \\
l+r-1=c \\
m+t=d   
\end{array}
\right.
\qquad\hbox{or}\qquad
\left\{
\begin{array}{l}
j+p=a   \\
k+q=b   \\
l+r-2=c \\
m+t-1=d 
\end{array}
\right. \quad . 
\end{eqnarray}
The vertical tune line $V_h(0,1)$ (identical considerations 
apply to the horizontal one) receives contributions from all 
terms proportional to $\zeta_{y,-}$: These may come from 
those selected by $abcd=0001$, but also from 
$abcd=1101$, which is proportional to $|\zeta_x|^2\zeta_{y,-}$, 
as well as from $abcd=0012$, which scales with 
$|\zeta_y|^2\zeta_{y,-}$. Only coupling and focusing errors 
may contribute to $abcd=0001$, whereas normal sextupoles 
excite the terms introduced by $abcd=1101$ and $abcd=0012$ 
(skew sextupoles RDTs also contribute, but are here 
neglected, as they are assumed to be generated by 
tilted normal sextupoles and hence to be a 
small fraction of the normal sextupole RDTs).
By solving the two systems of Eq.~\eqref{eq:selectA1} with 
these two later conditions, and repeating the same 
algebra for the horizontal tune line, it can be shown that 
\begin{eqnarray}
H_h(1,0)&=&\left\{1+T_{Hx}|\zeta_x|^2+T_{Hy}|\zeta_y|^2\right\}
	   \zeta_{x,-}\quad , \quad\label{eq:selectA2x} \\
V_h(0,1)&=&\left\{1+T_{Vx}|\zeta_x|^2+T_{Vy}|\zeta_y|^2\right\}
	   \zeta_{y,-}\quad , \quad\label{eq:selectA2y}
\end{eqnarray}
where $T_{Hx,Hy}$ and $T_{Vx,Vy}$ vary along 
the ring and are quadratic functions of the 
first-order normal sextupoles RDTs  
($f_{3000}^{(1)},\ f_{1002}^{(1)}$, and the 
like).  Similar expressions apply also for the lines 
$H_h(-1,0)$, proportional to $\zeta_{x,+}$, and 
$V_h(0,-1)$, proportional to $\zeta_{y,+}$. The tune 
lines of the real Courant-Snyder signals $\tilde{x}$ 
and $\tilde{y}$ are modulated too, since 
$H(1,0)=(H_h(1,0)+H_h(-1,0))/2$ and 
$V(0,1)=(V_h(0,1)+V_h(0,-1))/2$. 
Sextupoles therefore modify to the second order 
(i.e. quadratically in their RDTs and hence in 
their strengths) the tune-line amplitudes, which 
are no longer equivalent to $\sqrt{2I_{x,y}}/2$ 
(first two rows of Table~\ref{tab:line-selection}). 
Equations~\eqref{eq:selectA2x} and~\eqref{eq:selectA2y} 
reveal also the nonlinear coupling introduced by sextupoles 
which makes the tune-line amplitude in one plane 
dependent on the action $|\zeta|$ of the other. 
If not negligible, this nonlinear modulation may 
corrupt (or prevent) the evaluation of the 
invariants $(2I_{x,y})$ from the tune-line amplitudes  
and hence the overall analysis discussed here, which 
is based on the formulas of Table~\ref{tab:line-rdt}. 

\begin{figure}
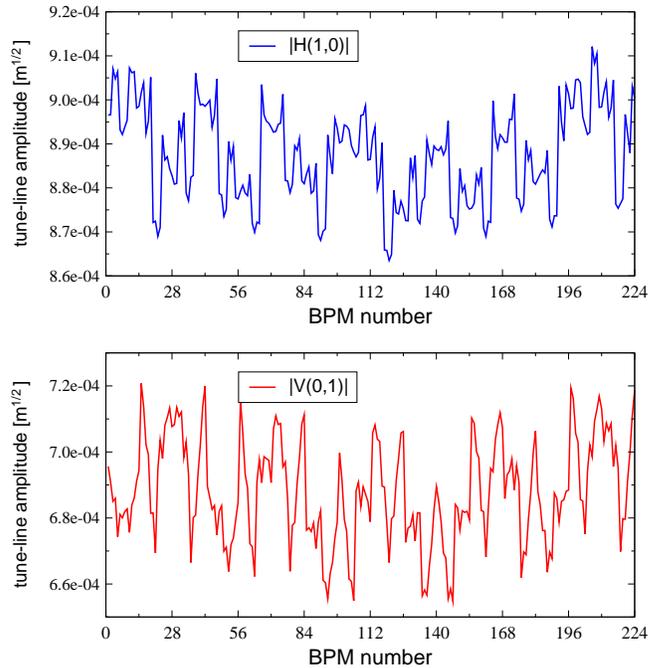

\rule{0mm}{0mm}
\centerline{
  \includegraphics[width=8.5cm,angle=0]
  {fig28A.eps}}
\vskip 0.3cm
\centerline{
  \includegraphics[width=8.5cm,angle=0]
  {fig28B.eps}}
  \caption{\label{fig_Tuneline} (Color) Example of variation of 
	the horizontal (top) and vertical (bottom) tune-line 
	amplitudes along the ESRF storage ring (224 BPMs, tracking 
	simulations and FFT with initial conditions $x_0=8$ mm 
	and $y_0=2.0$ mm, last row of table~\ref{tab:track-scan}). 
        Normal sextupoles induce a modulation 
	in both planes of about $2\%$, which would be zero in their 
	absence.}
\rule{0mm}{3mm}
\end{figure}

Given a certain sextupole setting, $T_{Hx,Hy}$ 
and $T_{Vx,Vy}$ are fixed. 
It is then of interest to estimate the largest 
beam oscillation amplitude that keeps the modulation 
below a tolerable level. To this end, 
preliminary single-particle tracking simulations 
at different initial conditions $x_0$ and $y_0$ (i.e. 
amplitudes) may be carried out. By storing the 
particle positions at all BPMs and performing the 
FFT of the Courant-Snyder coordinates $\tilde{x}$ and 
$\tilde{y}$, the tune-line amplitudes may be plotted 
along the ring and their modulation evaluated against
$x_0$ and $y_0$. In Fig.~\ref{fig_Tuneline} an example with 
betatron oscillations of $x_0=10$ mm and $y_0=2.2$ 
mm (large values by ESRF standards) is shown. 
Focusing errors are included in the model 
when computing the beta functions and the residual modulation 
of about $3\%$ is generated exclusively by sextupole RDTs. 
Results from a more complete scan of the ESRF storage 
ring are reported in Table~\ref{tab:track-scan}.  
They suggest to limit the transverse oscillation to 
$x_0\simeq3.5$~mm and $y_0\simeq1$~mm, corresponding to 
$\sqrt{2I_{x}}\simeq\sqrt{2I_{y}}\simeq3.1\times10^{-4}$~m$^{1/2}$ 
with relative variation of about $0.2\%$, 
well within the measurement statistical fluctuations of 
about $1\%$. Larger oscillations would induce a beneficial 
larger spectral resolution (higher ratio between sextupolar 
secondary lines and background noise), but also a  
larger modulation of the tune-line amplitudes, which is 
detrimental for the correct evaluation of the 
invariants $2I_{x,y}$, hampering in turn the 
correct measurements of sextupolar RDTs.

\begin{table}[t]\vskip -0.2 cm 
\caption{Dependence of tune-line amplitudes mean value and rms modulation 
	against initial conditions evaluated from single-particle 
	tracking and FFT of the particle position recorded at the 224 
	BPMs if the ESRF storage ring. The sextupole setting is the 
	same used in the experiments and the focusing errors are
	included in the lattice model when computing the $\beta$ functions 
	used to normalize $x$ and $y$.}
\centering
{
\begin{tabular}{l c c r}
\hline\hline\vspace{-0.25cm} \\ \vspace{0.1 cm}
$(x_0,y_0)$\ \ &$(|H(1,0)|,|V(0,1)|)$ & $(|H(1,0)|,|V(0,1)|)$ & rel.  \\
               &    mean value        &    rms modul.         & modul.\\
(mm)           &  m$^{1/2}$           &    m$^{1/2}$          &  \\
\hline
\hline\vspace{-0.28 cm} \\ \vspace{0.05 cm}
(0.4,\ 0.1) &$(0.3,\ 0.3)\times10^{-4}$ &$(3.1,\ 2.7)\times10^{-9}$& $\sim10^{-4}$\\ 
\hline\vspace{-0.28 cm} \\ \vspace{0.05 cm}
(1.2,\ 0.5) &$(1.0,\ 1.4)\times10^{-4}$ &$(4.6,\ 5.0)\times10^{-8}$ & $<10^{-3}$\\ 
\hline\vspace{-0.28 cm} \\ \vspace{0.05 cm}
(2.2,\ 0.5) &$(2.0,\ 1.4)\times10^{-4}$ &$(0.1,\ 0.1)\times10^{-6}$ & $<10^{-3}$  \\
\hline\vspace{-0.28 cm} \\ \vspace{0.05 cm}
(2.2,\ 0.7) &$(2.0,\ 2.0)\times10^{-4}$ &$(0.2,\ 0.2)\times10^{-6}$ & $\sim0.1\%$  \\
\hline\vspace{-0.28 cm} \\ \vspace{0.05 cm}
(2.8,\ 0.8) &$(2.6,\ 2.6)\times10^{-4}$ &$(0.4,\ 0.4)\times10^{-6}$ & $\sim0.2\%$  \\
\hline\vspace{-0.28 cm} \\ \vspace{0.05 cm}
(4.5,\ 0.8) &$(4.1,\ 2.6)\times10^{-4}$ &$(0.8,\ 0.9)\times10^{-6}$ & $\sim0.3\%$  \\
\hline\vspace{-0.28 cm} \\ \vspace{0.05 cm}
(4.4,\ 0.9) &$(4.1,\ 3.0)\times10^{-4}$ &$(0.9,\ 1.1)\times10^{-6}$ & $\sim0.4\%$  \\
\hline\vspace{-0.28 cm} \\ \vspace{0.05 cm}
(8.0,\ 2.0) &$(8.9,\ 6.9)\times10^{-4}$ &$(1.1,\ 1.5)\times10^{-5}$ & $\sim2\%$  \\
\hline
\end{tabular}}
\vskip 0.4 cm
\label{tab:track-scan}
\end{table}

As reported in Ref.\cite{Bartolini1}, the spectral 
line $H_h(-1,0)$ of the complex C-S signals  
$\tilde{x}-i\tilde{p}_x$ receives a contribution from 
the normal octupole term $x^4$ via two ORDTs, $g_{3100}$ 
and $g_{2011}$. It can be shown that 
\begin{eqnarray}
H_h(-1,0)&=&\left[3g_{3100}(2I_x)^{3/2} +4g_{2011}(2I_x)^{1/2}(2I_y)
            \right]\times \nonumber \\
          &&  (-2i)e^{-i(2\pi Q_xN+\psi_{x0})}\ .
\end{eqnarray}
This implies that, on top of the focusing errors 
not included in the model and second-order normal sextupole 
terms, the tune line $H(1,0)$ of the real 
C-S signal $\tilde{x}$ receives a contribution from 
octupolar terms, since $H(1,0)=1/2[H_h(1,0)+H_h(-1,0)]$.
Three problems arise at large excitation amplitude 
and or with large octupolar components in the lattice: 
$(i)$ The tune phase is affected by octupolar terms and the 
linear lattice modelling discussed in Sec.~\ref{sec_LinMeas} 
is corrupted; $(ii)$ It is no longer possible to 
extract any linear combination of $g_{3100}$ and 
$g_{2011}$, because of the different dependence on 
the action of the two terms, $\propto (2I_x)^{3/2}$ and 
$\propto (2I_x)^{1/2}(2I_y)$ respectively; $(iii)$ The tune 
line amplitude is no longer equal to $(2I_x)^{1/2}$ and 
the computation of the other CRDTs from 
Tables~\ref{tab:line-rdt} and~\ref{tab:line-rdt-octu} 
are affected by an intrinsic error. Equivalent considerations 
apply for the vertical tune line, affected by the 
$y^4$ octupolar terms, $g_{0013}$ and $g_{1102}$.

In conclusion, the turn-by-turn analysis of the 
nonlinear lattice model may be carried out and 
the formulas of Tables~\ref{tab:line-selection} 
and~\ref{tab:line-rdt} applied provided that: 
($i$) an independent measurement of linear lattice 
errors, and hence of coupling and beta-beat RDTs, is 
performed prior the harmonic analysis; ($ii$) focusing 
errors are included when computing the C-S 
parameters to be used when evaluating the first-order 
RDTs of Table~\ref{tab:lattice-rdt} and when 
normalizing the measured turn-by-turn data before 
performing the FFT, $\tilde{x}=x/\sqrt{\beta_x}$ and 
$\tilde{y}=y/\sqrt{\beta_y}$; If a residual important 
modulation of the tune-line amplitudes along the ring 
is observed, the BPM beta functions may be further 
corrected by requiring that each tune-line amplitude be 
equal to the mean value (among all BPMs); ($iii$) the 
transverse beam excitation is conveniently chosen as 
a trade off between maximizing the spectral 
signal-to-noise ratio and minimizing the tune-line 
amplitude modulation induced by sextupoles and 
octupoles. This will result in tune lines 
having the same amplitude along the ring 
($|H(1,0)|\simeq\sqrt{2I_{x}}/2$ and 
$|V(0,1)|\simeq\sqrt{2I_{y}}/2$), in zero beta-beat RDTs, 
and hence in the possibility of using first-order RDTs 
for the analysis of normal sextupole errors.

\section{Combined RDTs from single-BPM turn-by-turn data}
\label{app:4}
In this Appendix formulas for the single-BPM 
combined RDTs (CRDTs) of Tables~\ref{tab:line-selection} 
and~\ref{tab:line-selection-octu} are derived from the 
results of Appendix~\ref{app:2}. As discussed in 
Appendix~\ref{app:3} it is assumed that 
quadrupole errors are included in the computation 
of the C-S parameters, i.e. that the modulated 
beta functions are used in evaluating 
$\tilde{x}=x/\sqrt{\beta_x}$ and 
$\tilde{y}=y/\sqrt{\beta_y}$, and that transverse 
oscillations are sufficiently small to neglect the 
interference of sextupole and octupole RDTs on the tune 
lines and to avoid amplitude-dependent detuning: 
This will ensure that 
the two sums in Eqs.~\eqref{eq:NFtransform4x} 
and~\eqref{eq:NFtransform4y} do not contain the 
secondary lines $H_h(\pm1,0)$ and $V_h(0,\pm1)$. 
Results for the skew quadrupole CRDTs 
are presented for completeness, even though they 
are not considered suitable for a refined 
evaluation of betatron coupling (as mentioned 
in Appendix~\ref{app:3}). Equations~\eqref{eq:NFtransform4x} 
and~\eqref{eq:NFtransform4y} may be rewritten as 
\begin{eqnarray}\label{e:h_x}
&&\hspace{-0.3cm}h_{x}(N)=\sqrt{2I_x}e^{i(2\pi Q_xN+\psi_{x0})}\ - \\
&&\qquad\qquad 2i\sum_{jklm}{jg_{jklm}(2I_x)^{\frac{j+k-1}{2}}
                  (2I_y)^{\frac{l+m}{2}}} \ \times \nonumber\\
&&\qquad\qquad e^{i[(1-j+k)(2\pi Q_xN+\psi_{x0})+
                  (m-l)(2\pi Q_yN+\psi_{y0})]}\ ,\nonumber
\end{eqnarray}
and, equivalently for the vertical plane,
\begin{eqnarray}\label{e:h_y}
&&\hspace{-0.3cm}h_{y}(N)=\sqrt{2I_y}e^{i(2\pi Q_yN+\psi_{y0})}\ - \\
&&\qquad\qquad 2i\sum_{jklm}{lg_{jklm}(2I_x)^{\frac{j+k}{2}}
                 (2I_y)^{\frac{l+m-1}{2}}} \times \nonumber\\
&&\qquad\qquad e^{i[(k-j)(2\pi Q_xN+\psi_{x0})+
               (1-l+m)(2\pi Q_yN+\psi_{y0})]}\ \ .\nonumber
\end{eqnarray}
$g_{jklm}$ are the ORDT of Tables~\ref{tab:g_jklm_SQ},
~\ref{tab:g_jklm_NS} and~\ref{tab:g_jklm_SS}, measurable at 
the location of a generic BPM. The four index are selected 
according to the type of the magnetic elements exciting the 
RDTs and obey to the the selection rules of 
Table~\ref{tab:index-selection}. The real parts of 
the above expressions read
\begin{eqnarray}\label{e:tilde_x1}
\tilde{x}(N)&=&\frac{\sqrt{2I_x}}{2}
		\left[e^{i(2\pi Q_xN+\psi_{x0})}+c.c\right]\ + \\
&&\hskip-0.5cm 2\sum_{jklm}{j(2I_x)^{\frac{j+k-1}{2}}
                  (2I_y)^{\frac{l+m}{2}}} \ \times \nonumber\\
&&\hskip-0.5cm  \Im\left\{g_{jklm}e^{i[(1-j+k)(2\pi Q_xN+\psi_{x0})+
                  (m-l)(2\pi Q_yN+\psi_{y0})]}\right\}\ ,\nonumber
\end{eqnarray}
\begin{eqnarray}\label{e:tilde_y1}
\tilde{y}(N)&=&\frac{\sqrt{2I_y}}{2}
		\left[e^{i(2\pi Q_yN+\psi_{y0})}+c.c\right]\ + \\
&&\hskip-0.5cm 2\sum_{jklm}{l(2I_x)^{\frac{j+k}{2}}
                  (2I_y)^{\frac{l+m-1}{2}}} \ \times \nonumber\\
&&\hskip-0.5cm  \Im\left\{g_{jklm}e^{i[(k-j)(2\pi Q_xN+\psi_{x0})+
                  (1-l+m)(2\pi Q_yN+\psi_{y0})]}\right\}\ .\nonumber
\end{eqnarray}
In deriving the above equations, the relation 
$\Re\{iz\}=-\Im\{z\}$, valid for any complex 
number $z$, is used. $c.c.$ stands for complex 
conjugate. The following relations then apply
\begin{eqnarray}\label{e:tunelinex}
\left\{
\begin{aligned}
(2I_x)   &=&(2|H(1,0)|)^2 \\
\psi_{x0}&=&\hbox{arg}\{H(1,0)\}
\end{aligned}
\right.   \ , \\\label{e:tuneliney}
\left\{
\begin{aligned}
(2I_y)   &=&(2|V(0,1)|)^2 \\
\psi_{y0}&=&\hbox{arg}\{V(0,1)\}
\end{aligned}
\right.   \ .
\end{eqnarray}

\begin{table}[t]
\vskip 0.0 cm
\centering
\caption{Selection of index relative to the magnets.}
\begin{tabular}{l c c}
\hline\hline\vspace{-0.25cm} \\ \vspace{0.1 cm}
multipole kind  &\ \  magnetic term\ \   &  index relations \\
\hline
\hline\vspace{-0.28 cm} \\ \vspace{0.05 cm}
normal quadrupole  &  $x^2$           &  $j+k=2 \quad m+l=0$\\
\hline\vspace{-0.28 cm} \\ \vspace{0.05 cm}
normal quadrupole  &  $y^2$           &  $j+k=0 \quad m+l=2$\\
\hline\vspace{-0.28 cm} \\ \vspace{0.05 cm}
skew  quadrupole   &  $xy$           &  $j+k=1 \quad m+l=1$\\
\hline\vspace{-0.28 cm} \\ \vspace{0.05 cm}
normal sextupole   &   $x^3$           &  $j+k=3 \quad m+l=0$\\
\hline\vspace{-0.28 cm} \\ \vspace{0.05 cm}
normal sextupole   &   $xy^2$          &  $j+k=1 \quad m+l=2$\\
\hline\vspace{-0.28 cm} \\ \vspace{0.05 cm}
skew sextupole     &    $y^3$           &  $j+k=0 \quad m+l=3$\\
\hline\vspace{-0.28 cm} \\ \vspace{0.05 cm}
skew sextupole     &    $x^2y$          &  $j+k=2 \quad m+l=1$\\
\hline\vspace{-0.28 cm} \\ \vspace{0.05 cm}
normal octupole    &   $x^4$           &  $j+k=4 \quad m+l=0$\\
\hline\vspace{-0.28 cm} \\ \vspace{0.05 cm}
normal octupole    &   $y^4$           &  $j+k=0 \quad m+l=4$\\
\hline\vspace{-0.28 cm} \\ \vspace{0.05 cm}
normal octupole    &   $x^2y^2$        &  $j+k=2 \quad m+l=2$\\
\hline
\end{tabular}
\label{tab:index-selection}
\vskip 0.4 cm
\end{table}

{\sl\bf Coupling CRDTs. } By replacing $g_{jklm}$ with 
the first-order RDTs $f_{jklm}^{(1)}$ and selecting in 
Eq.~\eqref{e:tilde_x1} those terms satisfying the 
conditions $j+k=1$ and $m+l=1$ the following 
harmonics are generated
\begin{eqnarray}\label{e:tilde_x2}
&&2\sqrt{2I_y}\Im\left\{f_{1001}^{(1)}e^{i(2\pi Q_yN+\psi_{y0})]}+
		     f_{1010}^{(1)}e^{-i(2\pi Q_yN+\psi_{y0})]} 
	      \right\} =\nonumber \\
&&\frac{\sqrt{2I_y}}{i}\left[
	\left(f_{1001}^{(1)}  -f_{1010}^{(1)*}\right)e^{ i(2\pi Q_yN+\psi_{y0})}-
\right. \nonumber \\ && \hskip 1.1cm\left.
	\left(f_{1001}^{(1)*}-f_{1010}^{(1)}  \right)e^{-i(2\pi Q_yN+\psi_{y0})}
	\right]=\nonumber \\
&&\frac{\sqrt{2I_y}}{i}\left[
	F_{xy}e^{ i(2\pi Q_yN+\psi_{y0})}-
	c.c.
	\right]\ ,
\end{eqnarray}
where the coupling CRDT is defined as 
$F_{xy}=f_{1001}^{(1)}  -f_{1010}^{(1)*}$ and excites
the spectral line is $H(0,1)$, whose amplitude 
and phase are then
\begin{eqnarray}\label{e:H01}
\left\{
\begin{aligned}
|H(0,1)|            &=\sqrt{2I_y}|F_{xy}| \\
\hbox{arg}\{H(0,1)\}&=q_{F_{xy}}+\frac{3}{2}\pi+\psi_{y0}
\end{aligned}
\right.   \ . 
\end{eqnarray}
$(2I_y)$ and $\psi_{y0}$ are measurable from 
the vertical tune line $V(0,1)$, via   
Eq.~\eqref{e:tuneliney}, yielding
\begin{eqnarray}\label{e:F_xy}
\left\{
\begin{aligned}
F_{xy}    &=f_{1001}^{(1)}  -f_{1010}^{(1)*} \\
|F_{xy}|  &=|H(0,1)|/(2|V(0,1)|) \\
q_{F_{xy}}&=\hbox{arg}\{H(0,1)\}-\frac{3}{2}\pi-\hbox{arg}\{V(0,1)\}
\end{aligned}
\right.   \ . \hskip 1cm
\end{eqnarray}
The same algebra repeated for the vertical 
signal of Eq.~\eqref{e:tilde_y1} 
reads
\begin{eqnarray}\label{e:tilde_y2}
&&2\sqrt{2I_x}\Im\left\{f_{0110}^{(1)}e^{i(2\pi Q_xN+\psi_{x0})]}+
		     f_{1010}^{(1)}e^{-i(2\pi Q_xN+\psi_{x0})]} 
	      \right\} =\nonumber \\
&&\frac{\sqrt{2I_x}}{i}\left[
	\left(f_{0110}^{(1)}-f_{1010}^{(1)*}\right)e^{ i(2\pi Q_xN+\psi_{x0})}-
\right. \nonumber \\ && \hskip 1.1cm\left.
	\left(f_{0110}^{(1)*}-f_{1010}^{(1)}\right)e^{-i(2\pi Q_xN+\psi_{x0})}
	\right]=\nonumber \\
&&\frac{\sqrt{2I_x}}{i}\left[
	F_{yx}e^{ i(2\pi Q_xN+\psi_{x0})}-
	c.c.
	\right]\ ,
\end{eqnarray}
where the coupling CRDT is defined as 
$F_{yx}=f_{0110}^{(1)}-f_{1010}^{(1)*}$ and excites
the spectral line is $V(1,0)$, whose amplitude 
and phase are then
\begin{eqnarray}\label{e:V10}
\left\{
\begin{aligned}
|V(1,0)|            &=\sqrt{2I_x}|F_{yx}| \\
\hbox{arg}\{V(1,0)\}&=q_{F_{yx}}+\frac{3}{2}\pi+\psi_{x0}
\end{aligned}
\right.   \ . 
\end{eqnarray}
$(2I_x)$ and $\psi_{x0}$ are measurable from 
the horizontal tune line $H(1,0)$,  via 
Eq.~\eqref{e:tunelinex}, yielding
\begin{eqnarray}\label{e:F_yx}
\left\{
\begin{aligned}
F_{yx}    &=f_{0110}^{(1)} -f_{1010}^{(1)*} \\
|F_{yx}|  &=|V(1,0)|/(2|H(1,0)|) \\
q_{F_{yx}}&=\hbox{arg}\{V(1,0)\}-\frac{3}{2}\pi-\hbox{arg}\{H(1,0)\}
\end{aligned}
\right.   \ . \hskip 1cm
\end{eqnarray}
Equations~\eqref{e:F_xy} and~\eqref{e:F_yx} prove 
the skew quadrupole entries in  
Tables~\ref{tab:line-selection} and 
~\ref{tab:line-rdt}. 
Interestingly, one may think 
of inverting the linear system
\begin{eqnarray}\label{e:F_system}
\left\{
\begin{aligned}
F_{xy}    &=f_{1001}^{(1)} -f_{1010}^{(1)*} \\
F_{yx}    &=f_{1001}^{(1)*}-f_{1010}^{(1)*} \\
\end{aligned}
\right.   \ , \hskip 1cm
\end{eqnarray}
to extract the coupling RDTs from the two CRDTs. 
Unfortunately, the system is degenerate and this 
is not possible. Nevertheless, the following 
relation applies
\begin{eqnarray}
F_{xy0}=\Re\{F_{xy}\}-\Re\{F_{yx}\}\equiv 0\ .
\end{eqnarray}
$F_{xy0}$ may be then used to asses the reliability 
of the harmonic analysis, as far as coupling is 
concerned, in the same way $F_0$ of 
Eq.~\eqref{eq:F0} does it for the sextupolar 
analysis.

{\sl\bf  Normal sextupole CRDTs. } Normal sextupoles excite 
several spectral lines. As for the coupling case, ORDTs 
$g_{jklm}$ are replaced by first-order RDTs $f_{jklm}^{(1)}$. 
The potential term proportional to $x^3$ excites the lines 
$H(\pm2,0)$ and $H(0,0)$. The latter represents an offset 
and is not considered an Observable, as it contains 
dispersive terms, BPM offsets and contributions 
from other RDTs. By making use of 
Eq.~\eqref{eq:fft_Real} and selecting in 
Eq.~\eqref{e:tilde_x1} the RDTs exciting 
$H(\pm2,0)$, the following harmonics are generated
\begin{eqnarray}\label{e:tilde_x3}
&&4I_x\Im\left\{3f_{3000}^{(1)}e^{-2i(2\pi Q_xN+\psi_{x0})]}+
		     f_{1200}^{(1)}e^{2i(2\pi Q_xN+\psi_{x0})]} 
	      \right\} =\nonumber \\
&&\frac{(2I_x)}{i}\left[
	\left(3f_{3000}^{(1)}  -f_{1200}^{(1)*}\right)e^{-2i(2\pi Q_xN+\psi_{x0})}-
\right. \nonumber \\ && \hskip 1.1cm\left.
	\left(3f_{3000}^{(1)*}-f_{1200}^{(1)}  \right)e^{ 2i(2\pi Q_xN+\psi_{x0})}
	\right]=\nonumber \\
&&\frac{(2I_x)}{i}\left[
	F_{NS3}e^{-2i(2\pi Q_xN+\psi_{x0})}-
	c.c.
	\right]\ ,
\end{eqnarray}
where the corresponding sextupolar CRDT is 
defined as $F_{NS3}=3f_{3000}^{(1)}-f_{1200}^{(1)*}$ and 
excites the spectral line is $H(-2,0)$, whose 
amplitude and phase are then
\begin{eqnarray}\label{e:H-20}
\left\{
\begin{aligned}
|H(-2,0)|            &=(2I_x)|F_{NS3}| \\
\hbox{arg}\{H(-2,0)\}&=q_{F_{NS3}}+\frac{3}{2}\pi-2\psi_{x0}
\end{aligned}
\right.   \ . 
\end{eqnarray}
As for the coupling case, the tune line is 
used to extract amplitude and phase of 
$F_{NS3}$:
\begin{eqnarray}\label{e:F_NS3}
\left\{
\begin{aligned}
F_{NS3}    &=3f_{3000}^{(1)} -f_{1200}^{(1)*} \\
|F_{NS3}|  &=|H(-2,0)|/(4|H(1,0)|^2) \\
q_{F_{NS3}}&=\hbox{arg}\{H(-2,0)\}-\frac{3}{2}\pi+2\hbox{arg}\{H(1,0)\}
\end{aligned}
\right.   .\hskip 1cm
\end{eqnarray}

The mixing potential term proportional to 
$xy^2$ excite several lines: $H(0,\pm2)$ 
in the horizontal spectrum, $V(\pm1,\pm1)$ 
and $V(\pm1,\mp1)$ in the vertical. 
By repeating the same algebra in the sums of 
Eqs.~\eqref{e:tilde_x1} and~\eqref{e:tilde_y1} 
the following relations are derived: 
\begin{eqnarray}\label{e:F_NS2}
\left\{
\begin{aligned}
F_{NS2}    &=f_{1020}^{(1)} -f_{0120}^{(1)} \\
|F_{NS2}|  &=|H(0,-2)|/(4|V(0,1)|^2) \\
q_{F_{NS2}}&=\hbox{arg}\{H(0,-2)\}-\frac{3}{2}\pi+2\hbox{arg}\{V(0,1)\}
\end{aligned}
\right.   ,\hskip 1cm 
\end{eqnarray}
\begin{eqnarray}\label{e:F_NS1}
\left\{
\begin{aligned}
F_{NS1}    &=2f_{1020}^{(1)} -f_{0111}^{(1)*} \\
|F_{NS1}|  &=|V(-1,-1)|/(4|H(1,0)||V(0,1)|) \\
q_{F_{NS1}}&=\hbox{arg}\{V(-1,-1)\}-\frac{3}{2}\pi\ +\\
           &\hskip0.5cm\hbox{arg}\{H(1,0)\}+\hbox{arg}\{V(0,1)\}
\end{aligned}
\right.   ,\hskip 1.6cm
\end{eqnarray}
\begin{eqnarray}\label{e:F_NS0}
\left\{
\begin{aligned}
F_{NS0}    &=2f_{0120}^{(1)*} -f_{0111}^{(1)} \\
|F_{NS0}|  &=|V( 1,-1)|/(4|H(1,0)||V(0,1)|) \\
q_{F_{NS0}}&=\hbox{arg}\{V( 1,-1)\}-\frac{3}{2}\pi\ -  \\
           &\hskip0.5cm\hbox{arg}\{H(1,0)\} +\hbox{arg}\{V(0,1)\}
\end{aligned}
\right.    .\hskip 1.9cm
\end{eqnarray}

It is worthwhile noticing that the choice 
has been made here to select those lines 
laying within 0 and 0.5 in tune units, starting 
from betatron tunes of the ESRF storage ring, 
whose fractional parts are $Q_x=0.44$ and 
$Q_y=0.39$: $H(-2,0)$ appears at $1-2Q_x=0.12$, 
$H(0,-2)$ at $1-2Q_y=0.22$, $V(-1,-1$) at 
$1-Q_x-Q_y=0.17$ and $V(1,-1)$ at $Q_x-Q_y=0.05$. 
This choice is however arbitrary. Their complex 
conjugates may be used as well, $V(-1,1)$ instead of 
$V(1,-1)$ for instance, on condition to modify 
accordingly the computation of the CRDT phase. 

{\sl\bf Skew sextupole CRDTs. } 
The potential term proportional to $y^3$ excites 
the lines $V(0,\pm2)$ and $V(0,0)$ (not used). 
After selecting the RDTs in Eq.~\eqref{e:tilde_y1} 
exciting these lines and repeating the same algebra 
carried out for the normal sextupole, the following 
harmonics are generated
\begin{eqnarray}\label{e:tilde_y3}
&&4I_y\Im\left\{3g_{0030}e^{-2i(2\pi Q_yN+\psi_{y0})]}+
		     g_{0012}e^{2i(2\pi Q_yN+\psi_{y0})]} 
	      \right\} =\nonumber \\
&&\frac{(2I_y)}{i}\left[
	\left(3g_{0030}  -g_{0012}^*\right)e^{-2i(2\pi Q_yN+\psi_{y0})}-
\right. \nonumber \\ && \hskip 1.1cm\left.
	\left(3g_{0030}^*-g_{0012}  \right)e^{ 2i(2\pi Q_yN+\psi_{y0})}
	\right]=\nonumber \\
&&\frac{(2I_y)}{i}\left[
	F_{SS3}e^{-2i(2\pi Q_yN+\psi_{y0})}-
	c.c.
	\right]\ .
\end{eqnarray}
The corresponding skew sextupolar CRDT reads 
$F_{SS3}=3g_{0030}^*-g_{0012}^*$. $F_{SS3}$ 
excites the spectral line $V(0,-2)$, whose 
amplitude and phase are then

\begin{eqnarray}\label{e:V0-2}
\left\{
\begin{aligned}
|V(0,-2)|            &=(2I_y)|F_{SS3}| \\
\hbox{arg}\{V(0,-2)\}&=q_{F_{SS3}}+\frac{3}{2}\pi-2\psi_{y0}
\end{aligned}
\right.   \ , 
\end{eqnarray}
yielding

\begin{eqnarray}\label{e:F_SS3}
\left\{
\begin{aligned}
F_{SS3}    &=3g_{3000} -g_{1200}^* \\
|F_{SS3}|  &=|V(0,-2)|/(4|V(1,0)|^2) \\
q_{F_{SS3}}&=\hbox{arg}\{V(0,-2)\}-\frac{3}{2}\pi+2\hbox{arg}\{V(0,1)\}
\end{aligned}
\right.   .\hskip 1cm
\end{eqnarray}
The mixing potential term proportional to 
$x^2y$ excites several lines: $V(\pm2,0)$ 
in the vertical spectrum, $H(\pm1,\pm1)$ 
and $H(\pm1,\mp1)$ in the horizontal one. 
By repeating the same algebra in the sums of 
Eqs.~\eqref{e:tilde_x1} and~\eqref{e:tilde_y1},
the following relations are derived: \\
\begin{eqnarray}\label{e:F_SS2}
\left\{
\begin{aligned}
F_{SS2}    &=g_{2010,V}^* -g_{0210}^* \\
|F_{SS2}|  &=|V(-2,0)|/(4|H(1,0)|^2) \\
q_{F_{SS2}}&=\hbox{arg}\{V(-2,0)\}-\frac{3}{2}\pi+2\hbox{arg}\{H(1,0)\}
\end{aligned}
\right.   ,\hskip 1cm 
\end{eqnarray}
\begin{eqnarray}\label{e:F_SS1}
\left\{
\begin{aligned}
F_{SS1}    &=2g_{2010,H} -g_{1101}^* \\
|F_{SS1}|  &=|H(-1,-1)|/(4|H(1,0)||V(0,1)|) \\
q_{F_{SS1}}&=\hbox{arg}\{H(-1,-1)\}-\frac{3}{2}\pi\ +\\
           &\hskip0.5cm\hbox{arg}\{H(1,0)\}+\hbox{arg}\{V(0,1)\}
\end{aligned}
\right.   ,\hskip 1.6cm
\end{eqnarray}
\begin{eqnarray}\label{e:F_SS0}
\left\{
\begin{aligned}
F_{SS0}    &=g_{1110} -2g_{2001}^* \\
|F_{SS0}|  &=|H( 1,-1)|/(4|H(1,0)||V(0,1)|) \\
q_{F_{SS0}}&=\hbox{arg}\{H( 1,-1)\}-\frac{3}{2}\pi\ -  \\
           &\hskip0.5cm\hbox{arg}\{H(1,0)\} +\hbox{arg}\{V(0,1)\}
\end{aligned}
\right.    .\hskip 1.9cm
\end{eqnarray}

{\sl\bf Normal octupole CRDTs. } In Table~\ref{tab:g_jklm_NO}
 Ref.~\cite{Bartolini1} the list of spectral lines 
of the complex C-S signals $\tilde{x}-i\tilde{p}_x$ 
($\tilde{y}-i\tilde{p}_y$) 
excited by normal octupole terms $(x^4+y^4-6x^2y^2)$ 
is provided. The catalog is not complete, as two 
lines are missing, of which more later (the complete 
list may be found in Ref.\cite{Bartolini1}, though two 
terms $f_{1120}$ and $f_{1102}$ are mistakenly swapped). 
As far as the potential term $x^4$ is concerned the 
lines $H(\pm3,0)$ are excited. By making use of 
Eq.~\eqref{eq:fft_Real} and selecting in 
Eq.~\eqref{e:tilde_x1} the RDTs exciting this line, 
the following harmonics are generated
\begin{eqnarray}\label{e:tilde_x3}
&&2(2I_x)^{3/2}\times \nonumber \\
&& \Im\left\{4g_{4000}e^{-3i(2\pi Q_xN+\psi_{x0})]}+
		     g_{1300}e^{3i(2\pi Q_xN+\psi_{x0})]} 
	      \right\} =\nonumber \\
&&\frac{(2I_x)^{3/2}}{i}\left[
	\left(4g_{4000}  -g_{1300}^*\right)e^{-3i(2\pi Q_xN+\psi_{x0})}-
\right. \nonumber \\ && \hskip 1.1cm\left.
	\left(4g_{4000}^*-g_{1300}  \right)e^{ 3i(2\pi Q_xN+\psi_{x0})}
	\right]=\nonumber \\
&&\frac{(2I_x)^{3/2}}{i}\left[
	-F_{NO3}e^{3i(2\pi Q_xN+\psi_{x0})}-	c.c.
	\right]\ ,
\end{eqnarray}
where the CRDT is 
defined as $F_{NO3}=4g_{4000}^*-g_{1300}$ and 
excites the spectral line is $H(3,0)$, whose 
amplitude and phase are then
\begin{eqnarray}\label{e:H-20}
\left\{
\begin{aligned}
|H(3,0)|            &=(2I_x)^{3/2}|F_{NO3}| \\
\hbox{arg}\{H(3,0)\}&=q_{F_{NO3}}+\frac{\pi}{2}+3\psi_{x0}
\end{aligned}
\right.   \ ,
\end{eqnarray}
and
\begin{eqnarray}\label{e:F_NO3}
\left\{
\begin{aligned}
F_{NO3}    &=4g_{4000}^* -g_{1300} \\
|F_{NO3}|  &=|H(3,0)|/(8|H(1,0)|^3) \\
q_{F_{NO3}}&=\hbox{arg}\{H(3,0)\}-\frac{\pi}{2}-3\hbox{arg}\{H(1,0)\}
\end{aligned}
\right.   .\hskip 1cm
\end{eqnarray}
the spectral line $H(3,0)$ of the real C-S signal 
$\tilde{x}$ is used because for the ESRF tune 
working point this is the one laying within 0 and 
0.5\ . The vertical potential term $y^4$ excites 
the vertical lines $V(0,\pm3)$, resulting in 
\begin{eqnarray}\label{e:F_NO5}
\left\{
\begin{aligned}
F_{NO5}    &=4g_{0040}^* -g_{0013} \\
|F_{NO5}|  &=|V(0,3)|/(8|V(0,1)|^3) \\
q_{F_{NO5}}&=\hbox{arg}\{V(0,3)\}-\frac{\pi}{2}-3\hbox{arg}\{V(0,1)\}
\end{aligned}
\right.   .\hskip 1cm
\end{eqnarray}
As far as the mixing potential term $x^2y^2$ is 
concerned, the lines $H(\pm1,2)$ and $V(2,\pm1)$ 
are generated by four CRDTs:
\begin{eqnarray}\label{e:F_NO4}
\left\{
\begin{aligned}
F_{NO4}    &=2g_{2020,H}^* -g_{1102} \\
|F_{NO4}|  &=|H(1,2)|/(8|H(1,0)||V(0,1)|^2) \\
q_{F_{NO4}}&=           \hbox{arg}\{H(1,2)\}-\frac{\pi}{2} \\
           &\hskip0.5cm-\hbox{arg}\{H(1,0)\} -2\hbox{arg}\{V(0,1)\}
\end{aligned}
\right.   ,\hskip 1cm
\end{eqnarray}
\begin{eqnarray}\label{e:F_NO2}
\left\{
\begin{aligned}
F_{NO2}    &=2g_{2002} -g_{1120}^* \\
|F_{NO2}|  &=|H(-1,2)|/(8|H(1,0)||V(0,1)|^2) \\
q_{F_{NO2}}&=           \hbox{arg}\{H(-1,2)\}+\frac{\pi}{2} \\
           &\hskip0.5cm+\hbox{arg}\{H(1,0)\} -2\hbox{arg}\{V(0,1)\}
\end{aligned}
\right.   ,\hskip 1cm
\end{eqnarray}
\begin{eqnarray}\label{e:F_NO1}
\left\{
\begin{aligned}
F_{NO1}    &=2g_{0220} -g_{2011}^* \\
|F_{NO1}|  &=|V(2,-1)|/(8|H(1,0)|^2|V(0,1)|) \\
q_{F_{NO1}}&=           \hbox{arg}\{V(2,-1)\}+\frac{\pi}{2} \\
           &\hskip0.5cm-2\hbox{arg}\{H(1,0)\} +\hbox{arg}\{V(0,1)\}
\end{aligned}
\right.   ,\hskip 1cm
\end{eqnarray}
\begin{eqnarray}\label{e:F_NO0}
\left\{
\begin{aligned}
F_{NO0}    &=2g_{2020,V}^* -g_{0211} \\
|F_{NO0}|  &=|V(2,1)|/(8|H(1,0)|^2|V(0,1)|) \\
q_{F_{NO0}}&=           \hbox{arg}\{V(2,1)\}-\frac{\pi}{2} \\
           &\hskip0.5cm-2\hbox{arg}\{H(1,0)\} -\hbox{arg}\{V(0,1)\}
\end{aligned}
\right.   .\hskip 1cm
\end{eqnarray}
\clearpage
\ \\




\end{document}